\documentclass[mnsc,nonblindrev]{informs3_hide}
\OneAndAHalfSpacedXII % default for MS

\usepackage{subfigure}
\usepackage[english]{babel}
\usepackage[autostyle, english = american]{csquotes}
\MakeOuterQuote{"}
\usepackage{bbm}
\usepackage{bm}
\usepackage[colorlinks,linkcolor=blue, citecolor=blue]{hyperref}
\usepackage{hyperref}
\usepackage{cleveref}
\crefname{subsection}{subsection}{subsections}

\usepackage{eqnarray}

\newcommand{\eps}{\varepsilon}
\newcommand{\bI}{\mathbbm{1}}
\newcommand{\bE}{\mathbb{E}}
\newcommand{\bR}{\mathbb{R}}

\newcommand{\Ber}{\mathrm{Ber}}

\newcommand{\Bin}{\mathrm{Bin}}
\newcommand{\Pois}{\mathrm{Pois}}
\newcommand{\IID}{\mathrm{IID}}
\newcommand{\InnerLP}{\mathrm{innerLP}}
\newcommand{\IIDLP}{\mathrm{iidLP}}

\newcommand{\LP}{\mathsf{LP}}
\newcommand{\LPmon}{\LP^{\mathsf{mon}}}

\newcommand{\LPre}{\LP^{\mathsf{relax}}}

\newcommand{\OST}{\mathsf{OST}}
\newcommand{\ST}{\mathsf{ST}}
\newcommand{\DP}{\mathsf{DP}}
\newcommand{\Proph}{\mathsf{Proph}}
\newcommand{\ExAnte}{\mathsf{ExAnte}}

\newcommand{\cI}{\mathcal{I}}
\newcommand{\cK}{\mathcal{K}}
\newcommand{\cP}{\mathcal{P}}
\newcommand{\vG}{\mathbf{G}}
\newcommand{\vx}{\mathbf{x}}
\newcommand{\vy}{\mathbf{y}}

\newcommand{\otau}{\overline{\tau}}
\newcommand{\utau}{\underline{\tau}}

\usepackage[dvipsnames]{xcolor}

\newcommand{\blue}[1]{{\leavevmode\color{blue}#1}}

\usepackage{color}              % Need the color package
\usepackage{color-edits}

\usepackage{xparse,rotating}

\NewDocumentEnvironment{myproof}{o}
{\IfNoValueTF{#1}{\paragraph{{Proof.} }} {\paragraph{{#1.} }} }
{\hfill$\Halmos$}

\usepackage{natbib}
\usepackage{bbm,multirow,multicol}
 \bibpunct[, ]{(}{)}{,}{a}{}{,}%
 \def\bibfont{\small}%
\TheoremsNumberedThrough     % Preferred (Theorem 1, Lemma 1, Theorem 2)
\EquationsNumberedThrough    % Default: (1), (2), ...
\begin{document}
%%%%%%%%%%%%%%%%

%\setcounter{page}{1}
%\DOI{10.1287/xxxx.0000.0000}%

\RUNAUTHOR{Jiang, Ma and Zhang}

\RUNTITLE{Tightness without Counterexamples}

\TITLE{Tightness without Counterexamples: A New Approach and New Results for Prophet Inequalities
% Tight Prophet Inequalities without Lucky Coincidences
}

\ARTICLEAUTHORS{%
\AUTHOR{$\text{Jiashuo Jiang}^\dagger$ \quad $\text{Will Ma}^\ddagger$ \quad $\text{Jiawei Zhang}^\S$}

\AFF{\  \\
$\dagger~$Department of Industrial Engineering \& Decision Analytics, Hong Kong University of Science and Technology, \EMAIL{jsjiang@ust.hk}\\
$\ddagger~$Decision, Risk, and Operations Division, Graduate School of Business, Columbia University, \EMAIL{wm2428@gsb.columbia.edu}\\
$\S~$Department of Technology, Operations \& Statistics, Stern School of Business, New York University, \EMAIL{jz31@stern.nyu.edu}\\
}
}

\ABSTRACT{

Prophet inequalities consist of many beautiful statements that establish tight performance ratios between online and offline allocation algorithms.
Typically, tightness is established by constructing an algorithmic guarantee and a worst-case instance separately, whose bounds match as a result of some "ingenuity".
In this paper, we instead formulate the construction of the worst-case instance as an optimization problem, which directly finds the tight ratio without needing to construct two bounds separately.

Our analysis of this optimization problem involves identifying structure in a new "Type Coverage" dual problem.
It can be seen as akin to the celebrated Magician and OCRS (Online Contention Resolution Scheme) problems, except more general in that it can also provide tight ratios relative to the optimal offline allocation, whereas the earlier problems only establish tight ratios relative to the ex-ante relaxation of the offline problem.
% that the realization can be one of many ordered types, instead of being binary (either "active" or not).

Through this analysis, our paper provides a unified framework that derives new prophet inequalities and recovers existing ones, with our principal results being two-fold.
First, we show that the "oblivious" method of setting a static threshold due to \cite{chawla2020static}, surprisingly, is best-possible among all static threshold algorithms, under any number $k$ of selection slots.
We emphasize that this result is derived without needing to explicitly find any counterexample instances.
This implies
% for the first time
the tightness of the asymptotic convergence rate of $1-O(\sqrt{\log k/k})$ for static threshold algorithms
% , which dates back to
from
\cite{hajiaghayi2007automated}.
% , and establishes a separation with the convergence rate of adaptive algorithms,
%which is $1-\Theta(\sqrt{1/k})$ due to \cite{alaei2014bayesian}.
{Turning to the IID setting, our second principal result is to use our framework to characterize the tight guarantee (of adaptive algorithms) under any number $k$ of selection slots and any fixed number of agents $n$.}
% The tight guarantee has been previously characterized only under $k=1$ by \cite{correa2017posted},
% and our guarantees for $k>1$ exceed the state-of-the-art from the free-order setting due to \cite{beyhaghi2021improved}.

% We establish similar "no separation" results for static thresholds in the IID setting, which although previously known, required the construction of complicated counterexamples.
% Finally, our framework and in particular our Type Coverage problem yields a simplified derivation of the tight 0.745 ratio when $k=1$ in the IID setting.
}



\maketitle
%%%%%%%%%%%%%%%%%%%%%%%%%%%%%%%%%%%%%%%%%%%%%%%%%%%%%%%%%%%%%%%%%%%%%%
\section{Introduction}
Prophet inequalities describe fundamental problems in online decision making. In {the multi-unit variant of prophet inequalities}, there is a decision maker with $k$ slots and there are $n$ agents arriving sequentially.
Each agent has a valuation drawn from a known distribution, which will be convenient for us to denote in the following discrete form.
{(We note that general distributions can be approximated arbitrarily closely using discrete distributions.)}
There are $m$ possible \textit{types} and each agent $i$ belongs to one type $j$ independently with a given probability $p_{i,j}$. The types are indexed so that 1 is best and $m$ is worst, where the valuation (or "value") for an agent of type $j$ is given by $r_j$, with $r_1\ge\cdots\ge r_m\ge0$. After the type of an incoming agent is revealed, the reward of accepting this agent is known and the decision maker has to decide immediately whether to spend one slot to collect this reward, or reject the agent. The decision has to be made on-the-fly; i.e., the decision is irrevocable and cannot look ahead at the types of the agents who have not yet arrived. The performance of the decision maker is measured as the total value collected in expectation.

The guarantee of an online policy for the decision maker is defined by the ratio of its performance to that of a benchmark.
The typical benchmark for prophet inequalities is the expected value collected by a \textit{prophet}, who sees in advance the types of all agents and fills the slots prioritizing the agents with highest values. By comparing with the prophet, the classical work \citep{krengel1978semiamarts} obtains the tight guarantee of $1/2$ when there is one slot (i.e.\ $k=1$).  \cite{correa2017posted} show that the tight guarantee improves to $0.745$ under the $\IID$ special case, where the value distribution is identical for each agent. Another widely used benchmark is the \textit{ex-ante relaxation}, which {is obtained from taking expectation over the type distributions and} can be formulated via linear programming (LP) %\Willcomment{this sentence not grammatically correct}
({see formal definition in \Cref{def:ExAnte}}). The benefits of the ex-ante benchmark are twofold: (i) it is generally easier to analyze \citep{alaei2014bayesian} and possesses special structure such that the optimal policy and tight guarantee can be derived even when $k>1$ \citep{jiang2022tight}; (ii) it enables a decomposition technique where guarantees for a single resource can easily extend to online allocation problems with multiple resources (i.e.\ multiple types of slots) \citep{alaei2012online}, even allowing for very general allocation procedures involving customer choice \citep{gallego2015online,ma2021dynamic}.

Various online policies have been developed in the literature in order to achieve a good performance guarantee relative to the benchmarks. %For example, for the prophet benchmark, when there is one slot, \citenum{correa2017posted} develop an online policy that achieves the optimal guarantee under the $\IID$ special case. For the Ex-Ante benchmark, \citenum{jiang2022tight} derive the optimal guarantees even when there can be arbitrarily many slots. However, the policies used to achieve the guarantees in \cite{correa2017posted} and \cite{jiang2022tight} are necessarily dynamic,
% motivated from general dynamic programming and are dynamic in essence,
%i.e., the decision depends both on the realization of the current agent's value and the number of currently remaining slots.
Some of the tight performance guarantees are derived from optimal adaptive policies {found via dynamic programming}, where the decision depends both on the realization of the current agent's value and the number of remaining agents and slots.%\Willdelete{; see, e.g., \citenum{correa2017posted} and \citenum{jiang2022tight}}.
Simpler but more restrictive \textit{static} threshold policies also enjoy strong performance guarantees \citep{hajiaghayi2007automated}. A static threshold policy must fix a threshold value at the start and accept any agent with valuation exceeding the threshold, while there are empty slots. The further distinction of an \textit{oblivious} static threshold policy was introduced in \cite{chawla2020static}, where the threshold is allowed to depend on the probabilities $\{p_{i,j}\}_{\forall i\in[n], \forall j\in[m]}$ but not the values $\{r_j\}_{\forall j\in[m]}$.

The goal of this paper is to characterize tight performance guarantees for general online policies as well as (oblivious) static threshold policies, with respect to either the prophet benchmark or the ex-ante benchmark, in both the IID and non-IID settings.
%\Willdelete{We denote by $I$ the problem instance and $\mathcal{I}^{\IID}_{k,n}$ (resp, $\mathcal{I}_{k,n}$) the collection of all IID (resp. non-IID) problem instances with $n$ agents and $k$ slots.}
We denote by $\DP(I)$ the performance of the optimal dynamic program, $\ST(I)$ the performance of the optimal non-oblivious threshold policy, $\OST(I)$ the performance of the optimal oblivious threshold policy, $\Proph(I)$ the prophet benchmark, and $\ExAnte(I)$ the ex-ante benchmark, all for the problem instance $I$ (see \Cref{sec:problemclass} for formal definitions of these concepts).

\subsection{Main Technique} \label{sec:introNewContr}
The main difference between our approach and existing ones is that we formulate the problem of finding the worst-case instance as an optimization problem, while existing approaches rely on constructing an algorithmic guarantee and a worst-case instance separately.  {Our approach is similar in spirit to what is done in the pioneering work of \cite{buchbinder2014secretary} for the secretary problem.
In what follows, we elaborate on how we do this for the prophet problem and the new results it leads to for the literature.}

%The main difference between our approach and existing ones is that we try to ``cover'' each type of agents, instead of agents themselves. To be specific, for each type $j$ or better of agents, we show that the expected number of acceptance by our policy is at least $\theta$ times the expected number of acceptance by the benchmark, and this $\theta$ is our performance guarantee. In contrast, previous approaches consider how each individual agent is accepted. This new idea of ``covering'' the type allows us to achieve tight guarantee relative to both the prophet and the Ex-Ante benchmark. In what follows, we describe our approach, illustrate our new idea, and compare with the existing ones.
%\wnote{Jiashuo, I don't think the paragraph here flows logically.  Maybe in hindsight it is not needed.  I was thinking we could re-emphasize here WHY our approach of formulating the worst-case instance as an optimization problem, instead of separately deriving two guarantees, is different from what's done in the papers above.  Maybe one sentence (merged with the previous short paragraph) will do.  I leave it up to you.}

We introduce a new LP framework that explicitly computes the \textit{tight} guarantee of a class of online policy versus a benchmark. Note that in order to compute the guarantee of a policy, one needs to minimize the ratio of the expected value collected by the policy divided by the benchmark, over all instances, where an instance is described by $\{p_{i,j}, r_j\}_{i\in[n], j\in[m]}$. The new idea of our framework is to break down this minimization problem into two stages. Given a fixed number of slots $k$ and agents $n$, we first minimize over $m$ and the type distributions $\{p_{i,j}\}_{i\in[n], j\in[m]}$, and we then minimize over the type valuations $\{r_j\}_{j\in[m]}$.  For example, if we are computing the tight guarantee for adaptive online algorithms relative to the prophet over general non-IID instances, then our formulation would be
\begin{equation} \label{eqn:intro}
\inf_{m,\{p_{i,j}\}_{i\in[n], j\in[m]}}\inf_{\{r_j\}_{j\in[m]}}\frac{\DP(I)}{\Proph(I)}.
\end{equation}
The inner optimization problem over $\{r_j\}_{j\in[m]}$ can be formulated as an LP. The dual of this LP gives us a new problem that is defined for each $m$ and $\{p_{i,j}\}_{i\in[n], j\in[m]}$, whose optimal solution has a nice structure. This structure enables us to derive multiple new results, and recover several existing results in a clean way.

\paragraph{Type Coverage problem.}
The new dual LP can be interpreted as a ``Type Coverage'' problem, defined as follows. For each type $j$, let $Q_j$ denote the expected number of agents of \textit{type $j$ or better} accepted by the benchmark.
For example, if the benchmark is the prophet, then
\begin{equation} \label{eqn:introQj}
Q_j=\bE[\min\{\sum_{i=1}^n\Ber\Big(\sum_{j'=1}^jp_{i,j'}\Big),k\}],
\end{equation}
where $\Ber(p)$ denotes an independent Bernoulli random variable of mean $p$.
The RHS of~\eqref{eqn:introQj} then counts the total agents among $i=1,\ldots,n$ who realize to a type $j'$ whose value is at least as good as that of type $j$, and truncates this total by $k$ to indicate the number of such agents accepted by the prophet.
Finally, for an online policy to achieve a guarantee of $\theta$ in this dual Type Coverage problem, for every type $j\in[m]$,
the number of agents of type $j$ or better accepted by the online policy must be at least $\theta\cdot Q_j$.

Our framework shows that the guarantee under the worst-case instance in the original problem is equivalent to the guarantee under the worst-case instance for the Type Coverage problem, where now the parameters $r_j$ have been eliminated.
If the original prophet problem is restricted to an \textit{oblivious} static threshold policy, then the corresponding dual restriction is to have \textit{any} static threshold policy.
Meanwhile, if the original problem allowed non-oblivious static thresholds, then the dual allows \textit{randomized} static thresholds (which can achieve a strictly greater $\theta$ for the Type Coverage problem, as we show later in \Cref{prop:STbetterOST}).
These correspondences in the restrictions, which are not a priori obvious, are consequences of our framework and formalized in \Cref{sec:mainIdeaST}.
{Note that a $1/e$ ``oblivious'' is developed in \cite{fu2022oblivious}. However, the meaning of ``oblivious'' in their paper is different from ours. To be specific, the OCRS in \cite{fu2022oblivious} is oblivious to $\bm{x}$, where $\bm{x}$ is a vector of the probabilities for each query to be ``active'' in the OCRS. One common way to compute $\bm{x}$ is to solve the ex-ante relaxation, which actually requires the full distributional knowledge (including the densities and valuations) of each query. On the contrary, the ``oblivious'' in our paper means that our oblivious algorithm does not require knowing the valuations of the queries.}

% This holds for both IID and non-IID settings, and when restricting to subclasses of algorithms (although the analogues of (oblivious) static threshold policies are not the obvious ones in the Type Coverage problem; we elaborate on this in the main body).
Another interpretation of our approach in~\eqref{eqn:intro} is that we are decoupling the adversary's optimization into an outer problem over \textit{ordinal} distributions (of $m$ ranked types)
and an inner problem over \textit{cardinal} valuations consistent with these ordinal rankings.
General static thresholds can be set after the adversary decides these cardinal valuations, whereas oblivious static thresholds must be set before the adversary decides them.

\paragraph{Comparison: Magician/OCRS (Online Contention Resolution Scheme) problems.}
Magician (\cite{alaei2014bayesian}) and OCRS (\cite{feldman2021online}) are existing auxiliary problems used to derive prophet inequalities against the ex-ante benchmark. In these problems, each agent has an \textit{active} probability and must be guaranteed to be accepted with a probability at least $\theta$ conditional on being active, by the online policy. The goal is to achieve the maximum $\theta$.
% It has been shown in \cite{jiang2022tight} that the optimal guarantees for the $k$-unit Magician/OCRS problems are actually the same,
% % Ex-Ante prophet inequalities are equivalent to the optimal $\theta$ for the Magician/OCRS problems, which illustrates the power of Magician/OCRS problems for
% leading to tight Ex-Ante prophet inequalities.
{It has been shown in \cite{lee2018optimal} that the worst-case for ex-ante prophet inequalities always occurs on weighted Bernoulli distributions and using LP duality, they show that the tight guarantee matches that of a Magician/OCRS problem.}

Although ex-ante prophet inequalities are stronger and hence imply non-ex-ante prophet inequalities, no analogues of Magician/OCRS have been known for deriving \textit{tight non-ex-ante} prophet inequalities, until now.
Our new Type Coverage problem essentially generalizes the Magician/OCRS problems to allow for agents who take \textit{non-binary} states, which leads to a formulation of tight guarantees for non-ex-ante prophet inequalities.
Our Type Coverage problem is also more general in that it can recover the Magician/OCRS problems when we use ex-ante as the benchmark in our framework, as we derive in \Cref{sec:magicianOCRS}.

\subsection{New Results}

Beyond our new framework that derives tight guarantees without counterexamples, we would like to summarize our contributions in terms of new statements that were previously unknown.

\begin{enumerate}
\item In the non-IID setting, we show that the guarantees for static threshold policies from \cite{chawla2020static} are tight for every $k$ {(\Cref{thm:chawlaTight} and \Cref{cor:bernOpt})}, and moreover grow as $1-\Theta(\sqrt{\log k/k})$ when $k\to\infty$ {(\Cref{prop:StaticPolicy})}.  This establishes
% for the first time
% \footnote{Previous versions of some papers had cited \citenum{ghosh2016optimal} as already establishing the upper bound of $1-\Omega(\sqrt{\log k/k})$ for static threshold policies, but to the best of our knowledge, this was based on a misunderstanding and the citations have been retracted in the updated versions.}
the tightness of the asymptotic guarantee of $1-O(\sqrt{\log k/k})$ for static threshold policies from \cite{hajiaghayi2007automated}, and also separates static threshold policies from the asymptotic guarantee of $1-O(\sqrt{1/k})$ for adaptive policies due to \cite{alaei2014bayesian}.
\item Specializing our framework to the IID setting, we characterize the tight guarantee $\gamma_{k,n}$ of adaptive policies for any number of slots $k$ and fixed number of agents $n$.  Although our characterization is through a semi-infinite LP, our discretization procedure (\Cref{prop:DiscreteIID}) allows to numerically compute $\gamma_{k,n}$ within any prespecified level of accuracy.
\cite{hill1982comparisons} has previously characterized $\gamma_{k,n}$ in the case where $k=1$, with further work analytically determining $\lim_{n\to\infty}\gamma_{1,n}$  (\cite{samuel1984comparison}, \S2) and $\inf_{n\ge1}\gamma_{1,n}\approx0.745$ (\cite{correa2017posted}).  Although we are unable to analytically determine $\lim_{n\to\infty}\gamma_{k,n}$ or $\inf_{n\ge1}\gamma_{k,n}$ for general $k$, our method allows to numerically observe that $\lim_{n\to\infty}\gamma_{k,n}=\inf_{n\ge1}\gamma_{k,n}$, and we also display the numerical values of $\gamma_{k,n}$ at $n=1000$ at the end of \Cref{sec:DPProphIID}.  Very recently, $\lim_{n\to\infty}\gamma_{k,n}$ has been analytically determined for general $k$ using a variant of our framework (see "Concurrent and Subsequent work" in \Cref{sec:relatedWork}).  It remains a challenging open problem to analytically verify for all $k>1$ our numerical observation that $\lim_{n\to\infty}\gamma_{k,n}=\inf_{n\ge1}\gamma_{k,n}$, which would resolve the open question of $k$-unit IID prophet inequality.
\item In the non-IID setting, we show that non-oblivious static thresholds are better than oblivious ones for some type distributions (\Cref{prop:STbetterOST}), but that the four ratios $\ST/\Proph$, $\ST/\ExAnte$, $\OST/\Proph$, and $\OST/\ExAnte$ are all equivalent after taking an infimum over type distributions (\Cref{thm:STequalOST}).
In the IID setting, we show that non-oblivious static thresholds are never better than oblivious ones (\Cref{thm:iidSTNoBetter}), and that the five ratios $\DP/\ExAnte$, $\ST/\Proph$, $\ST/\ExAnte$, $\OST/\Proph$, and $\OST/\ExAnte$ are all equivalent (\Cref{thm:iidST} and \Cref{cor:OSTandSTsameInIID}).  These relationships are summarized in a table in \Cref{apx:worstCaseRatioRelationships}.
Our framework also seamlessly implies tight guarantees for all combinations of $k$ and $n$, some of which were previously only proven in the limit as $n\to\infty$.
\end{enumerate}

\subsection{Further Related Work}~ \label{sec:relatedWork}

\paragraph{Multi-unit Prophet Inequalities.} Since first posed by \cite{krengel1978semiamarts}, prophet inequalities have been extensively studied (see a survey by \cite{correa2019recent}), and in recent years, there has been a lot of working studying prophet inequalities with various feasibility constraints over the agents that can be accepted, including general matroid constraints (\cite{kleinberg2012matroid}), knapsack constraints (\cite{feldman2021online,jiang2022tight}), and matching constraints (\cite{ezra2022prophet}).
Our problem poses a uniform matroid constraint for the agents, and our problem is also referred to as \textit{multi-unit} prophet inequalities (\cite{hajiaghayi2007automated}). In a seminal work, \cite{alaei2014bayesian} obtains a $(1-\frac{1}{\sqrt{k+3}})$-guarantee when there are $k$ slots in total. \cite{jiang2022tight} improves the guarantee for every $k>1$ and shows that the improved guarantee is tight with respect to the Ex-Ante benchmark. Our work shows how to formulate an LP to compute the tight ratio for given type distributions even with respect to the prophet benchmark.

% brustle2022competition

%By minimizing the value distribution densities, we are able to obtain the tight guarantee, not only with respect to the Ex-Ante benchmark, but also with respect to the prophet benchmark. Our framework also enables us to study various types of policies, including the (oblivious) static threshold policies, at the same time, while the approaches developed in \citenum{alaei2014bayesian, jiang2022tight} are to analyze dynamic policies.

\paragraph{Static Thresholds.} Static threshold policies are important policies for both prophet inequalities and online pricing, and have been extensively studied. In particular, \cite{hajiaghayi2007automated} shows that a static threshold policy enjoys a guarantee of $1-O(\sqrt{\frac{\log k}{k}})$, where $k$ is the number of slots.
% , and the loss of $\Theta(\sqrt{\frac{\log k}{k}})$ has been shown to be asymptotically tight by \citenum{ghosh2016optimal}.
In the $\IID$ setting, \cite{yan2011mechanism} shows that a static threshold policy achieves a guarantee of $1-\frac{k^k}{e^kk!}$, which is tight with respect to the Ex-Ante benchmark. This guarantee has been extended in \cite{arnosti2021tight} to the prophet secretary setting where the agents arrive in a random order, and has been shown to be tight among the static threshold policies. Static policy has also been used in online pricing problem with strategic customers to achieve a guarantee of $1-1/e$ (\cite{chen2019efficacy}). Beyond explicitly computing the tight guarantees for the optimal static threshold policies, our framework also reveals that the optimal static threshold policy can be value-oblivious under various settings, which implies that only the type distributions are required to be known when designing the optimal static threshold policy. The benefits of being oblivious have been illustrated in \cite{arnosti2021tight} when only a monotone transformation of values (rather than the values themselves) can be observed.

\paragraph{Concurrent and Subsequent work.}
\cite{perez2022iid} study prophet inequalities for a class of policies whose power lies between $\ST$ (static thresholds) and $\DP$ (the optimal dynamic program), and concurrently come upon a \textit{quantile-based LP for prophet inequality} that is equivalent to our semi-infinite formulation in the IID setting, modulo the following.
\begin{enumerate}
\item They have an extra set of variables ($\eta$) in the dual.  We avoid this by comparing the algorithm to the prophet on the number of agents accepted with quantile $q$ \textit{or better}, when taking the dual.
\item Their dual variables prescribe a distribution over quantiles.  We show that randomized quantiles are unnecessary, and hence avoid having an infinite number of variables in the dual.
% \item Their formulation is for continuous distributions, whereas our formulation is for discrete distributions.  However, this is just a cosmetic difference for presentation purposes; in \Cref{sec:continuousDist} we show how to extend our results to continuous distributions.
\end{enumerate}

For these reasons, we believe our formulation to be simpler.
% In fact, we can derive their guarantees for prophet inequality with a small number of threshold changes using a \textit{much shorter proof}, as shown in \Cref{sec:limitedFlexibility}.
However, their formulation and in particular the randomized quantiles apparently better leads to an analytical determination of $\lim_{n\to\infty}\gamma_{k,n}$ (something we do not achieve), in the subsequent work (\cite{brustle2024splitting}).

Finally, we note that \cite{perez2022iid} focus on the IID setting, and derive the quantile-based LP from a (worst-case) minimax formulation of the IID prophet problem and a linearization reformulation using the inverse of the CDF. In contrast,
our work also considers independent non-identical distributions and our LPs are derived through a different, "two-step" duality approach, giving rise to our Type Coverage problem over quantiles.
Originally, quantile-based arguments have been used for prophet inequality in \cite{correa2021prophet}, and the idea of using LP duality to characterize tight guarantees was done for the secretary problem in \cite{buchbinder2014secretary}.
% \cite{perez2022iid} derives the quantile-based LP from a , while we utilize a ``two-step'' duality approach.

% We also use this quantile-based LP to study a different problem than \citenum{perez2022iid} (multiple units, instead of limited threshold changes), and come upon it (our "two-step" duality approach discussed in , instead of explicitly constructing distributions).

% There are other LP frameworks based on duality developed for prophet inequalities, specifically for the limited flexibility setting \citenum{li2022query,perez2022iid}, where the decision maker sets at most $k'$ thresholds for selecting agents, for some fixed $k'$.
% ekbatani2024prophet
% The fundamental limits of static threshold policies have also been discussed in \cite{ganesh2024fundamental}.

\section{Problem Classes Considered and General Framework}\label{sec:problemclass}

We consider the prophet inequality problem where $k$ out of $n>k$ agents can be accepted.
That is, initially there are $k$ \textit{slots}, and $n$ agents arrive in order $i=1,\ldots,n$, each with a \textit{valuation} $R_i\ge0$ that is revealed only when agent $i$ arrives.
One must then immediately decide whether to use a slot to accept agent $i$, or to reject agent $i$ forever, with this decision being irrevocable.
Once all $k$ slots have been used, no further agents can be accepted.
The valuations $R_i$ are drawn independently at random from known distributions.
The objective is to make accept/reject decisions on-the-fly, in a way that maximizes the expected sum of valuations of accepted agents.

We assume that each $R_i$ is drawn from a discrete distribution as follows.
There is a universe of $m$ possible valuations, sorted in the order $r_1\ge\cdots\ge r_m\ge0$.
For each agent $i$, we let $p_{ij}\ge0$ denote the probability that their valuation $R_i$ realizes to $r_j$, for all $j=1,\ldots,m$, with $\sum_{j=1}^mp_{ij}=1$.
We refer to index $j$ as the \textit{type} of an agent, with smaller indices $j$ said to be \textit{better}.
For simplicity, we also assume that agents arrive in exactly the order $i=1,\ldots,n$, which is known in advance.
Although many of the algorithms we discuss hold under certain adversarial manipulations of the arrival order, we do not attempt to make such distinctions comprehensively.

\begin{definition}[Instance, IID vs.\ Non-IID]
An \textit{instance} $I$ of the prophet inequality problem is defined by the number of slots $k$, agents $n$, types $m$, the valuations $r_1,\ldots,r_m$, and the probability vectors $(p_{1j})_{j=1}^m,\ldots,(p_{nj})_{j=1}^m$ under the fixed arrival order $i=1,\ldots,n$.
We let $\cI_{k,n}$ denote the class of all instances with $k$ slots and $n$ agents, with $\cI_k:=\bigcup_{n=1}^\infty \cI_{k,n}$.
If $p_{ij}$ is identical to some $p_j$ across all agents $i$, for each type $j$, then we say that the instance is \textit{IID} (independent and identically distributed),
and we let $\cI^\IID_{k,n}$ denote the class of all IID instances with $k$ slots and $n$ agents, with $\cI^\IID_k:=\bigcup_{n=1}^\infty\cI^\IID_{k,n}$.
We sometimes refer to general instances
as \textit{non-IID}.
\end{definition}

Under the assumptions mentioned earlier, given an instance, the optimal policy for making accept/reject decisions is easy to compute using dynamic programming (DP).
However, as discussed in the Introduction, there is great interest and applicability in comparing the performance of DP to that of a \textit{prophet} who knows the realizations of $R_i$ in advance, over different instances.
Furthermore, one can compare against the stronger \textit{ex-ante} benchmark in which the number of agents of each type always equals its expectation.
We now formally define these benchmarks.

\begin{definition}[Prophet]
The prophet's performance on an instance $I$, denoted $\Proph(I)$, is the expected sum of the $k$ largest realized valuations.  Formally, letting $[n]$ denote the set $\{1,\ldots,n\}$,
\begin{align*}
\Proph(I)=\bE\left[\max_{S\subseteq[n],|S|\le k}\sum_{i\in S}R_i\right].
\end{align*}
\end{definition}

\begin{definition}[Ex-ante Relaxation]\label{def:ExAnte}
The ex-ante relaxation is defined by the following LP:
\begin{eqnarray*}
\ExAnte(I)=&\max & \sum_{j=1}^m r_ja_j \\
&\mathrm{s.t.\ } & \sum_{j=1}^m a_j \le k \\
& & 0\le a_j \le\sum_{i=1}^n p_{ij} \quad \quad \forall j\in[m]
\end{eqnarray*}
in which variable $a_j$ can be interpreted as the number of agents of type $j$ accepted.
\end{definition}

The following standard result is easy to show, holding because on every sample path, the prophet's selection of $k$ largest valuations forms a feasible solution to the ex-ante LP.

\begin{proposition}[Folklore] \label{prop:exAnteUB}
$\Proph(I)\le\ExAnte(I)$ for all instances $I$.
\end{proposition}

Having explained the benchmarks $\Proph(I)$ and $\ExAnte(I)$, we are now ready to define the notion of a prophet inequality.  Let $\DP(I)$ denote the expected sum collected by the optimal DP algorithm on an instance $I$.
Let $\cI$ denote a class of instances.
%We let $\cI$ denote a class of instances, usually consisting of all $I$ that have a particular number of slots $k$, sometimes restricted to be IID, and sometimes also restricted to have a particular number of agents $n$.

\begin{definition}\label{defn:guarantee}
A \textit{prophet inequality (resp.\ ex-ante prophet inequality)} is a statement of the form
\begin{align} \label{eqn:defProphetIneq}
\frac{\DP(I)}{\Proph(I)}\ge\alpha \qquad (\text{resp. }\frac{\DP(I)}{\ExAnte(I)}\ge\alpha) &&\forall I\in\cI
\end{align}
for some constant $\alpha\le1$.
%for some class of instances $\cI$ and $\alpha\le1$.
We refer to $\alpha$ as the \textit{guarantee relative to the prophet (resp.\ ex-ante relaxation)}.
We say that $\alpha$ is \textit{tight} if it is the supremum value for which~\eqref{eqn:defProphetIneq} holds.
\end{definition}

In this paper we are interested in tight guarantees relative to both the prophet and ex-ante relaxation, for both the classes of IID and non-IID instances with a particular number of slots $k$.
That is, we are interested in the values of $\alpha=\inf_{I\in\cI}\frac{\DP(I)}{\Proph(I)}$ and $\alpha=\inf_{I\in\cI}\frac{\DP(I)}{\ExAnte(I)}$ when $\cI$ can be $\cI_k$ or $\cI^\IID_k$ for some $k$, which will affect the value of $\alpha$.
Many of our results also imply tight guarantees for more granular classes of instances, e.g.\ $\cI=\cI_{k,n}$ or $\cI^\IID=\cI_{k,n}$, which restrict to having exactly $n$ agents.
Guarantees relative to the ex-ante relaxation are worse than those relative to the prophet (due to \Cref{prop:exAnteUB}), and guarantees are also worse for larger classes of instances (e.g.\ non-IID instead of IID).
Finally, we sometimes consider the following subclass of policies that are not as powerful as the optimal DP, which would also make guarantees worse.

\begin{definition}[Static Threshold Policies]
A \textit{static threshold} policy accepts the first $k$ agents to arrive who have valuation at least $r_J$, or equivalently have a type $j$ with $j\le J$, for some fixed index $J\in[m]$.  A static threshold policy is also allowed to set a tie-break probability $\rho\in(0,1]$, where each agent of type exactly $J$ is accepted (while slots remain) according to an independent coin flip of probability $\rho$.
\end{definition}

We emphasize that a static threshold policy is not allowed to change $J$ on-the-fly based on the remaining number of slots or agents, making them less powerful than the optimal DP.
Static policies of this type have been previously studied in \cite{ehsani2018prophet,chawla2020static,arnosti2021tight}, who note that the tie-break probability is unnecessary (i.e.\ one can always set $\rho=1$) under alternative models where the valuation distributions are continuous.

We further distinguish between \textit{oblivious} static threshold algorithms that must set $J,\rho$ without knowing the cardinal values of $r_1,\ldots,r_m$ (but knowing $m$ and $\{p_{ij}:i\in[n],j\in[m]\}$), vs.\ an algorithm that can set $J,\rho$ optimally with full knowledge of the instance $I$.  This distinction was introduced by \cite{chawla2020static} and the benefits of being oblivious are elaborated on in \cite{arnosti2021tight}.
As an example of this distinction, a thresholding rule based on the \textit{median} value of $\Proph(I)$ (as proposed in \cite{samuel1984comparison}) is oblivious, but a thresholding rule based on the \textit{mean} value (as proposed in \cite{kleinberg2012matroid}) is not.

\subsection{Main Idea of General Framework} \label{sec:mainIdea}

The idea of our general framework is to \textit{explicitly formulate the adversary's optimization problem} of minimizing some prophet inequality ratio, e.g. $\DP(I)/\Proph(I)$, over all instances $I$ belonging to some class.
This allows us to \textit{directly compute the tight guarantee} $\alpha$ in different settings, without having to separately construct a lower bound (usually based on analyzing a simple algorithm) and hoping that it is possible to construct a matching upper bound.
Of course, the adversary's problem can be highly intractable, because it needs to optimize over the space of distributions, and encapsulate the best response from the (possibly restricted) algorithm on each instance $I$.

Our general framework overcomes this intractability by breaking down the adversary's problem into two stages.
Treating the number of slots $k$ and agents $n$ as fixed, we assume that the adversary first optimizes over $m$ and the type distributions $\{p_{ij}:i\in[n],j\in[m]\}$, and then optimizes over the specific valuations $\{r_j:j\in[m]\}$.
The inner optimization problem over $r_j$'s can then be formulated as an LP, which encapsulates the algorithm's best response using constraints that are linear when the $p_{ij}$'s are fixed.
The dual of this LP gives rise to a new problem that is expressed solely in terms of the type distributions, whose optimal solution has a nice structure.
This allows us to solve the outer optimization problem over type distributions in many settings.

Before proceeding we define some notation that will simplify the formulation of these problems.

\begin{definition}[$\Delta_j,G_{ij}$] \label{def:DeltaG}
Let $\Delta_j:=r_j-r_{j+1}$ for all $j\in[m]$, with $r_{m+1}$ understood to be 0.
We will equivalently write the adversary's inner optimization problem using decision variables $\Delta_j$.
Recalling that $r_1\ge\cdots\ge r_m\ge0$, we have $\Delta_j\ge0$ for all $j$, where $\Delta_j$ can be interpreted as the ``valuation gain'' when going from type $j+1$ to type $j$.

Also, let $G_{ij}:=\sum_{j'=1}^j p_{ij'}$, the probability that agent $i$'s valuation is at least $r_j$, for all $i\in[n]$ and $j=0,\ldots,m$.
Note that $0=G_{i0}\le\cdots\le G_{im}=1$ for all $i$.
\end{definition}

\begin{proposition} \label{prop:sumDeltaQ}
For any instance $I$,
\begin{align} \label{eqn:6662}
\Proph(I)=\sum_{j=1}^m\Delta_j\cdot\bE[\min\{\sum_{i=1}^n\Ber(G_{ij}),k\}]
\text{~~and~~}\ExAnte(I)=\sum_{j=1}^m\Delta_j\cdot\min\{\sum_{i=1}^nG_{ij},k\},
\end{align}
where $\Ber(G_{ij})$ denotes an independent Bernoulli random variable with success probability $G_{ij}$.
\end{proposition}

\Cref{prop:sumDeltaQ} is
proven in \Cref{sec:pfsec2}.
Letting $Q_j$ denote $\bE[\min\{\sum_{i=1}^n\Ber(G_{ij}),k\}]$ (resp.\ $\min\{\sum_{i=1}^nG_{ij},k\}$), $Q_j$ can be interpreted as the expected number of agents of type $j$ or better accepted by the prophet (resp.\ ex-ante relaxation), which explains the formulas in~\eqref{eqn:6662}.

We are now ready to formulate the adversary's inner problem of minimizing $\DP(I)/\Proph(I)$ or $\DP(I)/\ExAnte(I)$ over decision variables $\Delta_j$.
By \Cref{prop:sumDeltaQ}, as long as the type distributions given by $G_{ij}$ are fixed, both $\Proph(I)$ and $\ExAnte(I)$ can be expressed as linear combinations of $\Delta_j$, which the adversary normalizes to 1.  Subject to this, the adversary then tries to minimize $\DP(I)$.

\begin{definition}[Inner Problem for Minimizing $\DP$]
Consider the following linear program, with $Q_j$ set to $\bE[\min\{\sum_{i=1}^n\Ber(G_{ij}),k\}]$ (resp.\ $\min\{\sum_{i=1}^nG_{ij},k\}$) for all $j\in[m]$.
\begin{subequations} \label{lp:innerPrimal}
\begin{align}
\min\ &V^k_1& \label{eqn:primalObj}
\\ \mathrm{s.t.\ }&V_i^l=\sum_{j=1}^m p_{ij}U^l_{ij}+V^l_{i+1} &\forall i\in[n],l\in[k] \label{eqn:dpValueToGo}
\\ &U^l_{ij} \ge\sum_{j'=j}^m \Delta_{j'}-(V^l_{i+1}-V^{l-1}_{i+1}) &\forall i\in[n],j\in[m],l\in[k] \label{eqn:dpUtility}
\\ &\sum_{j=1}^m Q_j\Delta_j=1 & \label{eqn:optIs1}
\\ &\Delta_j,U^l_{ij} \ge0 &\forall i\in[n],j\in[m],l\in[k] \label{eqn:primalNonneg}
\end{align}
\end{subequations}
%\begin{subequations} \label{lp:innerPrimal}
%\begin{eqnarray}
%& \min\ \  & V^k_1 & \ \ \label{eqn:primalObj}\\
%&\mathrm{s.t. \ } & V_i^l =\sum_{j=1}^m p_{ij}U^l_{ij}+V^l_{i+1} & \forall i\in[n],l\in[k] \ \ \label{eqn:dpValueToGo}\\
%& & U^l_{ij} \ge\sum_{j'=j}^m \Delta_{j'}-(V^l_{i+1}-V^{l-1}_{i+1}) & \forall i\in[n],j\in[m],l\in[k]  \ \ \label{eqn:dpUtility} \\
%& &\sum_{j=1}^m Q_j\Delta_j =1  & \ \ \label{eqn:optIs1}\\
%& & \Delta_j, U^l_{ij} \ge0 &\forall i\in[n],j\in[m],l\in[k] \ \  \label{eqn:primalNonneg}
%\end{eqnarray}
%\end{subequations}
In constraint~\eqref{eqn:dpValueToGo},
free variable $V^l_i$ denotes the value-to-go of the DP when agent $i$ arrives with exactly $l$ slots remaining, with $V^l_i$ understood to be 0 if $i=n+1$ or $l=0$.
Meanwhile, auxiliary variable $U^l_{ij}$
denotes the utility gain when agent $i$ realizes to type $j$ with $l$ slots remaining.
The utility gain $U^l_{ij}$ is lower-bounded by both 0 and the expression $\sum_{j'=j}^m \Delta_{j'}-V^l_{i+1}+V^{l-1}_{i+1}$ in~\eqref{eqn:dpUtility}, which denotes the immediate gain from accepting agent $i$ (who has valuation $r_j=\sum_{j'=j}^m \Delta_{j'}$) minus the loss $(V^l_{i+1}-V^{l-1}_{i+1})$ from proceeding to agent $i+1$ with $l-1$ instead of $l$ slots remaining.
Finally, constraint~\eqref{eqn:optIs1} normalizes the value of $\Proph(I)$ (resp.\ $\ExAnte(I)$) to 1.
\end{definition}

Therefore, in the linear program $V^k_1$ will equal precisely the optimal performance $\DP(I)$ of dynamic programming (see e.g. \cite{de2003linear}, \cite{adelman2007dynamic}),
% \Willcomment{Is this the correct reference---can you please point to the theorem number?},
and hence LP~\eqref{lp:innerPrimal} correctly describes the adversary's inner problem of minimizing $\DP(I)/\Proph(I)$ (resp.\ $\DP(I)/\ExAnte(I)$) over all instances $I$ with some given type distributions.
We now take the dual of~\eqref{lp:innerPrimal} to uncover a new problem.

\begin{definition}[Dual of Inner Problem for Minimizing $\DP$]
Defining dual variables $x^l_i,y^l_{ij},\theta$ for constraints \eqref{eqn:dpValueToGo},\eqref{eqn:dpUtility},\eqref{eqn:optIs1} respectively, the following LP is dual to~\eqref{lp:innerPrimal}.
\begin{subequations} \label{lp:innerDual}
\begin{align}
\max\ & \theta & \  \label{eqn:dualObj}
\\
 \mathrm{s.t.\ \ } & \theta\cdot Q_j \le\sum_{i=1}^n\sum_{l=1}^k\sum_{j'=1}^j y^l_{ij'} &\forall j\in[m] \label{eqn:dualCover}
\\
& y^l_{ij} \le p_{ij} x^l_i &\forall i\in[n],j\in[m],l\in[k] \label{eqn:dualUB}
\\
& x^l_i =
\begin{cases}
1, &i=1,l=k \\
0, &i=1,l<k \\
x^l_{i-1}-\sum_{j=1}^m (y^l_{i-1,j}-y^{l+1}_{i-1,j}), &i>1 \\
\end{cases}
& \forall i\in[n],l\in[k] \label{eqn:dualUpdate}
\\
&  y^l_{ij} \ge0 &\forall i\in[n],j\in[m],l\in[k] \label{eqn:dualNonneg}
\end{align}
\end{subequations}
\end{definition}

Before trying to interpret the LP~\eqref{lp:innerDual}, we notice the following structure.
An optimal solution will always saturate constraints~\eqref{eqn:dualUB} for better types $j$ before setting $y^l_{ij'}>0$ for types $j'>j$.
This allows us to reformulate the LP using collapsed variables $y^l_i=\sum_j y^l_{ij}$, as formalized below.

\begin{lemma}[Dual Simplification] \label{lem:dualSimplification}
LP~\eqref{lp:innerDual} has the same optimal value as the following LP.
\begin{subequations} \label{lp:innerDualSimplified}
\begin{align}
\max\ &\theta &\label{eqn:dualObjSimplified}
\\ \mathrm{s.t.\ \ }&\theta\cdot Q_j \le\sum_{i=1}^n\sum_{l=1}^k\min\{y^l_i,G_{ij}x^l_i\} &\forall j\in[m] \label{eqn:dualCoverSimplified}
\\ &y^l_{i} \le x^l_i &\forall i\in[n],l\in[k] \label{eqn:dualUBSimplified}
\\ &x^l_i =
\begin{cases}
1, &i=1,l=k \\
0, &i=1,l<k \\
x^l_{i-1}-y^l_{i-1}+y^{l+1}_{i-1}, &i>1 \\
\end{cases}
& \forall i\in[n],l\in[k] \label{eqn:dualUpdateSimplified}
\\ &y^l_{i} \ge0 &\forall i\in[n],l\in[k] \label{eqn:dualNonnegSimplified}
\end{align}
\end{subequations}
\end{lemma}

\Cref{lem:dualSimplification} is proven in \Cref{sec:pfsec2}.
We refer to~\eqref{lp:innerDualSimplified} as an LP since the non-linear term $\min\{y^l_i,G_{ij}x^l_i\}$ can easily be represented using an auxiliary variable and linear constraints. Though we derive LP \eqref{lp:innerDualSimplified} assuming discrete distributions, the LP formulation could potentially extend to more general distributions by working with an infinite LP, with the discrete index $j$ being replaced by a continuous index.

We note that LP~\eqref{lp:innerDualSimplified}
has the following interpretation.
Free variable $x^l_i$ denotes the probability of having exactly $l$ slots remaining when agent $i$ arrives, and $y^l_{i}$ denotes the (unconditional) probability of accepting the agent in this state, which must lie in $[0,x^l_i]$ as enforced by~\eqref{eqn:dualUBSimplified}, \eqref{eqn:dualNonnegSimplified}.
Meanwhile, \eqref{eqn:dualUpdateSimplified} correctly updates the state probabilities $x^l_i$ based on the acceptance probabilities $y^l_j$, which are understood to be 0 if $l=k+1$.
Finally, $\min\{y^l_i,G_{ij}x^l_i\}$ represents the probability of accepting agent $i$ with type $j$ or better when there are $l$ slots remaining, which is bottlenecked by $G_{ij}x^l_i$ (the probability of agent $i$ having type $j$ or better in state $l$) and $y^l_i$ (the unconditional probability of accepting agent $i$ in state $l$).
Therefore, constraint~\eqref{eqn:dualCoverSimplified} says that the expected number of agents of type $j$ or better accepted must be at least $\theta$ in comparison to $Q_j$, which is the number of agents of type $j$ or better accepted by the prophet or ex-ante relaxation.
The algorithm's guarantee is then given by the maximum $\theta$ that can be uniformly achieved across all types $j$.

\begin{example}[Single-unit prophet inequality]
{We show that it is easy to recover the $1/2$ competitive ratio for the single-unit prophet inequality using our framework. We consider the ex-ante relaxation as the benchmark and then we have $Q_j=\min\{\sum_{i=1}^{n}G_{ij}, 1\}$. Without loss of generality, we assume that there exists $j'\in[m]$ such that $\sum_{i=1}^{n}G_{ij'}=1$ (otherwise we can break one type into two). Then, we can set $\theta=1/2$, $x_i=1-\frac{1}{2}\cdot\sum_{i'=1}^{i-1}G_{i'j'}$ and $y_i=\frac{1}{2}\cdot G_{ij'}$ for each $i\in[n]$. It is easy to see that we obtain a feasible solution to LP \eqref{lp:innerDualSimplified} by noting that $x_i\geq\frac{1}{2}$ for each $i\in[n]$. Such a procedure recovers the Magician's policy in \cite{alaei2012online} for $k=1$.}
\end{example}

We now formalize some more notation and summarize the developments of this \namecref{sec:mainIdea}.

\begin{definition}[$\vG$, Simplified Duals for $\DP$]\label{def:DPDual}
Let $\vG$ denote the collective information about the type distributions, which includes $m$ and the values of $G_{ij}$ that must satisfy $G_{i1}\le\cdots\le G_{im}=1$ for all $i$.
For any such valid $\vG$, let $\InnerLP^{\DP/\Proph}_{k,n}(\vG)$ (resp.\ $\InnerLP^{\DP/\ExAnte}_{k,n}(\vG)$) denote the LP~\eqref{lp:innerDualSimplified} where $Q_j$ is set to $\bE[\min\{\sum_{i=1}^n\Ber(G_{ij}),k\}]$ (resp.\ $\min\{\sum_{i=1}^nG_{ij},k\}$) for all $j\in[m]$.
\end{definition}

\begin{theorem}[Reformulation of Tight Guarantees for $\DP$] \label{thm:NoniidDP}
For any fixed $k$ and $n>k$,
\begin{align*}
\inf_{I\in\cI_{k,n}}\frac{\DP(I)}{\Proph(I)} =\inf_\vG\ \InnerLP^{\DP/\Proph}_{k,n}(\vG)
\qquad\text{and}\qquad \inf_{I\in\cI_{k,n}}\frac{\DP(I)}{\ExAnte(I)} =\inf_\vG\ \InnerLP^{\DP/\ExAnte}_{k,n}(\vG).
\end{align*}
\end{theorem}

Put in words, \Cref{thm:NoniidDP} says that the tight guarantee for the optimal DP relative to the prophet (resp.\ ex-ante relaxation) is given by $\inf_\vG\ \InnerLP^{\DP/\Proph}_{k,n}(\vG)$ (resp.\ $\inf_\vG\ \InnerLP^{\DP/\ExAnte}_{k,n}(\vG)$), which are based on our simplified dual formulation and have reduced the adversary's problem to be only over type distributions $\vG$.
Before delving into how to solve these problems over $\vG$, we develop analogues of \Cref{thm:NoniidDP} in the settings where the algorithm in the primal problem is restricted to static threshold policies.
We note that in the IID special case, there is a significant further simplification that expresses the adversary's entire problem as a single semi-infinite LP, which we derive in \Cref{sec:IID}.

\subsection{General Framework for (Oblivious) Static Threshold Algorithms} \label{sec:mainIdeaST}

When the algorithm is restricted to static threshold policies, we can similarly formulate an inner primal LP and take its dual to uncover a new problem that depends on the type distributions but not on the specific valuations of agents.
In fact, a nice \textit{distinction} emerges in the dual depending on whether the algorithm must set the static threshold while oblivious to the specific valuations.

\begin{definition}[Oblivious vs.\ Non-oblivious Static Thresholds]\label{Def:oblivious}
For any instance $I$, let $\ST(I)$ denote the expected performance of the best static threshold policy on $I$, whose parameters $J,\rho$ can be set knowing instance $I$.
By contrast, an \textit{oblivious} static threshold (OST) algorithm must set the parameters $J,\rho$ without knowing the specific agent valuations in the instance (but knowing everything else, including the type distributions).
\end{definition}

The tight guarantee for OST algorithms relative to the prophet (resp.\ ex-ante relaxation) is defined by the following sequence of optimizations.
First, for a fixed $k$ and $n$, the adversary sets $\vG$, which we recall is defined by $m$ and $\{G_{ij}:i\in[n],j\in[m]\}$.
Based on $\vG$, the algorithm fixes the parameters $J,\rho$ of the static threshold policy to be used.
Finally, the adversary sets the valuations, defined by $\Delta_j$, to minimize the policy's performance relative to $\Proph(I)$ (resp.\ $\ExAnte(I)$).

We first focus on tight guarantees for OST algorithms, which are simpler to capture using our framework.  For a fixed $\vG$ and parameters $J,\rho$ chosen by the OST, we write the adversary's inner optimization problem over $\Delta_j$.

\begin{definition}[Inner Problem for Minimizing OST]
Consider the following LP, where coefficient $Q_j$ can be set to either $\bE[\min\{\sum_{i=1}^n\Ber(G_{ij}),k\}]$ or $\min\{\sum_{i=1}^nG_{ij},k\}$ like before.
\begin{subequations} \label{lp:innerPrimalOST}
\begin{align}
\min\ &V^k_1 \label{eqn:primalObjOST}
\\ \mathrm{s.t.\ \ }&V_i^l =\sum_{j<J}p_{ij}U^l_{ij}+p_{iJ}\rho U^l_{iJ}+V^l_{i+1} &\forall i\in[n],l\in[k] \label{eqn:dpValueToGoOST}
\\ &U^l_{ij}=\sum_{j'=j}^m \Delta_{j'}-V^l_{i+1}+V^{l-1}_{i+1} &\forall i\in[n],j\in[m],l\in[k] \label{eqn:dpUtilityOST}
\\ &\sum_{j=1}^m Q_j\Delta_j =1 & \label{eqn:optIs1OST}
\\ &\Delta_j\ge0 &\forall j\in[m] \label{eqn:primalNonnegOST}
\end{align}
\end{subequations}
\end{definition}

Compared to the inner LP~\eqref{lp:innerPrimal} for the optimal DP algorithm, LP~\eqref{lp:innerPrimalOST} differs in two ways.
First, $U^l_{ij}$ is now a free variable that could be negative, representing the change in utility when agent $i$ takes type $j$ and is \textit{accepted} with $l$ slots remaining, as set in~\eqref{eqn:dpUtilityOST}.
Second, the policy is now \textit{forced to accept} an agent with type $j<J$ w.p.~1 and an agent with type $j=J$ w.p.~$\rho$, regardless of $l$ and $i$, as reflected in constraints~\eqref{eqn:dpValueToGoOST}.
This allows the adversary to create a smaller objective value in LP~\eqref{lp:innerPrimalOST}, ultimately yielding a maximization problem for the dual in which the collapsed variable $y^l_i$ (from the simplification in \Cref{lem:dualSimplification}) must satisfy essentially the same static threshold rule as defined by $J$ and $\rho$.
This is formalized in the definition and theorem below.

\begin{definition}[Simplified Duals for $\OST$]
For any type distributions $\vG$ and static threshold policy $J,\rho$, let $\InnerLP^{\OST(J,\rho)/\Proph}_{k,n}(\vG)$ (resp.\ $\InnerLP^{\OST(J,\rho)/\ExAnte}_{k,n}(\vG)$) be identical to LP $\InnerLP^{\DP/\Proph}_{k,n}(\vG)$ (resp.\ $\InnerLP^{\DP/\ExAnte}_{k,n}(\vG)$), except with the additional constraints
\begin{align} \label{eqn:3322}
y^l_i &=(\sum_{j<J}p_{ij}+p_{iJ}\rho)x^l_i=((1-\rho)G_{i,J-1}+\rho G_{iJ})x^l_i
&\forall i\in[n],l\in[k].
\end{align}
\end{definition}

\begin{theorem}[Reformulating the Tight Guarantees for $\OST$] \label{thm:NoniidOST}
For any fixed $k$ and $n>k$, the tight guarantee for OST algorithms relative to the prophet (resp.\ ex-ante relaxation) is equal to $\inf_\vG\sup_{J,\rho}\InnerLP^{\OST(J,\rho)/\Proph}_{k,n}(\vG)$ (resp.\ $\inf_\vG\sup_{J,\rho}\InnerLP^{\OST(J,\rho)/\ExAnte}_{k,n}(\vG)$).
\end{theorem}

The proof of \Cref{thm:NoniidOST} is similar to \Cref{sec:mainIdea} and deferred to \Cref{sec:pfsec2}.

We proceed to study tight guarantees for non-oblivious static thresholds.
Earlier, the way in which the static threshold restriction directly translated into dual constraint~\eqref{eqn:3322} was crucially dependent on the fact that in the primal LP~\eqref{lp:innerPrimalOST}, $J$ and $\rho$ were set before the $\Delta_j$'s.
We now show that if $J$ and $\rho$ are decided after the $\Delta_j$'s, then this translates into the dual algorithm being able to employ an \textit{arbitrary convex combination} of static threshold rules, which can change the dual objective.
First we formulate the adversary's inner problem for minimizing $\ST(I)$, which must set $\Delta_j$'s such that the performance of \textit{any} static threshold policy defined by $J,\rho$ is poor.

\begin{definition}[Inner Problem for Minimizing $\ST$]
Consider the following LP, where coefficient $Q_j$ can be set to either $\bE[\min\{\sum_{i=1}^n\Ber(G_{ij}),k\}]$ or $\min\{\sum_{i=1}^nG_{ij},k\}$ like before.
\begin{subequations} \label{lp:innerPrimalST}
\begin{align}
\min \ &\alpha & \label{eqn:primalObjST}\\
\mathrm{s.t.\ \ } &V_i^l(J,\rho) =\sum_{j<J}p_{ij}U^l_{ij}(J,\rho)+p_{iJ}\rho U^l_{iJ}(J,\rho)+V^l_{i+1}(J,\rho) &\forall i,l,J\in[m],\rho\in(0,1] \label{eqn:dpValueToGoST}\\
&U^l_{ij}(J,\rho) =\sum_{j'=j}^m \Delta_{j'}-V^l_{i+1}(J,\rho)+V^{l-1}_{i+1}(J,\rho) &\forall i,j,l,J\in[m],\rho\in(0,1] \label{eqn:dpUtilityST}\\
&\alpha \ge V^k_1(J,\rho) &\forall J\in[m],\rho\in(0,1] \label{eqn:alphaLB}\\
&\sum_{j=1}^m Q_j\Delta_j =1 & \label{eqn:optIs1ST}\\
&\Delta_j \ge0 &\forall j\in[m] \label{eqn:primalNonnegST}
\end{align}
\end{subequations}
\end{definition}

We note that LP~\eqref{lp:innerPrimalST} is similar to LP~\eqref{lp:innerPrimalOST}, except a copy of the variables has been created for every possible static threshold $J,\rho$, all of which must perform no better than $\alpha$.
%\wnote{Maybe write a sentence about measure-theoretic technicalities of such an LP?}
The simplified dual formulation corresponding to LP~\eqref{lp:innerPrimalST} will be easier to write using the following additional notation.

\begin{definition}[$\vx,\vy,\cP^k_n$] \label{def:xyP}
Let $\vx:=(x^l_i)_{i\in[n],l\in[k]},\vy:=(y^l_i)_{i\in[n],l\in[k]}$, and let $\cP^k_n$ denote the set of vectors $(\vx,\vy)$ that satisfy the simplified dual problem's state-updating constraints~\eqref{eqn:dualUBSimplified}--\eqref{eqn:dualNonnegSimplified}.
\end{definition}

\begin{definition}[Simplified Duals for $\ST$]
For any $\vG$, let $\InnerLP^{\ST/\Proph}_{k,n}(\vG)$ (resp.\ $\InnerLP^{\ST/\ExAnte}_{k,n}(\vG)$) denote the following LP, with coefficient $Q_j$ set to $\bE[\min\{\sum_{i=1}^n\Ber(G_{ij}),k\}]$ (resp.\ $\min\{\sum_{i=1}^nG_{ij},k\}$) for all $j\in[m]$.
\begin{subequations} \label{lp:innerDualST}
\begin{align}
\max\ &\theta\\
\mathrm{s.t.\ \ }&\theta\cdot Q_j \le\int_{J,\rho} \mu(J,\rho)\left(\sum_{i=1}^n\sum_{l=1}^k\min\{y^l_i(J,\rho),G_{ij}x^l_i(J,\rho)\}\right) &\forall j\in[m]  \label{constraintDualST2}\\ &y^l_i(J,\rho)=((1-\rho)G_{i,J-1}+\rho G_{iJ})x^l_i(J,\rho)
&\forall i,l,J\in[m],\rho\in(0,1] \label{constraintDualST}\\ &(\vx(J,\rho),\vy(J,\rho)) \in\cP^k_n &\forall J\in[m],\rho\in(0,1]\\ &\int_{J,\rho} \mu(J,\rho) =1 &\\
&\mu(J,\rho) \ge0 &\forall J\in[m],\rho\in(0,1]
\end{align}
\end{subequations}
\end{definition}

We note that there is no benefit for the primal algorithm using a convex combination of static thresholds (its expectation is maximized by choosing the best one), but since the dual problem has to uniformly cover each type $j$, there can be a benefit.

\begin{theorem}[Reformulating the Tight Guarantees for $\ST$] \label{thm:NoniidST}
For any fixed $k$ and $n>k$,
\begin{align*}
\inf_{I\in\cI_{k,n}}\frac{\ST(I)}{\Proph(I)} =\inf_\vG\ \InnerLP^{\ST/\Proph}_{k,n}(\vG)
\qquad\text{and}\qquad \inf_{I\in\cI_{k,n}}\frac{\ST(I)}{\ExAnte(I)} =\inf_\vG\ \InnerLP^{\ST/\ExAnte}_{k,n}(\vG).
\end{align*}
\end{theorem}

The proof of \Cref{thm:NoniidST} is similar to \Cref{sec:mainIdea} and also deferred to \Cref{sec:pfsec2}.

\section{General Framework applied to the Non-IID Setting} \label{sec:nonIID}

In this \namecref{sec:nonIID} we study tight guarantees over general non-IID instances, starting with those for the optimal DP.
Recall that for any number of slots $k$ and agents $n>k$, we have established in \Cref{thm:NoniidDP} that the tight guarantees relative to the prophet and ex-ante relaxation are given by $\inf_\vG\InnerLP^{\DP/\Proph}_{k,n}(\vG)$ and $\inf_\vG\InnerLP^{\DP/\ExAnte}_{k,n}(\vG)$ respectively.  As a recap,by using the notation from \Cref{def:xyP} that treats $\vx,\vy$ as vectors, $\InnerLP^{\DP/\Proph}_{k,n}(\vG)$ can be rewritten as
\begin{subequations} \label{lp:DpProphNoniid}
\begin{eqnarray}
&&\InnerLP^{\DP/\Proph}_{k,n}(\vG)\\
=&& \max ~~  \theta
\\
&& \mathrm{s.t.\ } ~~ \theta\cdot\bE[\min\{\sum_{i=1}^n\Ber(G_{ij}),k\}] \le\sum_{i=1}^n\sum_{l=1}^k \min\{y^l_i,G_{ij}x^l_i\} \ \ \forall j\in[m] \label{constr:ProphNoniid}
\\  && ~~~~~~~~ (\vx,\vy) \in\cP^k_n,
\end{eqnarray}
\end{subequations}
and $\InnerLP^{\DP/\ExAnte}_{k,n}(\vG)$ can be rewritten as
\begin{subequations} \label{lp:DpExanteNoniid}
\begin{eqnarray}
\InnerLP^{\DP/\ExAnte}_{k,n}(\vG)=&\max & \theta \\
&\mathrm{s.t.\ }& \theta\cdot\min\{\sum_{i=1}^n G_{ij},k\} \le\sum_{i=1}^n\sum_{l=1}^k \min\{y^l_i,G_{ij}x^l_i\} \ \ \forall j\in[m] \label{constr:ExanteNoniid}
\\
& & (\vx,\vy) \in\cP^k_n
\end{eqnarray}
\end{subequations}

\subsection{$\DP/\ExAnte$ in Non-IID Setting} \label{sec:magicianOCRS}
In general non-IID setting, we obtain the following result regarding $\DP/\ExAnte$.
\begin{theorem} \label{thm:rederiveWeightedBernoulli}
$\inf_\vG\ \InnerLP^{\DP/\ExAnte}_{k,n}(\vG)$ is equal to the optimal objective value of the following problem:
\begin{subequations} \label{lp:OCRS}
\begin{eqnarray}
\inf_{\substack{\sum_{i \in [n]} g_i\le k\\ 1\ge g_i\ge0\ \forall i}}\ \ \ \  & \max\ \theta
\\ & \mathrm{s.t.\ } & \theta\cdot g_i \le\sum_{l=1}^k \min\{y^l_i,g_ix^l_i\} \quad \forall i\in[n] \label{constr:OCRS}
\\ & & (\vx,\vy) \in\cP^k_n
\end{eqnarray}
\end{subequations}
\end{theorem}

\Cref{thm:rederiveWeightedBernoulli} is proved in \Cref{sec:pfsec3}.
The result derived in \Cref{thm:rederiveWeightedBernoulli} has a nice interpretation. The variable $g_i$ for each $i\in[n]$ can be interpreted as the marginal probability that agent $i$ got accepted in the ex-ante relaxation. Denote by $\theta^*$ the optimal value of LP \eqref{lp:OCRS} after taking minimum over $\mathbf{g}=(g_1,\ldots,g_n)$. Denote by $\{\theta^*, \vx_{\mathbf{g}}, \vy_{\mathbf{g}}\}$ a feasible solution to LP \eqref{lp:OCRS} for a fixed $\mathbf{g}$, where the value of $(\vx_{\mathbf{g}}, \vy_{\mathbf{g}})$ depends on $\mathbf{g}$. Then constraint \eqref{constr:OCRS} implies that each agent $i$ got accepted by the policy specified by $(\vx_{\mathbf{g}}, \vy_{\mathbf{g}})$ with a probability at least $\theta^*$ conditional on being accepted in the ex-ante relaxation, for any $\mathbf{g}$. Such an implication corresponds to the definition of $\theta^*$-balancedness online contention resolution scheme (OCRS) in \cite{feldman2021online}. Thus, \Cref{thm:rederiveWeightedBernoulli} implies that an OCRS achieves the tight guarantee of the $\DP$ policy, with respect to the ex-ante relaxation. Note that this point has been previously proved in \cite{jiang2022tight}. Here, we prove the same result in an alternative way by exploiting the structures of our LP framework $\inf_\vG\ \InnerLP^{\DP/\ExAnte}_{k,n}(\vG)$.

\subsection{Optimal Oblivious Static Thresholds in Non-IID Setting} \label{sec:NoniidOST}

Recall that for any fixed $k$ and $n>k$, we have established in \Cref{thm:NoniidOST} that the tight guarantees for OST algorithms relative to the prophet and ex-ante relaxation are given by $\inf_\vG\sup_{J,\rho}\InnerLP^{\OST(J,\rho)/\Proph}_{k,n}(\vG)$ and $\inf_\vG\sup_{J,\rho}\InnerLP^{\OST(J,\rho)/\ExAnte}_{k,n}(\vG)$ respectively, where the inner LP's correspond to~\eqref{lp:DpProphNoniid} and~\eqref{lp:DpExanteNoniid} respectively but both have the added constraints
\begin{align}
y^l_i &=((1-\rho)G_{i,J-1}+\rho G_{iJ})x^l_i
&\forall i\in[n],l\in[k]. \label{constr:OstNoniid}
\end{align}

The inner LP's for OST's are substantially easier to analyze because under constraints~\eqref{constr:OstNoniid}, the RHS that is common to~\eqref{constr:ProphNoniid} and~\eqref{constr:ExanteNoniid} can be rewritten as
\begin{align} \label{eqn:RHSsimplificationOST}
\sum_{i=1}^n\sum_{l=1}^k \min\{y^l_i,G_{ij}x^l_i\}=\sum_{i=1}^n\min\{((1-\rho)G_{i,J-1}+\rho G_{iJ}),G_{ij}\}\sum_{l=1}^k x^l_i,
\end{align}
where term $\min\{((1-\rho)G_{i,J-1}+\rho G_{iJ}),G_{ij}\}$ for each agent $i$ depends on the choices of $J,\rho$ but not on the number of remaining slots $l$.
Moreover, we have the following relationships, which are proved in \Cref{sec:pfsec3}.

\begin{lemma} \label{lem:analyzeOST}
Fix an OST $J,\rho$ and define $\tau_i=(1-\rho)G_{i,J-1}+\rho G_{i,J}$ for all $i\in[n]$.
Suppose vectors $\vx,\vy$ satisfy~\eqref{constr:OstNoniid}, i.e.\ $y^l_i=\tau_ix^l_i$ for all $i$ and $l$, as well as $(\vx,\vy)\in\cP^k_n$.
Then for all $i\in[n]$, we have
\begin{align*}
\sum_{l=1}^k x^l_i =\Pr[\sum_{i'<i}\Ber(\tau_{i'})<k]\qquad\text{and}\qquad\sum_{i'=1}^i \tau_{i'}\sum_{l=1}^k x^l_{i'} =\bE[\min\{\sum_{i'=1}^i\Ber(\tau_{i'}),k\}].
\end{align*}
\end{lemma}

%\Cref{lem:analyzeOST} is formally proven in \Cref{sec:pfsec3}, but natural under the interpretation that $x^l_i$ denotes the probability of having exactly $l$ slots remaining when agent $i$ arrives.
%Note that each agent $i$ ``clears the bar'' for acceptance independently with probability $\tau_i$.
%An agent $i$ is accepted if and only if they clear the bar and there is at least 1 slot remaining when they arrive, with the latter probability given by $\sum_{l=1}^k x^l_i$.
%The second part of \Cref{lem:analyzeOST} then follows because the number of agents accepted among $i'=1,\ldots,i$ is equal to the number of them who clear the bar, truncated by $k$.
%Meanwhile, the first part of \Cref{lem:analyzeOST} follows because there is a slot remaining for agent $i$ if and only if the number of previous agents $i'<i$ who cleared the bar is less than $k$.

Equipped with \Cref{lem:analyzeOST}, we are now ready to prove our result that the tight guarantees for OST algorithms relative to the stronger ex-ante benchmark are no worse than relative to the prophet.
We first show a lower bound of $\OST/\ExAnte$ in the following lemma.
\begin{lemma} \label{lem:exanteLB}
For any type distributions $\vG$ and static threshold policy $J,\rho$, we have
\begin{align*}
&\InnerLP^{\OST(J,\rho)/\ExAnte}_{k,n}(\vG)
\\ \ge&\min\left\{\Pr\left[\sum_{i<n}\Ber((1-\rho)G_{i,J-1}+\rho G_{iJ})<k\right],\frac{\bE[\min\{\sum_{i<n}\Ber((1-\rho)G_{i,J-1}+\rho G_{iJ}),k\}]}{k}\right\}.
\end{align*}
\end{lemma}

We then show an upper bound of $\OST/\Proph$ in the following lemma, which matches the lower bound established in \Cref{lem:exanteLB}.
\begin{lemma} \label{lem:prophUB}
For any fixed type distributions $\vG$, the value of $\inf_{\vG'}\sup_{J,\rho}\InnerLP^{\OST(J,\rho)/\Proph}_{k,n}(\vG')$ can be at most
\begin{equation}\label{eqn:2789}
\sup_{J,\rho}\ \min\left\{\Pr\left[\sum_{i<n}\Ber((1-\rho)G_{i,J-1}+\rho G_{iJ})<k\right],\frac{\bE[\min\{\sum_{i<n}\Ber((1-\rho)G_{i,J-1}+\rho G_{iJ}),k\}]}{k}\right\}.
\end{equation}
\end{lemma}

Note that the ex-ante benchmark is a stronger benchmark than the prophet. Combining \Cref{lem:exanteLB} and \Cref{lem:prophUB}, we have the following result, which is formally proved in \Cref{sec:pfsec3}.
\begin{theorem}\label{thm:chawlaTight}
For any fixed $k$ and $n$,
\begin{align}\label{lp:chawlaBernOpt}
&\inf_\vG\sup_{J,\rho}\ \InnerLP^{\OST(J,\rho)/\Proph}_{k,n}(\vG)
=\inf_\vG\sup_{J,\rho}\ \InnerLP^{\OST(J,\rho)/\ExAnte}_{k,n}(\vG)
\\ =&\inf_\vG\sup_{J,\rho}\ \min\left\{\Pr\left[\sum_{i<n}\Ber((1-\rho)G_{i,J-1}+\rho G_{iJ})<k\right],\frac{\bE[\min\{\sum_{i<n}\Ber((1-\rho)G_{i,J-1}+\rho G_{iJ}),k\}]}{k}\right\}.\nonumber
\end{align}
\end{theorem}

It is direct to see that for the formula in \eqref{lp:chawlaBernOpt}, the first term in the min operator is decreasing over the threshold $J,\rho$, while the second term in the min operator is increasing over the threshold $J,\rho$. Thus, in order to achieve the supremum, the two terms within the min operator must be equivalent, which yields the following result.
\begin{corollary} \label{cor:bernOpt}
For any number of slots $k$ and agents $n>k$, the best-possible guarantees for OST algorithms relative to the prophet
%$\inf_{I\in\cI_{k,n}}\frac{\OST(I)}{\Proph(I)}$
or ex-ante relaxation
%$\inf_{I\in\cI_{k,n}}\frac{\OST(I)}{\ExAnte(I)}$
are identically equal to
\begin{subequations} \label{lp:chawlaBernOpt2}
\begin{align}
\min\ \alpha
\\ \mathrm{s.t.\ }\alpha &=\Pr\left[\sum_{i=1}^{n-1}\Ber(q_i)<k\right]=\frac{\bE[\min\{\sum_{i=1}^{n-1}\Ber(q_i),k\}]}{k}
\\ q_i &\in[0,1] \quad \forall i\in[n-1]
\end{align}
\end{subequations}
\end{corollary}
\cite{chawla2020static} show that for a fixed $k$, the infimum value of problem~\eqref{lp:chawlaBernOpt2} over $n>k$ occurs as $n\to\infty$, and is equal to $\Pr[\Pois(\lambda)<k]=\frac{\bE[\min\{\Pois(\lambda),k\}]}{k}$, where $\lambda$ is the unique real number that makes these quantities identical.  They also show how to achieve this guarantee using an oblivious static threshold algorithm.  Our framework shows that their guarantees are tight, regardless of whether one is comparing to the prophet or ex-ante relaxation, and moreover never required explicitly computing its value or constructing a family of counterexamples to establish a matching upper bound! We further show that the infimum value of problem~\eqref{lp:chawlaBernOpt2} over $n>k$ grows at the order of $1-\Theta(\sqrt{\log k /k})$, establishing the tight order dependence on $k$ for the error term of static threshold policies. The following proposition is proved in \Cref{sec:pfsec3}.
% which is the first time in the literature that the tight order of optimal competitive ratio for static policies is \blue{formally} proved.

\begin{proposition}\label{prop:StaticPolicy}
For any integer $k$, let $\lambda_k$ be the solution to the equation
\[
\Pr[\Pois(\lambda_k)<k]=\frac{\bE[\min\{\Pois(\lambda_k),k\}]}{k}.
\]
Then, it holds that
\[
\Pr[\Pois(\lambda_k)<k]=\frac{\bE[\min\{\Pois(\lambda_k),k\}]}{k}=1-\Theta(\sqrt{\log k /k}).
\]
\end{proposition}

\subsection{Oblivious vs. Non-oblivious Static Thresholds in Non-IID Setting} \label{sec:oblVsNonobl}
We further study the performances of oblivious static threshold policies versus general non-oblivious static threshold policies in the non-IID setting. We first show that there exists an instance $\vG$ such that $\ST$ performs better than $\OST$, with respect to both the ex-ante benchmark and the prophet, which is formally proved in \Cref{sec:pfsec3}.

\begin{proposition}\label{prop:STbetterOST}
There exists an instance $\vG$ such that
\[
\InnerLP^{\ST/\Proph}_{k,n}(\vG)>\sup_{J,\rho}\InnerLP^{\OST(J,\rho)/\Proph}_{k,n}(\vG) \]
and\[ \InnerLP^{\ST/\ExAnte}_{k,n}(\vG)>\sup_{J,\rho}\InnerLP^{\OST(J,\rho)/\ExAnte}_{k,n}(\vG)
\]
\end{proposition}

%$\InnerLP^{\ST/\Proph}_{k,n}(\vG)\geq\frac{2}{3}$ and $\InnerLP^{\ST/\ExAnte}_{k,n}(\vG)\geq\frac{2}{3}$. For the same instance, we show that $\sup_{J,\rho}\InnerLP^{\OST(J,\rho)/\Proph}_{k,n}(\vG)\leq\frac{1}{2}$ and $\sup_{J,\rho}\InnerLP^{\OST(J,\rho)/\ExAnte}_{k,n}(\vG)\leq\frac{1}{2}$, which completes our proof of \Cref{prop:STbetterOST}.

% Note that in the proof of \Cref{prop:STbetterOST}, we essentially construct an example $\vG$ for which the expression from \Cref{thm:chawlaTight} is tight,
% \begin{align*}
% &\sup_{J,\rho}\InnerLP^{\OST(J,\rho)/\Proph}_{k,n}(\vG)
% =\sup_{J,\rho}\InnerLP^{\OST(J,\rho)/\ExAnte}_{k,n}(\vG)
% \\ &=\sup_{J,\rho}\min\left\{\Pr\left[\sum_{i<n}\Ber((1-\rho)G_{i,J-1}+\rho G_{iJ})<k\right],\frac{\bE[\min\{\sum_{i<n}\Ber((1-\rho)G_{i,J-1}+\rho G_{iJ}),k\}]}{k}\right\}.
% \end{align*}
%\textcolor{blue}{
%(achieving a value of 2/3 instead of 1/2).}

Therefore the method of \cite{chawla2020static} is not instance-optimal.  However, we now show that it is optimal in the worst case, hence their bound is tight even for the more powerful class of ST policies.
First we need the following analogue of \Cref{lem:prophUB} for non-oblivious static thresholds, which is proved in \Cref{sec:pfsec3}.

\begin{lemma} \label{lem:prophUBforST}
For any fixed type distributions $\vG$, the value of $\inf_{\vG'}\InnerLP^{\ST/\Proph}_{k,n}(\vG')$ can be at most the supremum of
\begin{align*}
&\min\left\{\int_{J,\rho}\Pr\left[\sum_{i<n}\Ber((1-\rho)G_{i,J-1}+\rho G_{iJ})<k\right]\mu(J,\rho),\right.\\
&\left.\int_{J,\rho}\frac{\bE[\min\{\sum_{i<n}\Ber((1-\rho)G_{i,J-1}+\rho G_{iJ}),k\}]}{k}\mu(J,\rho)\right\}
\end{align*}
over all measures $\mu$ over $J\in[m]$ and $\rho\in(0,1]$.
\end{lemma}

In \Cref{thm:chawlaTight}, we have shown that both the quantities $\inf_\vG\sup_{J,\rho}\ \InnerLP^{\OST(J,\rho)/\Proph}_{k,n}(\vG)$
and $\inf_\vG\sup_{J,\rho}\ \InnerLP^{\OST(J,\rho)/\ExAnte}_{k,n}(\vG)$ are identical to the quantity
\begin{align} \label{eqn:3151}
\inf_\vG\sup_{J,\rho}\ \min\left\{\Pr\left[\sum_{i<n}\Ber((1-\rho)G_{i,J-1}+\rho G_{iJ})<k\right],\frac{\bE[\min\{\sum_{i<n}\Ber((1-\rho)G_{i,J-1}+\rho G_{iJ}),k\}]}{k}\right\}.
\end{align}

The results in \cite{chawla2020static} imply that this infimum is in fact achieved in the simple case where there is only $m=1$ type and $G_{i1}=1$ for all $i$, in which case~\eqref{eqn:3151} reduces to
\begin{align} \label{eqn:2098}
\sup_{\rho\in(0,1]}\min\left\{\Pr\left[\Bin(n-1,\rho)<k\right],\frac{\bE[\min\{\Bin(n-1,\rho),k\}]}{k}\right\}.
\end{align}
We now show that $\inf_{\vG}\InnerLP^{\ST/\ExAnte}_{k,n}(\vG)$ cannot do any better, because even the larger quantity from \Cref{lem:prophUBforST}, which on this one-type instance reduces to
\begin{align} \label{eqn:TEMP}
\sup_{\mu:(0,1]\to\bR_{\ge0},\int_\rho \mu(\rho)=1}\min\left\{\int_{\rho}\Pr\left[\Bin(n-1,\rho)<k\right]\mu(\rho),\int_{\rho}\frac{\bE[\min\{\Bin(n-1,\rho),k\}]}{k}\mu(\rho)\right\},
\end{align}
is actually no larger than~\eqref{eqn:2098}. The above arguments are formalized in the following theorem.

\begin{theorem}\label{thm:STequalOST}
For any fixed $n$ and $k$, it holds that
\[\begin{aligned}
&\inf_{\vG}\ \InnerLP^{\ST/\Proph}_{k,n}(\vG)=\inf_{\vG}\ \InnerLP^{\ST/\ExAnte}_{k,n}(\vG)\\
=&\inf_\vG\sup_{J,\rho}\ \InnerLP^{\OST(J,\rho)/\Proph}_{k,n}(\vG)
=\inf_\vG\sup_{J,\rho}\ \InnerLP^{\OST(J,\rho)/\ExAnte}_{k,n}(\vG).
\end{aligned}\]
\end{theorem}

%\subsection{$\DP/\Proph$ in Non-IID Setting}

\section{Further Simplified General Framework, applied to the IID Setting} \label{sec:IID}

In this \namecref{sec:IID} we study tight guarantees for DP/OST/ST algorithms relative to the prophet/ex-ante relaxation, under the restriction that the type distributions $\vG$ must be IID.
That is, for each type $j\in[m]$, it is imposed that $G_{ij}$ is identically equal to some value $G_j$ across all agents $i\in[n]$.

When this is the case, the key dual constraints from \Cref{sec:nonIID} that compare to the prophet, e.g.~\eqref{constr:ProphNoniid} in $\InnerLP^{\DP/\Proph}_{k,n}(\vG)$, can be rewritten as
\begin{align} \label{eqn:3525}
\theta\cdot\bE[\min\{\sum_{i=1}^n\Ber(G_{j}),k\}] &\le\sum_{i=1}^n\sum_{l=1}^k \min\{y^l_i,G_{j}x^l_i\} &\forall j\in[m].
\end{align}
where the values $G_j$ no longer depend on $i$.  Consequently, these constraints will always be hardest to satisfy when the IID type distribution becomes infinitely granular, i.e.\ $G_j=j/m$ for all $j\in[m]$ with $m\to\infty$, simply because there are more constraints.  The overall problem of interest, $\inf_\vG\InnerLP^{\DP/\Proph}_{k,n}(\vG)$, can then have its outer optimization problem dropped, since the infimum always arises at the infinitely-granular $\vG$.  This leads to a drastic reduction where the tight guarantees are now described by a single semi-infinite LP, as formalized below.  We note that the same reduction can be made for the dual constraints from before that compare to the ex-ante relaxation:
\begin{align} \label{eqn:2740}
\theta\cdot\min\{\sum_{i=1}^n G_{j},k\} &\le\sum_{i=1}^n\sum_{l=1}^k \min\{y^l_i,G_{j}x^l_i\} &\forall j\in[m].
\end{align}

\begin{definition}[Semi-infinite LP's for $\DP$ in IID Setting]
Let $\Bin(n,q)$ denote a Binomial random variable with $n$ independent trials of success probability $q$.
Consider the semi-infinite families of constraints
\begin{align}
\theta\cdot\bE[\min\{\Bin(n,q),k\}] &\le\sum_{i=1}^n\sum_{l=1}^k \min\{y^l_i,qx^l_i\} &\forall q\in(0,1] \label{constr:ProphIid}
\\ \theta\cdot\min\{nq,k\} &\le\sum_{i=1}^n\sum_{l=1}^k \min\{y^l_i,qx^l_i\} &\forall q\in(0,1] \label{constr:ExanteIid}
\end{align}
which correspond to the limiting cases of~\eqref{eqn:3525} and~\eqref{eqn:2740} as $G_j=j/m=q$ and $m\to\infty$.
Then for any $k$ and $n>k$, let $\IIDLP^{\DP/\Proph}_{k,n}$ (resp.\ $\IIDLP^{\DP/\ExAnte}_{k,n}$) denote the semi-infinite LP defined by: maximize $\theta$, subject to constraints~\eqref{constr:ProphIid} (resp.~\eqref{constr:ExanteIid}) and $(\vx,\vy)\in\cP^k_n$.
% We often refer to $q$ as the \textit{quantile} in (0,1] to be covered.
\end{definition}

Static threshold policies $J,\rho$ also simplify nicely in this semi-infinite dual problem in the IID setting.  Namely, the previous dual constraints~\eqref{constr:OstNoniid} for a static threshold policy, $y^l_i=((1-\rho)G_{i,J-1}+\rho G_{iJ})x^l_i$, by setting $\tau=(1-\rho)G_{i,J-1}+\rho G_{iJ}$ which is identical across $i$, can be reduced to
\begin{equation} \label{constr:OstIid}
y^l_i =\tau x^l_i \quad \forall i\in[n],l\in[k]
\end{equation}
% now that $G_{ij}$ is identical across $i$ and infinitely granular.
Note that $\tau\in(0,1]$ is a single number in the IID setting; we no longer need an index $J\in[m]$ combined with a tiebreak probability $\rho\in(0,1]$.

\begin{definition}[Semi-infinite LP's for $\OST$ in IID Setting]
For any $k$, $n>k$, and fixed static threshold policy defined by $\tau$, let $\IIDLP^{\OST(\tau)/\Proph}_{k,n}$ (resp.\ $\IIDLP^{\OST(\tau)/\ExAnte}_{k,n}$) denote the semi-infinite LP defined by: maximize $\theta$, subject to~\eqref{constr:ProphIid} (resp.~\eqref{constr:ExanteIid}), constraints~\eqref{constr:OstIid} for a static threshold policy in the IID setting, and $(\vx,\vy)\in\cP^k_n$.
\end{definition}

Finally, for non-oblivious static thresholds, we define the analogue of~\eqref{lp:innerDualST} for the IID setting.

\begin{definition}[$\ST$ in IID Setting]
For any $k$ and $n>k$, let $\IIDLP^{\ST/\Proph}_{k,n}$ (resp.\ $\IIDLP^{\ST/\ExAnte}_{k,n}$) denote the following optimization problem, when $Q(q)=\bE[\min\{\Bin(n,q),k\}]$ (resp.\ $Q(q)=\min\{nq,k\}$) for all $q\in(0,1]$.
\begin{subequations} \label{lp:innerDualSTIID}
\begin{align}
\max\ &\theta&
\\ \mathrm{s.t.\ \ }&\theta\cdot Q(q) \le\int_{\tau\in(0,1]} \mu(\tau)\left(\sum_{i=1}^n\sum_{l=1}^k\min\{y^l_i(\tau),q x^l_i(\tau)\}\right) &\forall q\in(0,1] \label{constr:innerDualSTIIDCoverage}
\\ &y^l_i(\tau)=\tau x^l_i(\tau) \label{constr:innerDualSTIIDThreshold}
&\forall i\in[n],l\in[k],\tau\in(0,1]
\\ &(\vx(\tau),\vy(\tau))\in\cP^k_n &\forall\tau\in(0,1]
\\ &\int_{\tau\in(0,1]} \mu(\tau)=1 &
\\ &\mu(\tau) \ge0 &\forall \tau\in(0,1]
\end{align}
\end{subequations}
\end{definition}

We now formalize that all of these are the correct formulations, which compute tight guarantees for DP/OST/ST algorithms relative to the prophet/ex-ante relaxation in the IID special case.

\begin{theorem}[Reformulations in IID Setting] \label{thm:iidFramework}
For any $k$ and $n>k$, when type distributions $\vG$ are constrained to be IID, the adversary's optimization problems over $\vG$ can be reformulated as
\begin{align*}
&\inf_{\vG:G_{1j}=\cdots=G_{nj}\forall j}\InnerLP^{\DP/\Proph}_{k,n}(\vG) &&=\IIDLP^{\DP/\Proph}_{k,n}
\\
&\inf_{\vG:G_{1j}=\cdots=G_{nj}\forall j}\InnerLP^{\DP/\ExAnte}_{k,n}(\vG) &&=\IIDLP^{\DP/\ExAnte}_{k,n}
\\
&\inf_{\vG:G_{1j}=\cdots=G_{nj}\forall j}\sup_{J,\rho}\ \InnerLP^{\OST(J,\rho)/\Proph}_{k,n}(\vG) &&=\sup_{\tau}\ \InnerLP^{\OST(\tau)/\Proph}_{k,n}
\\
&\inf_{\vG:G_{1j}=\cdots=G_{nj}\forall j}\sup_{J,\rho}\ \InnerLP^{\OST(J,\rho)/\ExAnte}_{k,n}(\vG) &&=\sup_{\tau}\ \InnerLP^{\OST(\tau)/\ExAnte}_{k,n}
\\
&\inf_{\vG:G_{1j}=\cdots=G_{nj}\forall j}\InnerLP^{\ST/\Proph}_{k,n}(\vG) &&=\IIDLP^{\ST/\Proph}_{k,n}
\\
&\inf_{\vG:G_{1j}=\cdots=G_{nj}\forall j}\InnerLP^{\ST/\ExAnte}_{k,n}(\vG) &&=\IIDLP^{\ST/\ExAnte}_{k,n}
\end{align*}
\end{theorem}
\Cref{thm:iidFramework} is proven in \Cref{sec:pfsec4}.

\subsection{Equivalence of $\DP/\ExAnte$, $\OST/\Proph$, and $\OST/\ExAnte$ in IID Setting} \label{sec:IIDost}

Similar to before,
OST's are easier to analyze because under constraints~\eqref{constr:OstIid}, the RHS that is common to~\eqref{constr:ProphIid} and~\eqref{constr:ExanteIid} can be rewritten as
\begin{align} \label{eqn:RHSsimplificationOSTIID}
\sum_{i=1}^n\sum_{l=1}^k \min\{y^l_i,qx^l_i\}=\min\{\tau,q\}\sum_{i=1}^n\sum_{l=1}^k x^l_i,
\end{align}
where now term $\min\{\tau,q\}$ depends only on the choice of $\tau$ and not on agent $i$ nor the number of remaining slots $l$.
Moreover, the analogues of the relationships in \Cref{lem:analyzeOST} are that assuming $y^l_i=\tau x^l_i$ for all $i$ and $l$, for all $i\in[n]$, we have
\begin{align} \label{eqn:lemAnalogues}
\sum_{l=1}^k x^l_i =\Pr[\Bin(i-1,\tau)<k]\qquad\text{and}\qquad\tau\sum_{i'=1}^i\sum_{l=1}^k x^l_{i'} =\bE[\min\{\Bin(i,\tau),k\}].
\end{align}

This allows us to prove the following \namecref{thm:iidST}.
Although this result is well-known in the literature, {as explained in \Cref{sec:introNewContr}}, we emphasize that our framework simultaneously obtain both the lower bound and the upper bound of the ratios, without explicitly constructing a counterexample to show the upper bound and constructing a policy to show the lower bound.
\begin{theorem} \label{thm:iidST}
For any fixed $k$ and $n$,
\begin{align*}
\IIDLP^{\DP/\ExAnte}_{k,n}=\sup_{\tau}\ \IIDLP^{\OST(\tau)/\Proph}_{k,n}=\sup_{\tau}\ \IIDLP^{\OST(\tau)/\ExAnte}_{k,n}=\frac{\bE[\min\{\Bin(n,k/n),k\}]}{k}.
\end{align*}
\end{theorem}
\Cref{thm:iidST} is formally proved in \Cref{sec:pfsec4}.

\subsection{Equivalence of Oblivious and Non-oblivious Static Thresholds in IID Setting} \label{sec:IIDst}

We now show that even the best static threshold performance $\ST(I)$ knowing the full instance $I$ cannot beat the quantities in \Cref{thm:iidST}.
In fact, we prove a stronger result which was not true in the non-IID setting (\Cref{sec:NoniidOST})---$\ST(I)$ is no better than oblivious static thresholds on \textit{any} instance.

To prove this fact, we show that in the $\ST$ dual problem~\eqref{lp:innerDualSTIID}, one never benefits from using a convex combination of thresholds instead of a single threshold.
It suffices to show that given any two thresholds $\utau,\otau$ with $\utau<\otau$, there exists a $\tau$ lying between them which contributes more to the RHS of~\eqref{constr:innerDualSTIIDCoverage} than the average of $\utau$ and $\otau$, \textit{simultaneously} for every $q\in(0,1]$.
This is formalized in the \namecref{thm:iidSTNoBetter} below, which is formally proved in \Cref{sec:pfsec4}.

\begin{theorem} \label{thm:iidSTNoBetter}
Given any $\utau,\otau$ with $\utau<\otau$, there exists a $\tau\in[\utau,\otau]$ such that
\begin{align} \label{eqn:iidSTNoBetter}
\sum_{i=1}^n\sum_{l=1}^k\min\{y^l_i(\tau),q x^l_i(\tau)\} &\ge\frac{1}{2}\left(\sum_{i=1}^n\sum_{l=1}^k\min\{y^l_i(\utau),q x^l_i(\utau)\}+\sum_{i=1}^n\sum_{l=1}^k\min\{y^l_i(\otau),q x^l_i(\otau)\}\right)
\end{align}
holds for any $q\in(0, 1]$.
\end{theorem}

% \Cref{thm:iidSTNoBetter} is proven in \Cref{sec:pfsec4}.
% The main idea involves setting $\tau$ so that $\bE[\min\{\Bin(n,\tau),k\}]=(\bE[\min\{\Bin(n,\utau),k\}]+\bE[\min\{\Bin(n,\otau),k\}])/2$, and showing through a careful sequence of stochastic comparisons that this implies
% \begin{align} \label{eqn:totAvailProb}
% \sum_{i=1}^n\Pr[\Bin(i-1,\tau)<k]\ge\frac{1}{2}\left(\sum_{i=1}^n\Pr[\Bin(i-1,\utau)<k]+\sum_{i=1}^n\Pr[\Bin(i-1,\otau)<k]\right).
% \end{align}
% That is, if $\tau$ is set so that the number of agents accepted is the mean of that for $\utau$ and $\otau$, then its average probability of having a slot available across $i=1,\ldots,n$ can only be higher than the mean of that for $\utau$ and $\otau$.
% We note that simply setting $\tau=(\utau+\otau)/2$ does not work, which can be seen through a simple example where $k=1$, $n=3$, $\utau=1/3$, $\otau=1$.  Then the LHS of~\eqref{eqn:totAvailProb} when $\tau=2/3$ is $1+1/3+1/9=13/9$.  Meanwhile, the RHS of~\eqref{eqn:totAvailProb} is $\frac{(1+2/3+4/9)+1}{2}=14/9$, and hence setting $\tau=(\utau+\otau)/2$ would not suffice.

\Cref{thm:iidSTNoBetter} shows that when designing the convex combination of thresholds given by $\mu(\tau)$ in problem~\eqref{lp:innerDualSTIID}, if there is mass on two distinct thresholds $\utau,\otau$, then it is always better to move them to be a single mass at some intermediate threshold.  This means that ultimately it is better to place all the mass at one point, leading to the following \namecref{cor:OSTandSTsameInIID}.

\begin{corollary} \label{cor:OSTandSTsameInIID}
It holds that $$\IIDLP^{\ST/\Proph}_{k,n}=\sup_\tau\IIDLP^{\OST(\tau)/\Proph}_{k,n}$$ and $$\IIDLP^{\ST/\ExAnte}_{k,n}=\sup_\tau\IIDLP^{\OST(\tau)/\ExAnte}_{k,n}.$$
\end{corollary}

We note that the same argument works on a particular instance, defined by the type distribution given by $\{G_j:j\in[m]\}$ that is common across agents (in which case constraints~\eqref{constr:innerDualSTIIDCoverage} only need to be checked at the points $q=G_j$ for some $j\in[m]$).
Therefore, OST's are \textit{instance-optimal} for static threshold policies in the IID setting.

\subsection{$\DP/\Proph$ in IID Setting: a Numerical Procedure for General $k\geq1$}\label{sec:DPProphIID}

We now develop a numerical procedure to compute the guarantee of $\DP/\Proph$ for each fixed $k$ and $n$, i.e., we compute the value of $\IIDLP^{\DP/\Proph}_{k,n}$. %\Willdelete{By letting $n\rightarrow\infty$, we obtain the tight guarantee of adaptive algorithms for $k$ selection slots, and our guarantees for $k>1$ exceed the state-of-the-art from the free-order setting due to \cite{beyhaghi2021improved}.}
First, we note that $\IIDLP^{\DP/\Proph}_{k,n}$ can be formulated as follows.
\begin{align*}
\IIDLP^{\DP/\Proph}_{k,n}= \max\   \theta
\\ \mathrm{s.t.\ } \theta\cdot\bE[\min\{\Bin(n,q),k\}] &\le\sum_{i=1}^n\sum_{l=1}^k \min\{y^l_i,qx^l_i\} & \forall q\in(0,1]
\\ (\vx,\vy) &\in\cP^k_n.
\end{align*}
The main difficulty for obtaining the value of $\IIDLP^{\DP/\Proph}_{k,n}$ is that the above optimization problem is a semi-infinite LP where there are infinite number of constraints, which is due to the requirement that $\theta\cdot\bE[\min\{\Bin(n,q),k\}]\le\sum_{i=1}^n\sum_{l=1}^k \min\{y^l_i,qx^l_i\}$ needs to be satisfied for any $q\in(0,1]$. We now develop a way to discretize the interval $(0,1]$ into a finite number of points and we approximate the value of $\IIDLP^{\DP/\Proph}_{k,n}$ by solving a finite-dimensional LP, with promised theoretical guarantees.

To ease notations, we treat $k$ as a fixed value and omit $k$ in our expression unless otherwise specified. We denote by $B_n(q)=\bE[\min\{\Bin(n,q),k\}]$ and denote by $A_n(q,\vx,\vy)=\sum_{i=1}^n\sum_{l=1}^k \min\{y^l_i,q x^l_i\}$. For any set of constraints $\cK\subseteq(0,1]$, we denote by
\begin{align*}
\IIDLP_n(\cK):=\max_{(\vx,\vy)\in\cP_n}\inf_{q\in\cK}\frac{A_n(q,\vx,\vy)}{B_n(q)}.
\end{align*}
Our goal is to approximate $\IIDLP_n((0,1])$ within any additive error by specifying a $\cK\subseteq(0,1]$ that contains a finite number of elements. In the following lemma, we show how to specify $\cK\subseteq(0,1]$ to achieve this goal.
\begin{proposition}\label{prop:DiscreteIID}
Fix $n$. For any $\eps>0$, let $\cK=\{q_M,\ldots,q_1\}$ be a discretization of $(0,1]$ with $0<q_M<\cdots<q_1=1$, satisfying:
\begin{enumerate}
\item $B_n(q_{j+1})\ge(1-\eps)B_n(q_j)$, for all $j\in[M-1]$;
\item $\Pr[\Pois(n q_M)>k]\le n q_M\eps$.
\end{enumerate}
Then it holds
\begin{align} \label{eqn:bleh}
\IIDLP_n(\cK)\ge\IIDLP_n((0,1])\ge(1-\eps)\IIDLP_n(\cK).
\end{align}
\end{proposition}

Therefore, for any $k$ and $n$, one can fix the accuracy level $\eps$ and then obtain the set $\cK$ using \Cref{prop:DiscreteIID}. This allows to approximate $\IIDLP_n((0,1])$ within a multiplicative factor of $1- \eps$ by solving the finite-dimensional LP $\IIDLP_n(\cK)$.

In \Cref{tab:approguarantee}, we report the numerical results for $\DP/\Proph$ in the IID setting for $k$ from 1 to 5, following the approach described above. We set $\eps=0.001$ to approximate the value of $\IIDLP^{\DP/\Proph}_{k,n}$, and fix $n=1000$. For $k=1, \ldots,5$, 
we use an LP solver to compute $\IIDLP_{1000}(\cK)$ with accuracy level set to be $0.001$, which produces a guaranteed upper bound on $\IIDLP^{\DP/\Proph}_{k,1000}$. A guaranteed lower bound can be obtained by subtracting the numerical error from the LP solver solution and then multiplying by $1-\eps$, which by~\eqref{eqn:bleh} is a  lower bound on $\IIDLP^{\DP/\Proph}_{k,1000}$. %$\IIDLP_{1000}((0,1])$.
%for $k=1,\ldots,5$ and $n=1000$, plotting the value of $1-\eps$ times a lower bound on $\IIDLP_{1000}(\cK)$ guaranteed by the LP solver, which by~\eqref{eqn:bleh} is a lower bound on $\IIDLP_{1000}((0,1])$.
% Our numerical guarantees for $n = 1000$ are significantly higher than the previous-best bounds.

{Though we didn't explicitly compute the worst-case ratios for every $k$, we know that the tight order for the worst-case ratios should be $1-O(1/\sqrt{k})$. To see this point, the algorithm and the result from \cite{alaei2014bayesian} implies a $1-O(1/\sqrt{k})$ lower bound for the ratio. On the other hand, if we consider a special case where $k$ scales linearly with $n$, then the $\Omega(\sqrt{n})$ regret lower bound established in a series of work (e.g. Lemma 1 from \cite{arlotto2019uniformly} and Proposition 2 from \cite{jiang2020online}) implies the $1-\Omega(1/\sqrt{k})$ upper bound of the ratio under the IID setting.}

\begin{table}[!h]
    \centering
    {\footnotesize
    \begin{tabular}{|c|c|c|c|c|c|}
    \hline
     $k$    & 1 & 2 & 3 & 4 & 5  \\
     \hline
     Numerical upper bounds for $n=1000$    & 0.7475 & 0.8377 & 0.8748 & 0.8955 & 0.9093 \\
     \hline
     Numerical lower bounds for $n=1000$ & 0.7445 & 0.8347 & 0.8718 & 0.8925 & 0.9063 \\
     \hline
    $1-\frac{k^k}{k!e^k}$ & 0.6321 & 0.7293 & 0.7760 & 0.8046 & 0.8245 \\
    \hline
    \cite{beyhaghi2021improved} & 0.6543 & 0.7427 & 0.7857 & 0.8125 & 0.8311 \\
    \hline
    \end{tabular}
    }
    \caption{Numerically-verified upper and lower bounds on the tight guarantees for $k=1,\ldots,5$ and $n=1000$. We also report the previously best known lower bounds $1-\frac{k^k}{k!e^k}$ and the free-order guarantees from \cite{beyhaghi2021improved}.}
    \label{tab:approguarantee}
\end{table}

\subsection{$\DP/\Proph$ in IID Setting: A Simplified Derivation of the Tight $\alpha=0.745$ for $k=1$}\label{sec:IID745}
We now assume $k=1$ and show how our framework can be used to recover the tight guarantee $0.745$ for the optimal online policy relative to the prophet, by further exploiting the optimality structure of $\IIDLP^{\DP/\Proph}_{1,n}$. In the following part, we first derive a relaxation of $\IIDLP^{\DP/\Proph}_{1,n}$, denoted as $\LPre_n$, and we obtain an optimal solution of $\LPre_n$ with a closed form. We then establish that the relaxation does not improve the objective value, i.e. $\IIDLP^{\DP/\Proph}_{1,n}=\LPre_n$ , showing that an optimal solution of $\LPre_n$ can be transformed into an optimal solution of $\IIDLP^{\DP/\Proph}_{1,n}$. Finally, based on the closed form solution of $\LPre_n$, we show that the worst case occurs as $n\rightarrow\infty$ and we obtain $0.745$ as the tight guarantee.
We now elaborate on the key techniques in this three-step derivation, with all proofs of statements relegated to \Cref{sec:pfsec4}.

\subsubsection*{First step.}
We rewrite $\IIDLP^{\DP/\Proph}_{1,n}$ with superscript $l$ omitted:
\begin{subequations}
\begin{eqnarray}
\IIDLP^{\DP/\Proph}_{1,n}=&\max\  &\theta \label{lp:first}
\\ 
&\text{s.t. } & (1-(1-\kappa)^n)\theta \le\sum_{i=1}^n\min\{y_i,\kappa(1-\sum_{i'=1}^{i-1}y_{i'})\} \quad \forall\kappa\in(0,1] \label{eqn:sumOfMin}
\\ & & \sum_{i=1}^ny_i \le1 \nonumber
\\ & & y_i \ge0 \quad \forall i\in[n] \nonumber
\end{eqnarray}
\end{subequations}
This is a semi-infinite program. The constraint ~\eqref{eqn:sumOfMin} is equivalent to 
\begin{align} \label{eqn:6616}
(1-(1-\kappa)^n)\theta &\leq \sum_{i\in S}y_i+\kappa\cdot\sum_{i\in[n]\setminus S}(1-\sum_{i'=1}^{i-1}y_{i'})  \quad \quad \forall\kappa\in[0,1],S\subseteq[n],
\end{align}
which can be relaxed to
\begin{equation} \label{eqn:8107}
(1-(1-\kappa)^n)\theta \le\sum_{i=1}^I y_i+\kappa\sum_{i=I+1}^n(1-\sum_{i'=1}^{i-1}y_{i'}) \quad \quad \forall\kappa\in(0,1], I=0,1 \ldots, n.
\end{equation}
We later show that this relaxation does not improve the objective value of $\IIDLP^{\DP/\Proph}_{1,n}$.

Letting $Y_i=\sum_{i'=1}^iy_{i'}$ for all $i\in[n]$, we now consider the optimization problem
\begin{subequations}\label{lp:monotone}
\begin{eqnarray}
& \max\ & \theta \label{lp:second}
\\ &\text{s.t. } & \max_{\kappa \in (0,1]} \left\{(1-(1-\kappa)^n)\theta -\kappa\sum_{i=I}^{n-1}(1-Y_i) \right\}\le Y_I \quad  \forall  I=0,\ldots,n \label{eqn:minOverKappaI}
\\ & & Y_0\leq 0\le Y_1\le\cdots\le Y_n \le 1. \label{eqn:contrK}
\end{eqnarray}
\end{subequations}
Notice that the LHS of~\eqref{eqn:minOverKappaI}
involves solving a concave maximization problem with $\kappa$ being the decision variable, which has a closed-form solution. Thus the semi-infinite constraint \eqref{eqn:minOverKappaI} can be replaced by an equivalent nonlinear but finite-dimensional constraint. 
Thus, we have obtained a finite-dimensional nonlinear program as 
a relaxation of $\IIDLP^{\DP/\Proph}_{1,n}$. This discussion is formalized in  Lemma \ref{lem:LPmon} below. Its proof, which requires a variable substitution of the form $z_I=\frac{1}{n\theta}\sum_{i=I}^{n-1}(1-Y_i)$, is elementary.

\begin{lemma}[Relaxation after Eliminating $\kappa$ and Substituting Variables]\label{lem:LPmon}
It holds that \  $\IIDLP^{\DP/\Proph}_{1,n}\le \LPre_n$ where
\begin{eqnarray*}
\LPre_n:= &\max\ & \theta
\\ & \mathrm{s.t. } & (n-1)z_I^{n/(n-1)}  \le nz_{I+1}+\frac{1}{\theta}-1 \quad \quad \forall I=0,\ldots,n-1
\\ & & z_0=1, z_n =0
\\ & & z_i \in[0,1]  \quad \quad \forall i \in [n-1].
\end{eqnarray*}
\end{lemma}

 The optimization problem in \Cref{lem:LPmon} then has the following structured optimal solution where the $z_I$'s are decreasing from $z_0=1$ to $z_n=0$.

\begin{lemma}[Closed-Form Solution for $\LPre_n$]\label{Ostructurelemma}
Denote $\{\theta, z_I\}_{I=0}^n$ such that $z_0=1$, $z_n=0$ and
\[
z_{I+1}=\frac{n-1}{n}z_{I}^{n/(n-1)}-\frac{1}{n\theta}+\frac{1}{n},~~\forall I=0,\ldots,n-1
\]
Then $\{\theta, z_I\}_{I=0}^n$ is an optimal solution of $\LPre_n$.
\end{lemma}

\subsubsection*{Second step.}
We must show that the solution constructed in \Cref{Ostructurelemma} can be converted into a feasible solution of $\IIDLP^{\DP/\Proph}_{1,n}$ with identical objective value.
The challenge lies in verifying that the reverse substitution $y_i=n\theta(z_{i+1}-2z_i+z_{i-1})$ (with the $z_i$'s defined according to \Cref{Ostructurelemma}) satisfies all of the constraints in $\IIDLP^{\DP/\Proph}_{1,n}$, which can be distilled down to showing inequality (\ref{eqn:6616}).
A priori,~\eqref{eqn:6616} is only satisfied when $S$ takes the form $\{1,\ldots,I\}$ (since those are the constraints we kept in the first relaxation~\eqref{eqn:8107}); in fact~\eqref{eqn:6616} is even non-obvious when $\kappa=0$ and $S$ consists of a singleton $i$ (in which case~\eqref{eqn:6616} is equivalent to analytically checking that $y_i\ge0$).
To streamline the proof of~\eqref{eqn:6616}, we define a set function $f(S)$ that substitutes the pessimal value of $\kappa$ into~\eqref{eqn:6616} for each set $S$, and show this set function to be supermodular.
This allows us to ultimately show that it is maximized when $S$ takes the form of an interval $\{1,\ldots,I\}$, for which we already knew by construction that~\eqref{eqn:6616} is satisfied as equality.
This is all formalized in the proof of the \namecref{lem:LPequivalent} below.

\begin{lemma}\label{lem:LPequivalent}
It holds that $\IIDLP^{\DP/\Proph}_{1,n}=\LPre_n$ for each $n \ge 1$.
\end{lemma}

\subsubsection*{Third step.}

Having established $\IIDLP^{\DP/\Proph}_{1,n}=\LPre_n$, the proof is completed by showing that the objective value of $\LPre_n$ is minimized as $n\to\infty$.  Although monotonicity in $n$ is difficult to prove in general, we bypass this difficulty by comparing the values of $\LP_n$ with $\LP_{2n}$, which our closed-form solution allows us to do.
% By exploiting the closed formulation of the optimal solution found in \Cref{Ostructurelemma}, we have the following result, which shows that the worst case occurs when $n\rightarrow\infty$.
\begin{lemma}\label{Incnlemma}
For any $n \ge 1$, it holds that $\LPre_n\geq\LPre_{2n}$.
\end{lemma}
Thus, in order to obtain the tight worst case guarantee, it remains to analyze the behavior of the optimal solution $\{\theta, z_I\}_{I=0}^n$ of $\LPre_n$ when $n\to\infty$. By Lemma \ref{Ostructurelemma}, we have the recursive equation
\[
z_{I+1}=\frac{n-1}{n}z_{I}^{n/(n-1)}-\frac{1}{n\theta}+\frac{1}{n},~~\forall I=0,\ldots,n-1
\]
with $z_0=1$ and $z_n=0$. By treating $x$ as $I/n$ and $H(x)$ as $z_I$,the recursive equation above motivates the following differential equation,
\begin{equation}\label{equationH}
\frac{1}{\theta^*}-1 = H(x)(\ln H(x)-1)-H'(x), \qquad\forall x\in[0,1]
\end{equation}
with the boundary conditions $H(0)=1$ and $H(1)=0$, with $\theta^*$ being the precise constant that allows these relationships to hold. Then, we have the following result, which is the guarantee for $\DP/\Proph$ in the IID setting with $k=1$.
\begin{theorem}\label{theorem:limit}
It holds that $\inf_{n}\IIDLP^{\DP/\Proph}_{1,n}=\lim_{n\rightarrow\infty}\IIDLP^{\DP/\Proph}_{1,n}=\theta^*$, where $\theta^*$ is defined in (\ref{equationH}).
\end{theorem}
Note that from the definition of the function $H(x)$ and $\theta^*$, it holds that
\[
\int_1^0\frac{1}{1-\frac{1}{\theta^*}-H(1-\ln H)}dH =1.
\]
This recovers the same integral relationship as in \cite{hill1982comparisons,correa2017posted,liu2021variable} which is used to establish the numerical guarantee of $\alpha=\theta^*\approx0.745$.

\section{Concluding Remarks}

In this work, we derive tight guarantees for multi-unit prophet inequalities. The procedure of computing the tight guarantees is formulated as an optimization problem (LP) and by analyzing this optimization problem, we directly find the tight ratios without constructing any worst-case instances. Our analysis of this optimization problem is based on identifying a new ``Type Coverage'' dual problem from the optimization problem formulation. This new problem can be seen as akin to the celebrated Magician/OCRS problems, except more general in that it can also provide tight ratios relative to the prophet directly, whereas the Magician/OCRS problems only establish tight ratios relative to the ex-ante relaxation.
% that the realization can be one of many ordered types, instead of being binary (either "active" or not).
By further analyzing the structures of the ``Type Coverage'' dual problem, we derive new prophet inequalities and recover existing ones.
We first show that the "oblivious" static threshold policy achieves a best-possible guarantee among static threshold algorithms, under any number $k$ of starting units (even though it is not instance-optimal in the non-IID setting).
This result implies the tightness of the asymptotic convergence rate of $1-O(\sqrt{\log k/k})$ for static threshold algorithms, which dates back to \cite{hajiaghayi2007automated}, and establishes a separation with the convergence rate of adaptive algorithms, 
which is $1-\Theta(\sqrt{1/k})$ due to \cite{alaei2014bayesian}. Under the IID setting, we then use our framework to numerically illustrate the tight guarantee (of adaptive algorithms) under any number $k$ of starting units.  The tight guarantee has been previously characterized under $k=1$ by \cite{correa2017posted}, and our guarantees for $k>1$ exceed the state-of-the-art from the free-order setting due to \cite{beyhaghi2021improved}.

There are multiple new prospective directions for future research. First, there is a recent stream of work studying the performance of the online algorithm against the online optimum (DP) (e.g. \cite{papadimitriou2021online}). For the online optimum benchmark, one could still apply a similar LP duality approach by interpreting the coefficient $Q_j$ as the expected number of agents of type $j$ or better accepted by the online optimum. However, the caveat is that when making sense of the ``Type Coverage'' formulation, the value of $Q_j$ is implicitly defined through a DP (cf. the formulations in \Cref{prop:sumDeltaQ}), from which it may be difficult to prove results using the ``Type Coverage'' problem. Second, there are other variants of prophet inequalities defined on more general matroids beyond the $k$-uniform matroids (e.g. \cite{kleinberg2012matroid}). The challenge of extending our approach to general matroid prophet inequalities relies on lower bounding what an online policy can earn in the primal problem, which is also the adversary's problem of designing a worst-case instance. For $k$-uniform matroids, we could finish this step fairly cleanly for the DP/static threshold online policies by having $k$ state variables for when each agent arrives. However, for more general matroid constraints, if we still apply the same approach, there may need to be exponentially many variables for the state when each agent arrives, which might lead to an unwieldy dual. We leave these problems for future research.

\section*{Acknowledgments}
A one-page abstract of this work appeared in the proceedings of the 24th ACM Conference on \textit{Economics and Computation (EC'23)}.
The authors thank all anonymous reviewers whose comments helped improve the presentation of this paper.

\bibliographystyle{informs2014} % outcomment this and next line in Case 1
%\setcitestyle{numbers}
%\bibliographystyle{abbrv}
\bibliography{bibliography} % if more than one, comma separated

% FOR SUBMISSION
%\ECSwitch
%\ECDisclaimer
%\ECHead{E-Companion}

% FOR SSRN
\clearpage

% Appendix here
% Options are (1) APPENDIX (with or without general title) or
%             (2) APPENDICES (if it has more than one unrelated sections)
% Outcomment the appropriate case if necessary
%
% \begin{APPENDIX}{<Title of the Appendix>}
% \end{APPENDIX}
%
%   or
%

\begin{APPENDICES}
\crefalias{section}{appendix}
\crefalias{subsection}{appendix}

\section{Extending to Continuous Distributions}\label{sec:continuousDist}

In our paper, while we demonstrate our techniques and present our results under the assumption that each agent's reward follows a distribution with finite support, it is important to note that our methodology and findings are fully extendable to scenarios where the agent's reward follows a continuous distribution with infinite support. We can simply interpret each infinitesimal point in the reward support as a type and we can obtain a similar (semi-infinite) LP framework, where the dual LP represents the type coverage problem.
To be specific,
for fixed $n$, $k$, and for infinite support, we specify an index $q'\in(-\infty, +\infty)$ and the value of the agent can be expressed as $r(q')$ using the index $q'$, which holds even when the value distributions are unbounded. Here, $r(q')$ can be regarded as a function over $q'$, and this function is a decision variable to be optimized, restricted to be monotone increasing.
%we specify an interval $[0, \bar{r}]$ to cover the support of the value distribution of each agent and specify $r(q)=q\cdot\bar{r}$ for any $q\in[0,1]$. Therefore, we are able to express the value using the index $q$ for each agent, and 
Following this notation, we can express the value distribution for agent $i$ using CDF $F_i(q')$, which is defined on $q'\in(-\infty, +\infty)$. This formulation will cover the continuous support case.

We can further do the transformation noting that for any $q'\in(-\infty, \infty)$, there exists a unique $q\in[-1, 1]$ such that $q'=\tan(\frac{\pi}{2}\cdot q)$. Then, the value function $r(q')$ and the distribution function $F_i(q')$ for each agent $i$ can all be expressed using the index $q\in[-1, 1]$ as $r(q)$ and $F_i(q)$ for each agent $i$.
Thus, LP \eqref{lp:innerPrimal} can be rewritten in the following way. 
\begin{subequations} \label{lp:newinnerPrimal}
\begin{align}
\min\ &V^k_1& \label{eqn:primalObj}
\\ \mathrm{s.t.\ }&V_i^l=\int_{q=-\infty}^{+\infty} U^l_{i}(q)\cdot dF_{i}(q)+V^l_{i+1} &\forall i\in[n],l\in[k] \label{eqn:newdpValueToGo}
\\ &U^l_{i}(q) \ge\int_{q'=q}^{+\infty} dr(q)-(V^l_{i+1}-V^{l-1}_{i+1}) &\forall i\in[n],q\in[-1, 1], l\in[k] \label{eqn:newdpUtility}
\\ &\int_{q=-\infty}^{+\infty} Q(q)dr(q)=1 & \label{eqn:newoptIs1}
\\ &dr(q),U^l_{i}(q) \ge0 &\forall i\in[n],q\in[-1, 1],l\in[k] \label{eqn:newprimalNonneg}
\end{align}
\end{subequations}
Here, we would regard $dr(q)$ and $U_i^l(q)$ as decision variables, which play the same role as $\Delta_j$ and $U^l_{ij}$ in the original finite support formulation in LP \eqref{lp:innerPrimal}, except we are replacing the index $j$ by the index $q$. Also, $Q(q)$ plays the same role as $Q_j$ in the original finite support formulation in LP \eqref{lp:innerPrimal}, and $Q(q)$ denotes the expected number for the agents with valuation index among $[q, +\infty]$ to be accepted by the benchmark.

Note that the above problem \eqref{lp:newinnerPrimal} is infinite linear programming and we can write out the (simplified) dual exactly in the same way as we write out the dual for the finite support formulation. As long as the optimal objective value is finite, one can easily verify that \textit{Slater's conditions} hold for infinite LP \eqref{lp:newinnerPrimal} by setting $U^l_i(q)$ to always be larger than the right-hand side of constraint \eqref{eqn:newdpUtility}, which implies that strong duality holds between infinite LP \eqref{lp:newinnerPrimal} and its dual. The dual is as follows
\begin{subequations} \label{lp:newinnerDualSimplified}
\begin{align}
\max\ &\theta &\label{eqn:newdualObjSimplified}
\\ \mathrm{s.t.\ \ }&\theta\cdot Q(q) \le\sum_{i=1}^n\sum_{l=1}^k\min\{y^l_i,F_{i}(q)\cdot x^l_i\} &\forall q\in[-1, 1] \label{eqn:newdualCoverSimplified}
\\ &y^l_{i} \le x^l_i &\forall i\in[n],l\in[k] \label{eqn:newdualUBSimplified}
\\ &x^l_i =
\begin{cases}
1, &i=1,l=k \\
0, &i=1,l<k \\
x^l_{i-1}-y^l_{i-1}+y^{l+1}_{i-1}, &i>1 \\
\end{cases}
& \forall i\in[n],l\in[k] \label{eqn:newdualUpdateSimplified}
\\ &y^l_{i} \ge0 &\forall i\in[n],l\in[k] \label{eqn:newdualNonnegSimplified}
\end{align}
\end{subequations}
As we can see, the only difference between the dual \eqref{lp:newinnerDualSimplified} and the dual for the finite support formulation (LP \eqref{lp:innerDualSimplified}) is that the discrete index $j$ has been replaced by the continuous index $q\in[-1, 1]$. Following this line, we can write the other duals for $\OST$ and $\ST$ by replacing the discrete index with the continuous index and we will be able to obtain the same result for the continuous index formulation where the value distributions for the agents can take infinite support.

For example, we can prove the same result as \Cref{thm:rederiveWeightedBernoulli} in \Cref{sec:magicianOCRS}. Note that in the proof of \Cref{thm:rederiveWeightedBernoulli}, we have already shown that for a given sequence $g_1,\dots,g_n$ for the outer problem in \eqref{lp:OCRS} such that $\sum_{i=1}^n g_i\leq k$, we are able to construct an instance $\vG$ for which $\InnerLP^{\DP/\ExAnte}_{k,n}(\vG)$ is no greater than the optimal objective value of the inner problem in~\eqref{lp:OCRS}. This direction holds even when the distributions can be unbounded (we only need to find one such instance $\vG$ which could be the same as the one in the proof of \Cref{thm:rederiveWeightedBernoulli}). To prove the other direction, for an instance $\vG$ with the continuous index $q\in[-1, 1]$, we let the threshold $\rho$ be such that $\sum_{i=1}^nG_{i,\rho}=k$, which uniquely exists. Then we define $g_i=G_{i,\rho}$ for all $i$. Following the same procedure as the proof of \Cref{thm:rederiveWeightedBernoulli}, we can show that such $g_1,\dots, g_n$ forms a feasible solution to $\InnerLP^{\DP/\ExAnte}_{k,n}(\vG)$, which completes our proof of \Cref{thm:rederiveWeightedBernoulli}. Our other results can also be derived in the same way for unbounded distributions, with the instance now given by $G_{i,q}$ for continuous index $q\in[-1, 1]$, for each agent $i$.

% \section{Numerical Results for $\DP/\Proph$ in IID Setting}\label{sec:appendNu}
% In this section, we plot the numerical results for $\DP/\Proph$ in IID setting for general $k\geq1$, following the approach described in \Cref{sec:DPProphIID}. For each fixed $k$, we set $\eps=0.0001$ to approximate the value of $\IIDLP^{\DP/\Proph}_{k,n}$ with an error within $0.01\%$, for any $n$. We increase the value of $n$ until the value of $\IIDLP^{\DP/\Proph}_{k,n}$ appears to have stablized, as illustrated in \Cref{fig:approguarantee} for $k=1$ up to $k=10$. The numbers reported earlier in \Cref{tab:approguarantee} are for $n=8000$.

% \begin{figure}
% \centering
% \subfigure[$k=1$]{\includegraphics[width=0.3\textwidth]{01.jpg}}
% \subfigure[$k=2$]{\includegraphics[width=0.3\textwidth]{02.jpg}}
% \subfigure[$k=3$]{\includegraphics[width=0.3\textwidth]{03.jpg}}
% \subfigure[$k=4$]{\includegraphics[width=0.3\textwidth]{04.jpg}}
% \subfigure[$k=5$]{\includegraphics[width=0.3\textwidth]{05.jpg}}
% \subfigure[$k=6$]{\includegraphics[width=0.3\textwidth]{06.jpg}}
% \subfigure[$k=7$]{\includegraphics[width=0.3\textwidth]{07.jpg}}
% \subfigure[$k=8$]{\includegraphics[width=0.3\textwidth]{08.jpg}}
% \subfigure[$k=9$]{\includegraphics[width=0.3\textwidth]{09.jpg}}
% \subfigure[$k=10$]{\includegraphics[width=0.3\textwidth]{10.jpg}}
% \caption{The guarantee as $n\rightarrow\infty$, for $k=1$ up to $k=10$, with x-axis being $\log_2(n)$.}
% \label{fig:approguarantee}
% \end{figure}

\section{\Cref{tab:worstCaseRatioRelationships}} \label{apx:worstCaseRatioRelationships}

\begin{table}[!h]
\centering
\begin{tabular}{|cccc|}
\hline
\multicolumn{4}{|c|}{Stronger Benchmarks $\longrightarrow$} \\
& $\frac{\DP}{\Proph}$ & $>$ & $\frac{\DP}{\ExAnte}$ \\
More & & & $\vee$ \\
Restrictive & $\frac{\ST}{\Proph}$ & $=$ & $\frac{\ST}{\ExAnte}$ \\
Policies & \begin{turn}{90}$=$\end{turn} & & \begin{turn}{90}$=$\end{turn} \\
$\downarrow$ & $\frac{\OST}{\Proph}$ & $=$ & $\frac{\OST}{\ExAnte}$ \\
\hline
\end{tabular}$\qquad$
\begin{tabular}{|cccc|}
\hline
\multicolumn{4}{|c|}{Stronger Benchmarks $\longrightarrow$} \\
& $\frac{\DP}{\Proph}$ & $>$ & $\frac{\DP}{\ExAnte}$ \\
More & $\vee$ & & \begin{turn}{90}$=$\end{turn} \\
Restrictive & $\frac{\ST}{\Proph}$ & $=$ & $\frac{\ST}{\ExAnte}$ \\
Policies & \begin{turn}{90}$=$\end{turn} & & \begin{turn}{90}$=$\end{turn} \\
$\downarrow$ & $\frac{\OST}{\Proph}$ & $=$ & $\frac{\OST}{\ExAnte}$ \\
\hline
\end{tabular}
\caption{Relationships that we establish between worst-case ratios in the non-IID (left) and IID (right) settings.}
\label{tab:worstCaseRatioRelationships}
\end{table}

\section{Missing Proofs for \Cref{sec:problemclass}} \label{sec:pfsec2}

\begin{myproof}[Proof of \Cref{prop:sumDeltaQ}]
We denote by $X_j$ the number of agents that have a type $j$ and are accepted by the prophet. Clearly, we have
\[
\Proph(I)=\sum_{j=1}^m r_j\cdot\mathbb{E}[X_j]=\sum_{j=1}^m\Delta_j\cdot\mathbb{E}[\sum_{j'=1}^jX_{j'}]
\]
Moreover, note that the distribution of $\sum_{j'=1}^jX_{j'}$ is given by $\min\{\sum_{i=1}^n\Ber(G_{ij}), k\}$. Thus, it holds that
\[
\Proph(I)=\sum_{j=1}^m\Delta_j\cdot\mathbb{E}[\min\{\sum_{i=1}^n\Ber(G_{ij}), k\}]
\]
We now denote by $\{a^*_j\}_{j=1}^m$ one optimal solution to $\ExAnte(I)$. Clearly, for any $j_1<j_2$, if $a^*_{j_1}<\sum_{i=1}^np_{ij_1}$, it must hold that $a^*_{j_2}=0$. Otherwise, we construct another set of solution $\{a'_{j}\}_{j=1}^m$ by letting
\[\begin{aligned}
a'_{j}=a^*_{j}\text{~for~}j\neq j_1\text{~and~} j_2,~~ a'_{j_1}=a^*_{j_1}+\epsilon, ~~a'_{j_2}=a^*_{j_2}-\epsilon
\end{aligned}\]
where $0<\epsilon\leq\min\{a^*_{j_2}, \sum_{i=1}^np_{ij_1}-a^*_{j_1}\}$. Clearly, $\{a'_{j}\}_{j=1}^m$ is feasible to $\ExAnte(I)$ and yields a higher objective value than $\{a^*_j\}_{j=1}^m$, which contradicts the optimality of $\{a^*_j\}_{j=1}^m$. Thus, it holds that
\[
\sum_{j'=1}^ja^*_{j'}=\min\{\sum_{i=1}^nG_{ij}, k\}\text{~for~any~}j
\]
and
\[
\ExAnte(I)=\sum_{j=1}^mr_j\cdot a^*_j=\sum_{j=1}^m\Delta_j\cdot \sum_{j'=1}^ja^*_{j'}=\sum_{j=1}^m\Delta_j\cdot\min\{\sum_{i=1}^nG_{ij}, k\}
\]
which completes our proof of \eqref{eqn:6662}.
\end{myproof}

\begin{myproof}[Proof of \Cref{lem:dualSimplification}]
Given a feasible solution to~\eqref{lp:innerDual}, we construct the following feasible solution to~\eqref{lp:innerDualSimplified} with the same value of $\theta$.
Let $y^l_i=\sum_{j=1}^m y^l_{ij}$ for all $i\in[n]$ and $l\in[k]$.
Clearly constraints~\eqref{eqn:dualUpdateSimplified}--\eqref{eqn:dualNonnegSimplified} are satisfied.
Constraints~\eqref{eqn:dualUBSimplified} are satisfied due to the fact that $\sum_{j=1}^m p_{ij}=1$ for all $i$.
Finally, it remains to show that $\min\{y^l_i,G_{ij} x^l_i\}\ge \sum_{j'=1}^j y^l_{ij'}$ which would make constraints~\eqref{eqn:dualCoverSimplified} satisfied.
To see this, note that $\min\{y^l_i,G_{ij} x^l_i\}=\min\{\sum_{j=1}^m y^l_{ij},x^l_i\sum_{j'=1}^j p_{ij'}\}\ge\min\{\sum_{j=1}^m y^l_{ij},\sum_{j'=1}^j y^l_{ij'}\}$ where the inequality applies~\eqref{eqn:dualUB}.  Since both arguments in the $\min$ are at least $\sum_{j'=1}^j y^l_{ij'}$, this completes the proof.

Conversely, given a feasible solution to~\eqref{lp:innerDualSimplified}, we construct the following feasible solution to~\eqref{lp:innerDual} with the same value of $\theta$, which is the harder direction.
For each $i\in[n]$ and $l\in[k]$, we iteratively define
\begin{align*}
y^l_{i1} &=\min\{y^l_i,p_{i1}x^l_i\}
\\ y^l_{i2} &=\min\{y^l_i-y^l_{i1},p_{i2}x^l_i\}
\\ \cdots
\\ \\ y^l_{im} &=\min\{y^l_i-\sum_{j=1}^{m-1} y^l_{ij},p_{im}x^l_i\}.
\end{align*}
Constraints~\eqref{eqn:dualUB} hold from the second argument in the $\min$, while constraints~\eqref{eqn:dualNonneg} hold because by the first argument in the $\min$, the sum $\sum_j y^l_{ij}$ can never exceed $y^l_i$.
Meanwhile, it can be inductively established that $\sum_{j'=1}^j y^l_{ij'}=\min\{y^l_i,x^l_i\sum_{j'=1}^jp_{ij'}\}=\min\{y^l_i,G_{ij}x^l_i\}$ for all $j=1,\ldots,m$, establishing constraints~\eqref{eqn:dualCover}.  Finally, by the same fact $\sum_{j=1}^m y^l_{ij}=\min\{y^l_i,x^l_i\}=y^l_i$, establishing constraints~\eqref{eqn:dualUpdate} and completing the proof.
\end{myproof}

\begin{myproof}[Proof of \Cref{thm:NoniidOST}]
We consider the dual of LP \eqref{lp:innerPrimalOST}. We introduce $x_i^l$ as the dual variable for constraint \eqref{eqn:dpValueToGoOST}, the dual variable $y_{ij}^l$ for constraint \eqref{eqn:dpUtilityOST} and dual variable $\theta$ for constraint \eqref{eqn:optIs1OST}. Then, we get the following LP as the dual of LP \eqref{lp:innerPrimalOST}.
\begin{subequations} \label{lp:OSTinnerDual}
\begin{align}
\max\ \theta \label{eqn:OSTdualObj}
\\ \mathrm{s.t.\ }\theta\cdot Q_j &\le\sum_{i=1}^n\sum_{l=1}^k\sum_{j'=1}^j y_{ij'}^l &\forall j\in[m] \label{eqn:OSTdualCover}
\\ y^l_{ij} &=\begin{cases}
p_{ij} x^l_i, & j<J\\
p_{iJ}\rho x_i^l, & j=J\\
0, & j>J\\
\end{cases}&\forall i\in[n],l\in[k] \label{eqn:OSTdualUB}
\\ x^l_i &=
\begin{cases}
1, &i=1,l=k \\
0, &i=1,l<k \\
x^l_{i-1}-\sum_{j=1}^m y^l_{i-1, j}+\sum_{j=1}^m y^{l+1}_{i-1,j}, &i>1 \\
\end{cases}
& \forall i\in[n],l\in[k] \label{eqn:OSTdualUpdate}
\\ \theta &, x_{i}^l, y_{ij}^l\in\mathbb{R} &\forall i\in[n], \forall j\in[m], \forall l\in[k]\label{eqn:OSTdualNonneg}
\end{align}
\end{subequations}
For any $i\in[n]$, $l\in[k]$, we define $y_i^l=\sum_{j=1}^m y_{ij}^l$. Then, constraint \eqref{eqn:OSTdualUB} implies that
\[
y^l_i =(\sum_{j<J}p_{ij}+p_{iJ}\rho)x^l_i=((1-\rho)G_{i,J-1}+\rho G_{iJ})x^l_i,~~
\forall i\in[n],l\in[k].
\]
Moreover, for any $j\in[m]$, constraint \eqref{eqn:OSTdualUB} implies that
\[
\sum_{j'=1}^j y_{ij'}^l=\min\{y_i^l, G_{ij}x_i^l\},~~\forall i\in[n], l\in[k]
\]
Thus, a feasible solution to LP \eqref{lp:OSTinnerDual} can be translated into a feasible solution to the following LP, with the same objective value,
\begin{subequations} \label{lp:OSTinnerDualSimplified}
\begin{align}
\max\ \theta \label{eqn:OSTdualObjSimplified}
\\ \mathrm{s.t.\ }\theta\cdot Q_j &\le\sum_{i=1}^n\sum_{l=1}^k\min\{y^l_i,G_{ij}x^l_i\} &\forall j\in[m] \label{eqn:OSTdualCoverSimplified}
\\ y^l_{i} & =(\sum_{j<J}p_{ij}+p_{iJ}\rho)x^l_i=((1-\rho)G_{i,J-1}+\rho G_{iJ})x^l_i &\forall i\in[n],l\in[k] \label{eqn:OSTdualUBSimplified}
\\ x^l_i &=
\begin{cases}
1, &i=1,l=k \\
0, &i=1,l<k \\
x^l_{i-1}-y^l_{i-1}+y^{l+1}_{i-1}, &i>1 \\
\end{cases}
& \forall i\in[n],l\in[k] \label{eqn:OSTdualUpdateSimplified}
\\ y^l_{i} &\ge0 &\forall i\in[n],l\in[k]. \label{eqn:OSTdualNonnegSimplified}
\end{align}
\end{subequations}
We now show that a feasible solution to LP \eqref{lp:OSTinnerDualSimplified}, denoted by $\{\theta, x_i^l, y_i^l\}$, can be translated into a feasible solution to LP \eqref{lp:OSTinnerDual} with the same objective value, which implies that the objective value of LP \eqref{lp:OSTinnerDual} is equivalent to the objective value of LP \eqref{lp:OSTinnerDualSimplified}. To be specific, we define
\[
y_{ij}^l=\left\{\begin{aligned}
&p_{ij}x_i^l, &j<J\\
&p_{iJ}\rho x_i^l, &j=J\\
&0, &j>J
\end{aligned}  \right.
\]
Clearly, $\{\theta, x_i^l, y_{ij}^l\}$ satisfy the constraint \eqref{eqn:OSTdualUB} and \eqref{eqn:OSTdualUpdate}. We also have $\sum_{j'=1}^j y_{ij'}^l=\min\{y_i^l, G_{ij}x_i^l\}$ for any $i\in[n], l\in[k], j\in[m]$, which implies that constraint \eqref{eqn:OSTdualCover} is satisfied. Thus, $\{\theta, x_i^l, y_{ij}^l\}$ is a feasible solution to LP \eqref{lp:OSTinnerDual}.

For the problem instance $I=(\mathbf{G}, \mathbf{\Delta})$, we denote by $\OST_{J,\rho}(I)$ the total expected reward collected by the oblivious static threshold policy $J, \rho$ on problem instance $I$. Then, from the definition of LP~\eqref{lp:innerPrimalOST}, we have that
\[
\inf_{\mathbf{\Delta}}\frac{\OST_{J,\rho}(I)}{\Proph(I)}=\text{LP~}\eqref{lp:innerPrimalOST}\text{~with~variable~}Q_j=\mathbb{E}[\min\{\sum_{i=1}^n\Ber(G_{ij}),k\}]
\]
Similarly, we have
\[
\inf_{\mathbf{\Delta}}\frac{\OST_{J,\rho}(I)}{\ExAnte(I)}=\text{LP~}\eqref{lp:innerPrimalOST}\text{~with~variable~}Q_j=\min\{\sum_{i=1}^nG_{ij},k\}
\]
Note that LP \eqref{lp:OSTinnerDual} is the dual of LP \eqref{lp:innerPrimalOST}, and we have shown the objective value of LP \eqref{lp:OSTinnerDual} is equivalent to the objective value of LP \eqref{lp:OSTinnerDualSimplified}, which gives the expression of $\InnerLP^{\OST(J,\rho)/\Proph}_{k,n}(\vG)$ (resp. $\InnerLP^{\OST(J,\rho)/\ExAnte}_{k,n}(\vG)$) when $Q_j=\mathbb{E}[\min\{\sum_{i=1}^n\Ber(G_{ij}),k\}]$ (resp. $Q_j=\min\{\sum_{i=1}^nG_{ij},k\}$).
Thus, we have
\[
\inf_{\mathbf{\Delta}}\frac{\OST_{J,\rho}(I)}{\Proph(I)}=\InnerLP^{\OST(J,\rho)/\Proph}_{k,n}(\vG)\text{~and~}\inf_{\mathbf{\Delta}}\frac{\OST_{J,\rho}(I)}{\ExAnte(I)}=\InnerLP^{\OST(J,\rho)/\ExAnte}_{k,n}(\vG)
\]
Note that according to \Cref{Def:oblivious}, the choice of $J, \rho$ for $\OST$ can depend on $\mathbf{G}$. Thus, we know that the tight guarantee for $\OST$ relative to the prophet (resp. ex-ante relaxation) is given by
\[
\inf_{\mathbf{G}}\sup_{J,\rho}\InnerLP^{\OST(J,\rho)/\Proph}_{k,n}(\vG)\text{~(resp.~}\inf_{\mathbf{G}}\sup_{J,\rho}\InnerLP^{\OST(J,\rho)/\ExAnte}_{k,n}(\vG))
\]
which completes our proof.
%\begin{align*}
%y^l_{ij} &=
%\begin{cases}
%p_{ij} x^l_i, &j<J \\
%p_{ij}\rho x^l_i, &j=J \\
%0, &j>J \\
%\end{cases}
%\end{align*}
\end{myproof}

\begin{myproof}[Proof of \Cref{thm:NoniidST}]
We consider the dual of LP \eqref{lp:innerPrimalST}. We introduce the dual variable $x_i^l(J,\rho)$ for constraint \eqref{eqn:dpValueToGoST}, the dual variable $y_{ij}^l(J,\rho)$ for constraint \eqref{eqn:dpUtilityST}, the dual variable $\mu(J,\rho)$ for constraint \eqref{eqn:alphaLB} and the dual variable $\theta$ for constraint \eqref{eqn:optIs1ST}. Then, we get the following LP as the dual of LP \eqref{lp:innerPrimalST}.
\begin{subequations} \label{lp:STinnerDualST}
\begin{align}
\max\ &\theta&
\\ \mathrm{s.t.\ \ }&\theta\cdot Q_j \le\int_{J,\rho} \mu(J,\rho)\left(\sum_{i=1}^n\sum_{l=1}^k\sum_{j'=1}^j y_{ij'}^l(J,\rho)\right) ~~~~~\forall j\in[m] \ \label{PFSTDual1}
\\ &y^l_{ij}(J,\rho)=\begin{cases}
p_{ij} x^l_i(J,\rho), & j<J\\
p_{iJ}\rho x_i^l(J,\rho), & j=J ~~~~\forall i\in[n],l\in[k],J\in[m],\rho\in(0,1]\\
0, & j>J\\
\end{cases}
& \ \label{PFSTDual2}
\\ &x_i^l(J,\rho)=\begin{cases}
1, &i=1,l=k \\
0, &i=1,l<k ~~~~\forall i\in[n],l\in[k]\\
x^l_{i-1}(J,\rho)-\sum_{j=1}^m y^l_{i-1, j}(J,\rho)+\sum_{j=1}^m y^{l+1}_{i-1,j}(J,\rho), &i>1 \\
\end{cases}
& \ \label{PFSDual3}
\\ &\mu(J,\rho) \ge0 ~~~~~~~~~~~~\forall J\in[m],\rho\in(0,1]
\end{align}
\end{subequations}
Note that we can select a positive $\{x_i^l(J,\rho), y_{ij}^l(J,\rho)\}$ satisfying constraints \eqref{PFSTDual2} and \eqref{PFSDual3}, select $\mu(J,\rho)$ to be a uniform distribution over $J\in[m], \rho\in(0,1]$, and set $\theta=0$. Then, all the inequality constraints in LP \eqref{lp:STinnerDualST} can be satisfied as strict inequalities by $\{\theta, x_i^l(J,\rho), y_{ij}^l(J,\rho), \mu(J,\rho)\}$. Thus, the Slater's condition is satisfied and strong duality hols between LP \eqref{lp:innerPrimalST} and LP \eqref{lp:STinnerDualST} (Theorem 2.3 in \cite{shapiro2009semi}).

Now we define $y_i^l(J,\rho)=\sum_{j=1}^m y_{ij}^l(J,\rho)$ for any $i\in[n], l\in[k], J\in[m], \rho\in(0,1]$. Then, constraint \eqref{PFSTDual2} implies that
\[
y^l_i(J,\rho)=((1-\rho)G_{i,J-1}+\rho G_{iJ})x^l_i(J,\rho)
,~~\forall i\in[n],l\in[k],J\in[m],\rho\in(0,1]
\]
and
\[
\sum_{j'=1}^j y_{ij'}^l(J,\rho)=\min\{ y_i^l(J,\rho), G_{ij}x_i^l(J,\rho) \},~~\forall i\in[n], l\in[k], j, J\in[m], \rho\in(0,1]
\]
Thus, a feasible solution to LP \eqref{lp:STinnerDualST} can be translated into a feasible solution to LP \eqref{lp:innerDualST} with the same objective value.

On the other hand, we denote by $\{\theta, \vx(J,\rho), \vy(J,\rho), \mu(J,\rho)\}$ a feasible solution to LP \eqref{lp:innerDualST}. Then, for any $i\in[n],l\in[k],J\in[m],\rho\in(0,1]$, we define
\[
y_{ij}^l(J,\rho)=\begin{cases}
p_{ij} x^l_i(J,\rho), & j<J\\
p_{iJ}\rho x_i^l(J,\rho), & j=J\\
0, & j>J\\
\end{cases}
\]
Clearly, we have
$\sum_{j'=1}^j y_{ij'}^l(J,\rho)=\min\{ y_i^l(J,\rho), G_{ij}x_i^l(J,\rho) \}$ for any $i\in[n],l\in[k], j, J\in[m], \rho\in(0,1]$ and constraint \eqref{constraintDualST} implies that $\sum_{j=1}^m y_{ij}^l(J,\rho)=y_i^l(J,\rho)$ for any $i\in[n], l\in[k], J\in[m], \rho\in(0,1]$. Then, we have $\{\theta, x_i^l(J,\rho), y_{ij}^l(J,\rho), \mu(J,\rho)\}$ a feasible solution to LP \eqref{lp:STinnerDualST} with the same objective value. Thus, the objective value of LP \eqref{lp:STinnerDualST} is equivalent to the objective value of LP \eqref{lp:innerDualST}.

For the problem instance $I=(\mathbf{G}, \mathbf{\Delta})$, we denote by $\ST_{J,\rho}(I)$ the total expected reward collected by the static threshold policy $J, \rho$ on problem instance $I$. Then, from the definition of LP~\eqref{lp:innerPrimalST}, we have that
\[
\inf_{\mathbf{\Delta}}\sup_{J,\rho}\frac{\ST_{J,\rho}(I)}{\Proph(I)}=\text{LP~}\eqref{lp:innerPrimalST}\text{~with~variable~}Q_j=\mathbb{E}[\min\{\sum_{i=1}^n\Ber(G_{ij}),k\}]
\]
Similarly, we have
\[
\inf_{\mathbf{\Delta}}\sup_{J,\rho}\frac{\ST_{J,\rho}(I)}{\ExAnte(I)}=\text{LP~}\eqref{lp:innerPrimalST}\text{~with~variable~}Q_j=\min\{\sum_{i=1}^nG_{ij},k\}
\]
Note that LP \eqref{lp:STinnerDualST} is the dual of LP \eqref{lp:innerPrimalST}, where strong duality holds, and we have shown the objective value of LP \eqref{lp:STinnerDualST} is equivalent to the objective value of LP \eqref{lp:innerDualST}, which gives the expression of $\InnerLP^{\ST(J,\rho)/\Proph}_{k,n}(\vG)$ (resp. $\InnerLP^{\ST(J,\rho)/\ExAnte}_{k,n}(\vG)$) when $Q_j=\mathbb{E}[\min\{\sum_{i=1}^n\Ber(G_{ij}),k\}]$ (resp. $Q_j=\min\{\sum_{i=1}^nG_{ij},k\}$).
Thus, we have
\[
\inf_{\mathbf{\Delta}}\sup_{J,\rho}\frac{\ST_{J,\rho}(I)}{\Proph(I)}=\InnerLP^{\ST(J,\rho)/\Proph}_{k,n}(\vG)\text{~and~}\inf_{\mathbf{\Delta}}\sup_{J,\rho}\frac{\ST_{J,\rho}(I)}{\ExAnte(I)}=\InnerLP^{\ST(J,\rho)/\ExAnte}_{k,n}(\vG)
\]
Note that according to \Cref{Def:oblivious}, the choice of $J, \rho$ for $\ST$ can depend both on $\mathbf{G}$ and $\mathbf{\Delta}$. Thus, we know that the tight guarantee for $\ST$ relative to the prophet is given by
\[
\inf_{\mathbf{G}}\inf_{\mathbf{\Delta}}\sup_{J,\rho}\frac{\ST_{J,\rho}(I)}{\Proph(I)}=\inf_{\mathbf{G}}\InnerLP^{\ST(J,\rho)/\Proph}_{k,n}(\vG)
%\text{~(resp.~}\inf_{\mathbf{G}}\sup_{J,\rho}\InnerLP^{\OST(J,\rho)/\ExAnte}_{k,n}(\vG))
\]
and the tight guarantee for $\ST$ relative to the ex-ante relaxation is given by
\[
\inf_{\mathbf{G}}\inf_{\mathbf{\Delta}}\sup_{J,\rho}\frac{\ST_{J,\rho}(I)}{\ExAnte(I)}=\inf_{\mathbf{G}}\InnerLP^{\ST(J,\rho)/\ExAnte}_{k,n}(\vG).
\]
which completes our proof.
\end{myproof}

\section{Missing Proofs for \Cref{sec:nonIID}}\label{sec:pfsec3}

\begin{myproof}[Proof of \Cref{thm:rederiveWeightedBernoulli}]
First we show that $\inf_\vG \InnerLP^{\DP/\ExAnte}_{k,n}(\vG)\ge\eqref{lp:OCRS}$.
Given a $\vG$ for the LHS, we construct an instance defined by $g_1,\ldots,g_n$ for the inner maximization problem on the RHS and show that its optimal solution forms a feasible solution to $\InnerLP^{\DP/\ExAnte}_{k,n}(\vG)$, which would be sufficient.
To accomplish this, let $J\in[m],\rho\in(0,1]$ be such that $\sum_{i=1}^n (G_{i,J-1}+p_{iJ}\rho)=k$, which must uniquely exist since $G_{im}=1$ for all $i\in[n]$ and $n>k$.  Define $g_i=G_{i,J-1}+p_{iJ}\rho$ for all $i$, which we take to be our instance for the RHS, and consider an optimal solution defined by $\theta,\vx,\vy$ for its inner problem. We claim that this forms a feasible solution to $\InnerLP^{\DP/\ExAnte}_{k,n}(\vG)$.  To see why, note that for any $i$, the expression
$$
\frac{\sum_{l=1}^k \min\{y^l_i,G_{ij}x^l_i\}}{G_{ij}}=\sum_{l=1}^k \min\{\frac{y^l_i}{G_{ij}},x^l_i\}
$$
is decreasing in $G_{ij}$.  Therefore, for all $j<J$, since $G_{ij}\le g_i$, we have
$$
\frac{\sum_{l=1}^k \min\{y^l_i,G_{ij}x^l_i\}}{G_{ij}}\ge\frac{\sum_{l=1}^k \min\{y^l_i,g_ix^l_i\}}{g_i}\ge\theta.
$$
where the final inequality applies~\eqref{constr:OCRS}.  Therefore, for all $j<J$, we deduce
\begin{align*}
\frac{\sum_{i=1}^n\sum_{l=1}^k \min\{y^l_i,G_{ij}x^l_i\}}{\min\{\sum_{i=1}^nG_{ij},k\}}
&=\frac{\sum_{i=1}^n\sum_{l=1}^k \min\{y^l_i,G_{ij}x^l_i\}}{\sum_{i=1}^nG_{ij}}\ge\theta
\end{align*}
as required for~\eqref{constr:ExanteNoniid}.
Meanwhile, for all $j\ge J$, since $G_{ij}\ge g_i$, we directly have
\begin{align*}
\frac{\sum_{i=1}^n\sum_{l=1}^k \min\{y^l_i,G_{ij}x^l_i\}}{\min\{\sum_{i=1}^nG_{ij},k\}}
&\ge\frac{\sum_{i=1}^n\sum_{l=1}^k \min\{y^l_i,g_ix^l_i\}}{\sum_{i=1}^n g_i}\ge\theta
\end{align*}
which completes the proof that $\inf_\vG \InnerLP^{\DP/\ExAnte}_{k,n}(\vG)\ge\eqref{lp:OCRS}$.

To show that $\inf_\vG \InnerLP^{\DP/\ExAnte}_{k,n}(\vG)\le\eqref{lp:OCRS}$, which is the harder direction, given an instance defined by $g_1,\ldots,g_n$ for the outer problem in~\eqref{lp:OCRS} such that $\sum_{i=1}^ng_i\leq k$, we construct an instance $\vG$ for which $\InnerLP^{\DP/\ExAnte}_{k,n}(\vG)$ is no greater than the optimal objective value of the inner problem in~\eqref{lp:OCRS}.

The construction goes as follows. Define $m=n+1$. We let
\begin{equation}\label{ConstructG}
G_{ij}=g_i\cdot\bI(i\leq j)\text{~for~all~}j<m\text{~and~}G_{im}=1, \forall i\in[n].
\end{equation}
Note that the feasibility constraints of $G_{i1}\le\cdots\le G_{im}=1$ are satisfied for all $i$. Under this construction, it holds that
\[
Q_j=\min\{\sum_{i=1}^nG_{ij},k\}=\min\{\sum_{i=1}^jg_i,k \}=\sum_{i=1}^jg_i, \forall j\in[m] %\text{~and~}Q_m=\min\{n,k\}
\]
where $g_{m}=k-\sum_{i=1}^ng_i$.
%From \Cref{lem:dualSimplification}, it holds that $\InnerLP^{\DP/\ExAnte}_{k,n}(\vG)=\LP\eqref{lp:innerDual}$, which further implies that $\InnerLP^{\DP/\ExAnte}_{k,n}(\vG)=\LP\eqref{lp:innerPrimal}$.
Now, denote by $\{\Delta_j^*, U_{ij}^{l*}, V_i^{k*}\}$ one optimal solution of the following LP:
\begin{subequations} \label{Theorem4innerPrimal}
\begin{align}
\min\ &V^k_1 \label{thm4primalObj}
\\ \mathrm{s.t.\ \ }&V_i^l=\sum_{j=1}^m p_{ij}U^l_{ij}+V^l_{i+1} &\forall i\in[n],l\in[k] \label{thm4dpValueToGo}
\\ &U^l_{ij} \ge\sum_{j'=j}^m \Delta_{j'}-(V^l_{i+1}-V^{l-1}_{i+1}) &\forall i\in[n],j\in[m],l\in[k] \label{thm4dpUtility}
\\ &\sum_{j'=j}^m \Delta_{j'}\ge 0 &\forall j\in[m]
\\ &\sum_{j=1}^m Q_j\Delta_j=1 & \label{thm4optIs1}
\\ &\Delta_j\in\mathbb{R}, \Delta_m=0, U^l_{ij} \ge0 &\forall i\in[n],j\in[m],l\in[k] \label{thm4primalNonneg}
\end{align}
\end{subequations}
with $p_{ij}=G_{ij}-G_{i,j-1}$.
We further denote by $r^*_j=\sum_{j'=j}^m \Delta_{j'}^*$ for each $j\in[m]$ and denote by $\{\sigma(j), \forall j=1,\ldots,m\}$ a permutation of $\{1,\ldots,m\}$ such that $r^*_{\sigma(1)}\geq r^*_{\sigma(2)}\geq\ldots\geq r^*_{\sigma(m)}$. For each $j\in[m]$, we further denote %$\hat{G}_{ij}=g_{\sigma(i)}\cdot\bI(i\leq j)$
$\hat{G}_{ij}=\sum_{j'=1}^jp_{i\sigma(j')}$, for all $i\in[n]$. Then we have the following claim.
\begin{claim}\label{claim:deltaequal0}
It holds that $\LP\eqref{Theorem4innerPrimal} \geq\InnerLP^{\DP/\ExAnte}_{k,n}(\hat{\vG})$.
\end{claim}
On the other hand, the dual of LP\eqref{Theorem4innerPrimal} is given as
\begin{subequations} \label{Thm4innerDual}
\begin{align}
\max\ &\theta  &\label{Thm4dualObj}
\\ \mathrm{s.t.\ \ }&\theta\cdot Q_j+\sum_{j'=1}^j\alpha_{j'} =\sum_{i=1}^n\sum_{l=1}^k\sum_{j'=1}^j y^l_{ij'} &\forall j\in[n] \label{Thm4dualCover}
\\ &y^l_{ij}\le p_{ij} x^l_i &\forall i\in[n],j\in[m],l\in[k] \label{Thm4dualUB}
\\ &x^l_i=
\begin{cases}
1, &i=1,l=k \\
0, &i=1,l<k \\
x^l_{i-1}-\sum_{j=1}^m (y^l_{i-1,j}-y^{l+1}_{i-1,j}), &i>1 \\
\end{cases}
& \forall i\in[n],l\in[k] \label{Thm4dualUpdate}
\\ &y^l_{ij}\ge0, \alpha_j\geq0, &\forall i\in[n],j\in[m],l\in[k] \label{Thm4dualNonneg}
\end{align}
\end{subequations}
Denote by $\{\theta^*, \alpha_j^*, x_i^{l*}, y_{ij}^{l*}\}$ one optimal solution of LP\eqref{Thm4innerDual} and define $y_i^{l*}=\sum_{j=1}^m y_{ij}^{l*}$ for each $i\in[n]$, $l\in[k]$. We now show that $\{\theta^*, x_i^{l*}, y_i^{l*}\}$ is a feasible solution to LP\eqref{lp:OCRS}.

Clearly, $(\vx^*, \vy^*)\in \mathcal{P}_n^k$. From constraint \eqref{Thm4dualCover}, for $j\in[n]$, we have that
\[
\theta^*\cdot (Q_{j}-Q_{j-1})+\alpha_j^*=\sum_{l=1}^k y_{jj}^{l*}\leq\sum_{l=1}^k\min\{y_j^{l*}, g_j x_i^{l*}\}
\]
where the last inequality follows from the construction of $\vG$ and constraint \eqref{Thm4dualUB}. Further note that for $j\in[n]$, we have $\theta^*\cdot g_j\leq \theta^*\cdot (Q_{j}-Q_{j-1})+\alpha_j^*$. We conclude that $\{\theta^*, x_i^{l*}, y_i^{l*}\}$ is a feasible solution to LP\eqref{lp:OCRS}. Thus, from \Cref{claim:deltaequal0}, we have that
\[
\InnerLP^{\DP/\ExAnte}_{k,n}(\hat{\vG})\leq\LP\eqref{Theorem4innerPrimal}=\LP\eqref{Thm4innerDual}=\theta^*\leq \LP\eqref{lp:OCRS}
\]
which completes our proof.
\end{myproof}
\begin{myproof}[Proof of \Cref{claim:deltaequal0}]
Note that from \Cref{lem:dualSimplification},  $\InnerLP^{\DP/\ExAnte}_{k,n}(\hat{\vG})$ is given as the optimal objective value of the following LP:
\begin{subequations} \label{Claim1innerPrimal}
\begin{align}
\min\ &V^k_1 \label{claimprimalObj}
\\ \mathrm{s.t.\ \ }&V_i^l=\sum_{j=1}^m \hat{p}_{ij}U^l_{ij}+V^l_{i+1} &\forall i\in[n],l\in[k] \label{claimdpValueToGo}
\\ &U^l_{ij}\ge\sum_{j'=j}^m \Delta_{j'}-(V^l_{i+1}-V^{l-1}_{i+1}) &\forall i\in[n],j\in[m],l\in[k] \label{claimdpUtility}
\\ &\sum_{j=1}^m \hat{Q}_j\Delta_j=1 & \label{claimoptIs1}
\\ &\Delta_j\geq0, U^l_{ij}\ge0 &\forall i\in[n],j\in[m],l\in[k] \label{claimprimalNonneg}
\end{align}
\end{subequations}
where $\hat{p}_{ij}=p_{i\sigma(j)}$ and
\[
\hat{Q}_j=\min\{\sum_{i=1}^n \hat{G}_{ij}, k\}, \forall j<m\text{~and~}\hat{Q}_m=\min\{n,k\}
\]
Note that $\sigma(m)=m$. We have $\hat{Q}_j=\sum_{i=1}^j g_{\sigma(i)}$.
We now construct a feasible solution to \eqref{Claim1innerPrimal} from $\{\Delta^*_j, U_{ij}^{l*}, V_i^{k*}\}$, which is the optimal solution of \eqref{Theorem4innerPrimal} that gives rise to the definition of the permutation $\{\sigma(j), \forall j=1,\ldots,m\}$. To be specific, we define
\[
\hat{V}_i^k=V_i^{k^*}, \hat{U}_{ij}^l=U^{l*}_{i\sigma(j)}, \text{~and~}\hat{\Delta}_j=r^*_{\sigma(j)}-r^*_{\sigma(j+1)}, \forall i\in[n], l\in[k], j\in[m]
\]
where we define $r^*_{\sigma(m+1)}=0$. Clearly, constraints \eqref{thm4dpValueToGo} and \eqref{thm4dpUtility} imply that constraints \eqref{claimdpValueToGo} and \eqref{claimdpUtility} are satisfied by $\{\hat{\Delta}_j, \hat{U}_{ij}^{l}, \hat{V}_i^{k}\}$. Also, we have
\[
\sum_{j=1}^m\hat{Q}_j\Delta_j=\sum_{j=1}^m r^*_{\sigma(j)}\cdot(\hat{Q}_j-\hat{Q}_{j-1})=\sum_{j=1}^m r^*_{\sigma(j)}\cdot g_{\sigma(j)}=1
\]
where the last equation follows from constraint \eqref{thm4optIs1}. Thus, our proof is completed.
\end{myproof}

\begin{myproof}[Proof of \Cref{lem:analyzeOST}]
We prove stronger results by induction. First, we show that for any $i$, it holds that
\begin{equation}\label{eqnlemma21}
\sum_{l'=l}^k x_i^{l'}=\Pr[\sum_{i'<i}\Ber(\tau_{i'})<k-l+1],~~\forall l=1,\ldots,k
\end{equation}
We prove \eqref{eqnlemma21} by induction on $i$. When $i=1$, clearly, for any $l=1,\dots,k$, we have
\[
\sum_{l'=l}^k x_1^{l'}=1=\Pr[0<k-l+1]=\Pr[\sum_{i'<1}\Ber(\tau_{i'})<k-l+1]
\]
which implies that \eqref{eqnlemma21} holds. We now assume that \eqref{eqnlemma21} holds for $i$, and we consider $i+1$. For any $l=1,\ldots,k$, we have that
\[
x_{i+1}^l=x_i^l-y_i^l+y_i^{l+1}=(1-\tau_i)\cdot x_i^l+\tau_i\cdot x_i^{l+1}
\]
where we denote $x_i^{k+1}=0$. Thus, we have
\[
\sum_{l'=l}^kx_{i+1}^{l'}=(1-\tau_i)\cdot\sum_{l'=l}^kx_i^{l'}+\tau_i\cdot\sum_{l'=l+1}^k x_i^{l'}
\]
On the other hand, conditioning whether $\Ber(\tau_i)=1$, we have
\[\begin{aligned}
\Pr[\sum_{i'<i+1}\Ber(\tau_{i'})<k-l+1]&=\Pr(\sum_{i'<i}\Ber(\tau_{i'})<k-l+1)\cdot\Pr[\Ber(\tau_i)=0]\\
&~~~+\Pr(\sum_{i'<i}\Ber(\tau_{i'})<k-l)\cdot\Pr[\Ber(\tau_i)=1]\\
&=(1-\tau_i)\cdot\Pr(\sum_{i'<i}\Ber(\tau_{i'})<k-l+1)+\tau_i\cdot\Pr(\sum_{i'<i}\Ber(\tau_{i'})<k-l)
\end{aligned}\]
From induction hypothesis, we know that
\[
\sum_{l'=l}^kx_i^{l'}=\Pr(\sum_{i'<i}\Ber(\tau_{i'})<k-l+1)\text{~and~}\sum_{l'=l+1}^k x_i^{l'}=\Pr(\sum_{i'<i}\Ber(\tau_{i'})<k-l)
\]
Thus, we have that
\[
\sum_{l'=l}^kx_{i+1}^{l'}=\Pr[\sum_{i'<i+1}\Ber(\tau_{i'})<k-l+1]
\]
From induction, we know that \eqref{eqnlemma21} holds for any $i\in[n]$, which proves the first equation in \Cref{lem:analyzeOST}.

We now prove the second equation. We prove by induction to show that for any $i$, it holds
\begin{equation}\label{eqnlemma22}
\sum_{i'=1}^i\tau_{i'}\cdot\sum_{l=1}^kx_{i'}^l=\mathbb{E}[\min\{\sum_{i'=1}^i\Ber(\tau_{i'}),k\}]
\end{equation}
When $i=1$, clearly, \eqref{eqnlemma22} holds. We now assume that \eqref{eqnlemma22} holds for $i$ and we consider $i+1$. We have
\[
\sum_{i'=1}^{i+1}\tau_{i'}\cdot\sum_{l=1}^kx_{i'}^l=\sum_{i'=1}^i\tau_{i'}\cdot\sum_{l=1}^kx_{i'}^l+\tau_{i+1}\cdot\sum_{l=1}^kx_{i+1}^l=\mathbb{E}[\min\{\sum_{i'=1}^i\Ber(\tau_{i'}),k\}]+\tau_{i+1}\cdot\sum_{l=1}^kx_{i+1}^l
\]
On the other hand, denote by $\mathcal{A}_i$ the event $\{\sum_{i'=1}^i\Ber(\tau_{i'})<k\}$. We have
\[
\mathbb{E}[\min\{\sum_{i'=1}^{i+1}\Ber(\tau_{i'}),k\}]=\mathbb{E}[\sum_{i'=1}^i\Ber(\tau_{i'})+\Ber(\tau_{i+1})|\mathcal{A}_i]\cdot\Pr(\mathcal{A}_i)+k\cdot(1-\Pr(\mathcal{A}_i))
\]
Note that the random variable $\Ber(\tau_{i+1})$ is independent of the event $\mathcal{A}_i$. We have
\[\begin{aligned}
\mathbb{E}[\min\{\sum_{i'=1}^{i+1}\Ber(\tau_{i'}),k\}]&=\mathbb{E}[\sum_{i'=1}^i\Ber(\tau_{i'})|\mathcal{A}_i]\cdot\Pr(\mathcal{A}_i)+k\cdot(1-\Pr(\mathcal{A}_i))+\mathbb{E}[\Ber(\tau_{i+1})]\cdot\Pr(\mathcal{A}_i)\\
&=\mathbb{E}[\min\{\sum_{i'=1}^{i}\Ber(\tau_{i'}),k\}]+\tau_{i+1}\cdot\Pr(\mathcal{A}_i)
\end{aligned}\]
From \eqref{eqnlemma21}, we know that
\[
\Pr(\mathcal{A}_i)=\sum_{l=1}^kx_{i+1}^l.
\]
Thus, we have
\[
\sum_{i'=1}^{i+1}\tau_{i'}\cdot\sum_{l=1}^kx_{i'}^l=\mathbb{E}[\min\{\sum_{i'=1}^i\Ber(\tau_{i'}),k\}]+\tau_{i+1}\cdot\sum_{l=1}^kx_{i+1}^l=\mathbb{E}[\min\{\sum_{i'=1}^{i+1}\Ber(\tau_{i'}),k\}]
\]
By induction, \eqref{eqnlemma22} holds for any $i\in[n]$, which completes our proof.
\end{myproof}

\begin{myproof}[Proof of \Cref{lem:exanteLB}]
By~\eqref{eqn:RHSsimplificationOST}, the objective value of $\InnerLP^{\OST(J,\rho)/\ExAnte}_{k,n}(\vG)$ for any OST defined by $J,\rho$ will equal
\begin{align*}
\theta
&=\min_{j\in[m]}\frac{\sum_{i=1}^n\min\{(1-\rho)G_{i,J-1}+\rho G_{iJ},G_{ij}\}\sum_{l=1}^kx^l_i}{\min\{\sum_{i=1}^nG_{ij},k\}}
\\ &=\min\left\{\min_{j<J}\frac{\sum_{i=1}^nG_{ij}\sum_{l=1}^kx^l_i}{\min\{\sum_{i=1}^nG_{ij},k\}},\min_{j\ge J}\frac{\sum_{i=1}^n((1-\rho)G_{i,J-1}+\rho G_{iJ})\sum_{l=1}^kx^l_i}{\min\{\sum_{i=1}^nG_{ij},k\}}\right\}
\\ &\ge\min\left\{\min_{j<J}\frac{\sum_{i=1}^nG_{ij}\sum_{l=1}^kx^l_i}{\sum_{i=1}^nG_{ij}},\frac{\sum_{i=1}^n((1-\rho)G_{i,J-1}+\rho G_{iJ})\sum_{l=1}^kx^l_i}{k}\right\}
\\ &\ge\min\left\{\min_{i\in[n]}\sum_{l=1}^kx^l_i,\frac{\sum_{i=1}^{n-1}((1-\rho)G_{i,J-1}+\rho G_{iJ})\sum_{l=1}^kx^l_i}{k}\right\}
\\ &=\min\left\{\min_{i\in[n]}\Pr\left[\sum_{i'<i}\Ber((1-\rho)G_{i',J-1}+\rho G_{i'J})<k\right],\frac{\bE[\min\{\sum_{i=1}^{n-1}\Ber((1-\rho)G_{i,J-1}+\rho G_{iJ}),k\}]}{k}\right\}
\end{align*}
where the first argument in the second inequality holds because $\frac{\sum_{i=1}^nG_{ij}\sum_{l=1}^kx^l_i}{\sum_{i=1}^nG_{ij}}\ge\min_{i\in[n]}\sum_{l=1}^kx^l_i$ for all $j<J$, and the final equality applies \Cref{lem:analyzeOST} throughout the terms.
The proof is then completed by the observation that the $\min$ over $i\in[n]$ is always achieved when $i=n$.
\end{myproof}

\begin{myproof}[Proof of \Cref{lem:prophUB}]
Given any distributions $\vG$ over $m$ types for the outer problem in~\eqref{eqn:2789}, we construct distributions $\vG'_1,\ldots,\vG'_n$ over $m+2$ types such that for some small $\eps>0$, we have that $\sup_{J',\rho'}\InnerLP^{\OST(J',\rho')/\Proph}_{k,n}(\vG'_1,\ldots,\vG'_n)$ is at most
\begin{align} \label{eqn:7892}
\frac{\eps}{k}+\sup_{J,\rho}\min\left\{\Pr\left[\sum_{i<n}\Ber((1-\rho)G_{i,J-1}+\rho G_{iJ})<k\right],\frac{\bE[\min\{\sum_{i<n}\Ber((1-\rho)G_{i,J-1}+\rho G_{iJ}),k\}]}{k}\right\}.
\end{align}
Taking $\eps\to0$ would then complete the proof.

The construction entails defining type distributions for agents $i=1,\ldots,n-1$ as
\begin{align*}
G'_{ij}
&=\begin{cases}
1, &j=m+2; \\
G_{i,j-1}, &j=2,\ldots,m+1; \\
0, &j=1. \\
\end{cases}
\end{align*}
For the last agent, we define $G'_{nj}=\eps$ for all $j\le m+1$ and $G'_{n,m+2}=1$.
Note that this feasibly satisfies $0\le G'_{i1}\le\cdots\le G'_{i,m+2}=1$ for all agents $i\in[n]$.

Recall that for $\theta$ to be feasible in $\InnerLP^{\OST(J',\rho')/\Proph}_{k,n}(\vG'_1,\ldots,\vG'_n)$, applying~\eqref{eqn:RHSsimplificationOST}, we need
\begin{align} \label{eqn:0981}
\theta\cdot\bE[\min\{\sum_{i=1}^n\Ber(G'_{ij}),k\}] &\le\sum_{i=1}^n\min\{(1-\rho')G'_{i,J'-1}+\rho' G'_{iJ'},G'_{ij}\}\sum_{l=1}^kx^l_i &\forall j\in[m+2].
\end{align}
Taking $j=1$, we have $\bE[\min\{\sum_{i=1}^n\Ber(G'_{ij}),k\}]=\bE[\min\{\Ber(\eps),k\}]=\eps$.
Meanwhile, the RHS of~\eqref{eqn:0981} can be at most $\eps\sum_lx^l_n$ when $j=1$.  Thus, we know that $\theta\le\sum_lx^l_n$.
Since the feasible vectors $\vx,\vy$ in $\InnerLP^{\OST(J',\rho')/\Proph}_{k,n}(\vG'_1,\ldots,\vG'_n)$ satisfies the static threshold constraint~\eqref{constr:OstNoniid} and $(\vx,\vy)\in\cP^k_n$,
by \Cref{lem:analyzeOST}, we know $\sum_lx^l_n=\Pr[\sum_{i<n}\Ber((1-\rho')G'_{i,J'-1}+\rho' G'_{iJ'})<k]$.

On the other hand, taking $j=m+1$, we have $\bE[\min\{\sum_{i=1}^n\Ber(G'_{ij}),k\}]=\bE[\min\{(n-1)+\Ber(\eps),k\}]=k$ since $G'_{i,m+1}=G_{im}=1$ for all $i\in[n-1]$ and $n>k$.
Meanwhile, the RHS of~\eqref{eqn:0981} can be at most
$
\sum_{i=1}^{n-1}((1-\rho')G'_{i,J'-1}+\rho' G'_{iJ'})\sum_{l=1}^kx^l_i+\eps
$
when $j=m+1$.  Thus, we also know that
\begin{align*}
\theta
&\le\frac{\eps}{k}+\frac{\sum_{i=1}^{n-1}((1-\rho')G'_{i,J'-1}+\rho' G'_{iJ'})\sum_{l=1}^kx^l_i}{k}
 &=\frac{\eps}{k}+\frac{\bE[\min\{\sum_{i=1}^{n-1}\Ber((1-\rho')G'_{i,J'-1}+\rho' G'_{iJ'}),k\}]}{k}
\end{align*}
where we have again applied \Cref{lem:analyzeOST}.

Finally, by setting $J=J'-1$ and $\rho=\rho'$, due to the construction of $\vG'_1,\ldots,\vG'_n$ based on $\vG$, expression~\eqref{eqn:7892} evaluates to exactly
$$
\frac{\eps}{k}+\min\left\{\Pr\left[\sum_{i<n}\Ber((1-\rho')G'_{i,J'-1}+\rho' G'_{iJ'})<k\right],\frac{\bE[\min\{\sum_{i<n}\Ber((1-\rho')G'_{i,J'-1}+\rho' G'_{iJ'}),k\}]}{k}\right\}.
$$
(Setting $J=J'-1$ is only valid if $J'\notin\{1,m+2\}$, but it is easy to see that the $\sup$ over $J',\rho'$ never requires these values of $J'$ to achieve.)
This completes the proof of \Cref{lem:prophUB}.
\end{myproof}

\begin{myproof}[Proof of \Cref{thm:chawlaTight}]
By \Cref{prop:exAnteUB}, $\Proph(I)\le\ExAnte(I)$ for any instance $I$, and hence
 $\InnerLP^{\OST(J,\rho)/\ExAnte}_{k,n}(\vG)\le\InnerLP^{\OST(J,\rho)/\Proph}_{k,n}(\vG)$.
Meanwhile, by \Cref{lem:exanteLB,lem:prophUB}, we have
\begin{align*}
&\inf_\vG\sup_{J,\rho}\ \InnerLP^{\OST(J,\rho)/\ExAnte}_{k,n}(\vG)
\\ \ge&\inf_\vG\sup_{J,\rho}\ \min\left\{\Pr\left[\sum_{i<n}\Ber((1-\rho)G_{i,J-1}+\rho G_{iJ})<k\right],\frac{\bE[\min\{\sum_{i<n}\Ber((1-\rho)G_{i,J-1}+\rho G_{iJ}),k\}]}{k}\right\}
\\ \ge&\inf_\vG\sup_{J,\rho}\ \InnerLP^{\OST(J,\rho)/\Proph}_{k,n}(\vG).
\end{align*}
Combining these inequalities completes the proof of \Cref{thm:chawlaTight}.
\end{myproof}

\begin{myproof}[Proof of \Cref{prop:StaticPolicy}]
We denote by
\[
A_{1,k}:=P(\Pois(\lambda_k)\leq k-1)\text{~and~}B_{1,k}:=\frac{\mathbb{E}[\min\{\Pois(\lambda_k), k\}]}{k}.
\]
We also let $A_{2,k}=1-A_{1,k}$ and $B_{2,k}=1-B_{1,k}$. Following the definition of $\lambda_k$, we clearly have that
\begin{equation}\label{eqn:011601}
A_{1,k}=B_{1,k}\text{~and~}A_{2,k}=B_{2,k}.
\end{equation}
We give an estimate of $\lambda_k$ using the relationship \eqref{eqn:011601}.

We note that
\[\begin{aligned}
1-B_{2,k}=B_{1,k}&=\sum_{d=1}^{k-1}\frac{d}{k}\cdot P(\Pois(\lambda_k)=d)+P(\Pois(\lambda_k)\geq k)=\sum_{d=1}^{k-1}\frac{d}{k}\cdot\frac{e^{-\lambda_k}\cdot\lambda_k^d}{d!}+A_{2,k}
\end{aligned}\]
We also note that
\[
\sum_{d=1}^{k-1}\frac{d}{k}\cdot\frac{e^{-\lambda_k}\cdot\lambda_k^d}{d!}=\frac{\lambda_k}{k}\cdot\sum_{d=1}^{k-1}\frac{e^{-\lambda_k}\cdot\lambda_k^{d-1}}{(d-1)!}=\frac{\lambda_k}{k}\cdot\sum_{d=0}^{k-2}\frac{e^{-\lambda_k}\cdot\lambda_k^{d}}{d!}=\frac{\lambda_k}{k}\cdot P(\Pois(\lambda_k)\leq k-2).
\]
Therefore, we have
\[\begin{aligned}
1-B_{2,k}=B_{1,k}&=\frac{\lambda_k}{k}\cdot P(\Pois(\lambda_k)\leq k-2)+A_{2,k}=\frac{\lambda_k}{k}\cdot (1-A_{2,k}-P(\Pois(\lambda_k)=k-1))+A_{2,k}
\end{aligned}\]
which implies that
\begin{equation}\label{eqn:122303}
  B_{2,k}=\frac{\lambda_k}{k}\cdot P(\Pois(\lambda_k)=k-1)+(1-\frac{\lambda_k}{k})\cdot(1-A_{2,k})
\end{equation}
We now let $\lambda_k=k-\sqrt{k\cdot\alpha_k}$ and we characterize $\alpha_k$. From Theorem 7 of \cite{hajiaghayi2007automated}, we know that
\begin{equation}\label{eqn:122304}
\Theta\left(\sqrt{\frac{1}{k}}\right)\leq A_{2,k}=B_{2,k}\leq \Theta\left(\sqrt{\frac{\log k}{k}}\right).
\end{equation}
It is clear to see that $\frac{1}{2}\leq1-A_{2,k}\leq 1$ when $k$ is large. Therefore, it is enough to consider two terms:
\[
\text{I}=P(\Pois(\lambda_k)=k-1)\text{~~and~~}\text{II}=1-\frac{\lambda_k}{k}.
\]
We first bound the term I. By stirling's approximation, we have
\[\begin{aligned}
P(\Pois(\lambda_k)=k-1)&=\frac{e^{-\lambda_k}\cdot\lambda_k^{k-1}}{(k-1)!}=\frac{e^{\sqrt{k\cdot\alpha_k}}\cdot(k-\sqrt{k\alpha_k})^{k-1}}{\sqrt{2\pi (k-1)}\cdot(k-1)^{k-1}}\\
&=e^{\sqrt{k\cdot\alpha_k}}\cdot \left(1-\sqrt{\frac{\alpha_k}{k}}\right)^{k-1}\cdot\frac{1}{\sqrt{2\pi(k-1)}\cdot\left(1-\frac{1}{k}\right)^{k-1}}.
\end{aligned}\]
We note that $\lim_{k\rightarrow\infty}\frac{\alpha_k}{k}=0$, otherwise, for any $\eps>0$ and any integer $N$, there exists $k>N$ such that $\text{II}=\sqrt{\frac{\alpha_k}{k}}\geq\epsilon$, which would imply $B_{2,k}\geq\frac{\epsilon}{2}$ following \eqref{eqn:122303} and thus violates \eqref{eqn:122304}.
Then, we have
$\left(1-\sqrt{\frac{\alpha_k}{k}}\right)^{k-1}=\Theta(e^{-\sqrt{k\cdot\alpha_k}})$
and
$e^{\sqrt{k\cdot\alpha_k}}\cdot \left(1-\sqrt{\frac{\alpha_k}{k}}\right)^{k-1}=\Theta(1)$.
Moreover, note that $\lim_{k\rightarrow\infty}\left(1-\frac{1}{k}\right)^{k-1}=\frac{1}{e}$.
Therefore, we get
$\text{I}=P(\Pois(\lambda_k)=k-1)=\Theta\left(\sqrt{\frac{1}{k}}\right)$,
from which we have
\begin{equation}\label{eqn:122305}
B_{2,k}=\Theta\left( \max\{\sqrt{\frac{1}{k}}, \sqrt{\frac{\alpha_k}{k}}\} \right),
\end{equation}
where term $\text{II}=\sqrt{\frac{\alpha_k}{k}}$.
We now bound term II and we apply the following result.
\begin{lemma}[Corollary 7.2 in \cite{harremoes2016bounds}]\label{lem:boundviaG}
Let
\[
G(x)=\sqrt{2(x\log(\frac{x}{\lambda_k})+\lambda_k-x)}\cdot\text{sign}(x-\lambda_k)
\]
where $\text{sign}(y)$ denotes the positiveness or the negativeness of $y$,
and let $\Phi$ be the cumulative distribution function of a standard normal distribution. Then, it holds that
\[
\Phi(G(k-2))\leq P(\Pois(\lambda_k)\leq k-1)=1-A_{2,k}\leq \Phi(G(k-1)).
\]
\end{lemma}
Following \Cref{lem:boundviaG}, we have
\begin{equation}\label{eqn:122306}
A_{2,k}=\Theta\left( 1-\Phi(G(k)) \right).
\end{equation}
Note that by Taylor's expansion for $G(k)$, we have
\[
G(k)=\sqrt{2(k\log(\frac{k}{k-\sqrt{k\cdot\alpha_k}})-\sqrt{k\cdot\alpha_k})}=\sqrt{\alpha_k}+o(1).
\]
Also, {see, e.g.,} \cite{duembgen2010bounding}, it holds
$1-\Phi(x)=\Theta\left( \frac{\exp(-x^2/2)}{x\sqrt{2\pi}}\right).$
Then, we get
\begin{equation}\label{eqn:122307}
A_{2,k}=\frac{1}{\sqrt{\alpha_k}}\cdot\exp(-\frac{\alpha_k}{2}).
\end{equation}
Combining \eqref{eqn:122305} and \eqref{eqn:122307}, from $A_{2,k}=B_{2,k}$, we have
\begin{equation}\label{eqn:122308}
\frac{\sqrt{\alpha_k}}{\sqrt{k}}=\frac{1}{\sqrt{\alpha_k}}\cdot\exp(-\alpha_k/2).
\end{equation}
Therefore, $\alpha_k$ is the solution to the Lambert W function, and it holds that $\alpha_k=\Theta(\log k)$, which implies that term $\text{II}=\Theta\left(\sqrt{\frac{\log k}{k}}\right)$. Our proof is thus completed.
\end{myproof}

\begin{myproof}[Proof of \Cref{prop:STbetterOST}]
We construct the instance $\vG$ as follows: there are two agents, $n=2$, four types, $m=4$, and one slot, $k=1$. The distribution for the first agent is given by $\vG_1=(0,\frac{1}{2},\frac{1}{2},1)$, and the distribution for the second agent is given by $\vG_2=(\eps,\eps,1,1)$.

For any $J,\rho$, we denote $C_{J,\rho}$ as a vector such that $C_{J,\rho}(j)=\sum_{i=1}^2\min\{y_i(J,\rho), G_{ij}x_i(J,\rho) \}$ for each $j\in[m]$, where $\vx(J,\rho), \vy(J,\rho)$ satisfy
\[\begin{aligned}
&x_1(J,\rho)=1, &y_1(J,\rho)=((1-\rho)G_{1,J-1}+\rho G_{1J})x_1(J,\rho)\\
&x_2(J,\rho)=x_1(J,\rho)-y_1(J,\rho), &y_2(J,\rho)=((1-\rho)G_{2,J-1}+\rho G_{2J})x_2(J,\rho).
\end{aligned}\]
Clearly, we have $C_{1,1}=(\eps,\eps,\eps,\eps)$ and $C_{3,1}=(\frac{\eps}{2}, \frac{1}{2}+\frac{\eps}{2}, 1, 1)$. Moreover, note that
\[
Q^{\Proph}=(\eps+o(\eps), \frac{1}{2}+O(\eps), 1, 1)\text{~and~}Q^{\ExAnte}=(\eps+o(\eps), \frac{1}{2}+O(\eps), 1, 1).
\]
By setting $\mu(1,1)=\frac{1}{3}$, $\mu(3,1)=\frac{2}{3}$, and $\mu(J,\rho)=0$ for all other $J,\rho$ in $\InnerLP^{\ST/\Proph}_{k,n}(\vG)$ and $\InnerLP^{\ST/\ExAnte}_{k,n}(\vG)$, we know that
\[
\InnerLP^{\ST/\Proph}_{k,n}(\vG)\geq\frac{2}{3}\text{~and~}\InnerLP^{\ST/\ExAnte}_{k,n}(\vG)\geq\frac{2}{3}.
\]
On the other hand, we show that an $\OST$ cannot achieve a guarantee better than $\frac{1}{2}$ with respect to both the ex-ante benchmark and the prophet benchmark.

If $J\leq 2$, irregardless of the value of $\rho$, we have that $y_2(J,\rho)=\eps x_2(J,\rho)\leq\eps$. Then we have that $C_{J,\rho}(3)\leq\min\{y_1(J,\rho), \frac{x_1(J,\rho)}{2}\}+\eps\leq\frac{1}{2}+\eps$. Compared with $Q^{\Proph}(3)=Q^{\ExAnte}(3)=1$, we know $\theta$ cannot be better than $\frac{1}{2}$ as $\eps\rightarrow0$.

If $J\geq3$, irregardless of the value of $\rho$, we have that $x_2(J,\rho)=x_1(J,\rho)-y_2(J,\rho)\leq \frac{1}{2}$, and thus $C_{J,\rho}(1)\leq\eps\cdot x_2(J,\rho)\leq\frac{\eps}{2}$. Compared with $Q^{\Proph}(1)=Q^{\ExAnte}(1)=\eps$, we know $\theta$ cannot be better than $\frac{1}{2}$ as $\eps\rightarrow0$. Thus, we have
\[
\sup_{J,\rho}\InnerLP^{\OST(J,\rho)/\Proph}_{k,n}(\vG)\leq\frac{1}{2}\text{~and~}\sup_{J,\rho}\InnerLP^{\OST(J,\rho)/\ExAnte}_{k,n}(\vG)\leq\frac{1}{2}
\]
as $\eps\rightarrow\infty$, which completes our proof.
\end{myproof}

\begin{myproof}[Proof of \Cref{lem:prophUBforST}]
The proof mainly follows the proof of \Cref{lem:prophUB}.
Given any distributions $\vG$ over $m$ types, we construct distributions $\vG'=(\vG'_1,\ldots,\vG'_n)$ over $m+2$ types such that for some small $\eps>0$, we have that
\begin{align} \label{eqn:ST7892}
\InnerLP^{\ST/\Proph}_{k,n}(\vG')\leq\frac{\eps}{k}+\min&\left\{\int_{J,\rho}\Pr\left[\sum_{i<n}\Ber((1-\rho)G_{i,J-1}+\rho G_{iJ})<k\right]\mu(J,\rho),\right.\nonumber\\
&\left.\int_{J,\rho}\frac{\bE[\min\{\sum_{i<n}\Ber((1-\rho)G_{i,J-1}+\rho G_{iJ}),k\}]}{k}\mu(J,\rho)\right\}.
\end{align}
for a probability measure $\mu$ over $(J,\rho)$.
Taking $\eps\to0$ would then complete the proof.

The construction entails defining type distributions for agents $i=1,\ldots,n-1$ as
\begin{align*}
G'_{ij}
&=\begin{cases}
1, &j=m+2; \\
G_{i,j-1}, &j=2,\ldots,m+1; \\
0, &j=1. \\
\end{cases}
\end{align*}
For the last agent, we define $G'_{nj}=\eps$ for all $j\le m+1$ and $G'_{n,m+2}=1$.
Note that this feasibly satisfies $0\le G'_{i1}\le\cdots\le G'_{i,m+2}=1$ for all agents $i\in[n]$.

Recall that for $\theta$ to be feasible in $\InnerLP^{\ST/\Proph}_{k,n}(\vG'_1,\ldots,\vG'_n)$, from constraints \eqref{constraintDualST} and \eqref{constraintDualST2}, we have
\begin{align} \label{eqn:ST0981}
\theta\cdot\mathbb{E}[\min\{\sum_{i=1}^n\Ber(G'_{ij}),k\}] &\le\int_{J',\rho'}\mu'(J',\rho')\left(\sum_{i=1}^n\min\{(1-\rho')G'_{i,J'-1}+\rho G'_{iJ'},G'_{ij}\}\sum_{l=1}^kx^l_i(J',\rho') \right),
\end{align}
for all $j\in[m+2]$, for some $(\vx(J',\rho'), \vy(J',\rho'))\in\mathcal{P}_n^k$ and a measure $\mu'$ over $J'\in[m+2]$ and $\rho'\in(0,1]$.

Taking $j=1$, we have $\mathbb{E}[\min\{\sum_{i=1}^n\Ber(G'_{i1}),k\}]=\eps$.
Meanwhile, when $j=1$, the RHS of~\eqref{eqn:ST0981} can be at most
$\int_{J',\rho'}\mu'(J',\rho')\cdot\left( \sum_{i=1}^nG'_{i1}\sum_{l=1}^kx_i^l(J',\rho')\right)=\eps\cdot\int_{J',\rho'}\mu'(J',\rho')\cdot\sum_{l=1}^k x_n^l(J',\rho')$.
Thus, we know that
$\theta\leq\int_{J',\rho'}\mu'(J',\rho')\cdot\sum_{l=1}^k x_n^l(J',\rho')$.
Since the feasible vectors $\vx(J',\rho'),\vy(J',\rho')$ in $\InnerLP^{\ST/\Proph}_{k,n}(\vG'_1,\ldots,\vG'_n)$ satisfies the static threshold constraint~\eqref{constr:OstNoniid} and $(\vx(J',\rho'), \vy(J',\rho'))\in\cP^k_n$,
by \Cref{lem:analyzeOST}, we know $\sum_lx^l_n(J',\rho')=\Pr[\sum_{i<n}\Ber((1-\rho')G'_{i,J-1}+\rho' G'_{iJ})<k]$ for each $J',\rho'$, which implies that
$\theta\leq \int_{J',\rho'}\mu'(J',\rho')\cdot\Pr[\sum_{i<n}\Ber((1-\rho')G'_{i,J-1}+\rho' G'_{iJ})<k]$.

On the other hand, taking $j=m+1$, we have $\mathbb{E}[\min\{\sum_{i=1}^n\Ber(G'_{i,m+1}),k\}]=k$ since $G'_{i,m+1}=G_{im}=1$ for all $i\in[n-1]$ and $n>k$.
Meanwhile, the RHS of~\eqref{eqn:ST0981} can be at most
$\int_{J'.\rho'}\mu'(J',\rho')\cdot\left(\sum_{i=1}^{n-1}((1-\rho')G'_{i,J'-1}+\rho' G'_{iJ'})\sum_{l=1}^kx^l_i(J',\rho')\right)+\eps$
when $j=m+1$.  Thus, we also know that
\begin{align*}
\theta
&\le\frac{\eps}{k}+\frac{\int_{J'.\rho'}\mu'(J',\rho')\cdot\left(\sum_{i=1}^{n-1}((1-\rho')G'_{i,J'-1}+\rho' G'_{iJ'})\sum_{l=1}^kx^l_i(J',\rho')\right)}{k}
\\ &=\frac{\eps}{k}+\int_{J',\rho'}\mu(J',\rho')\cdot\frac{\bE[\min\{\sum_{i=1}^{n-1}\Ber((1-\rho')G'_{i,J'-1}+\rho' G'_{iJ'}),k\}]}{k}
\end{align*}
where we have again applied \Cref{lem:analyzeOST} to derive the last equality.

Finally, by setting $J=J'-1$ and $\rho=\rho'$, due to the construction of $\vG'_1,\ldots,\vG'_n$ based on $\vG$, expression~\eqref{eqn:ST7892} evaluates to exactly
$$\begin{aligned}
\frac{\eps}{k}+\min&\left\{\int_{J',\rho'}\mu'(J',\rho')\Pr\left[\sum_{i<n}\Ber((1-\rho')G'_{i,J'-1}+\rho' G'_{iJ'})<k\right],\right.\\
&\left.\int_{J',\rho'}\mu'(J',\rho')\frac{\bE[\min\{\sum_{i<n}\Ber((1-\rho')G'_{i,J'-1}+\rho' G'_{iJ'}),k\}]}{k}\right\}.
\end{aligned}$$
(Setting $J=J'-1$ is only valid if $J'\notin\{1,m+2\}$, but it is easy to see that the $\sup$ over $J',\rho'$ never requires these values of $J'$ to achieve.)
This completes the proof of \Cref{lem:prophUBforST}.
\end{myproof}

\begin{myproof}[Proof of \Cref{thm:STequalOST}]
In \Cref{thm:chawlaTight}, we showed that
\[\begin{aligned}
&\inf_\vG\sup_{J,\rho}\ \InnerLP^{\OST(J,\rho)/\Proph}_{k,n}(\vG)=\inf_\vG\sup_{J,\rho}\ \InnerLP^{\OST(J,\rho)/\ExAnte}_{k,n}(\vG)\\
&=\inf_\vG\sup_{J,\rho}\ \min\left\{\Pr\left[\sum_{i<n}\Ber((1-\rho)G_{i,J-1}+\rho G_{iJ})<k\right],\frac{\bE[\min\{\sum_{i<n}\Ber((1-\rho)G_{i,J-1}+\rho G_{iJ}),k\}]}{k}\right\}.
\end{aligned}\]
From Lemma $11$ in \cite{chawla2020static}, the infimum is achieved when there is only $m=1$ type and $G_{i1}=1$ for all $i$, which implies that
\begin{align} \label{eqn:proof1098}
&\inf_{\vG'}\sup_{J,\rho}\ \InnerLP^{\OST(J,\rho)/\Proph}_{k,n}(\vG')=\inf_{\vG'}\sup_{J,\rho}\ \InnerLP^{\OST(J,\rho)/\ExAnte}_{k,n}(\vG')\nonumber \\
%&=\sup_{\rho\in(0,1]}\min\left\{\Pr\left[\Bin(n-1,\rho)<k\right],\frac{\bE[\min\{\Bin(n-1,\rho),k\}]}{k}\right\}\nonumber\\
&=\sup_{\rho\in(0,1]}\min_{\beta\in[0,1]}\beta\cdot\Pr\left[\Bin(n-1,\rho)<k\right]+(1-\beta)\cdot \frac{\bE[\min\{\Bin(n-1,\rho),k\}]}{k}.
\end{align}
Then, fixing $\vG$ such that $m=1$ and $G_{i1}=1$ for each $i\in[n]$, from \Cref{lem:prophUBforST}, we have
\begin{align} \label{eqn:STOST01}
&\inf_{\vG'}\InnerLP^{\ST/\ExAnte}_{k,n}(\vG')\leq\inf_{\vG'}\InnerLP^{\ST/\Proph}_{k,n}(\vG')\nonumber\\
&\leq\sup_{\mu:(0,1]\to\bR_{\ge0},\int_\rho \mu(\rho)=1}\min\left\{\int_{\rho}\Pr\left[\Bin(n-1,\rho)<k\right]\mu(\rho),\int_{\rho}\frac{\bE[\min\{\Bin(n-1,\rho),k\}]}{k}\mu(\rho)\right\}\nonumber\\
%&=\min_{\beta\in[0,1]} \sup_{\mu:(0,1]\to\bR_{\ge0},\int_\rho \mu(\rho)=1}\beta\cdot\int_{\rho}\Pr\left[\Bin(n-1,\rho)<k\right]\mu(\rho)+(1-\beta)\cdot\int_{\rho}\frac{\bE[\min\{\Bin(n-1,\rho),k\}]}{k}\mu(\rho)\nonumber\\
&=\min_{\beta\in[0,1]}\sup_{\rho\in(0,1]} \beta\cdot\Pr\left[\Bin(n-1,\rho)<k\right]+(1-\beta)\cdot \frac{\bE[\min\{\Bin(n-1,\rho),k\}]}{k}.
\end{align}
Note that $\OST$ is a special case of $\ST$, the final result follows as long as the value of \eqref{eqn:proof1098} equals the value of \eqref{eqn:STOST01}. Denote by $\lambda(\beta,\rho)$ the function:
$\lambda(\beta,\rho):=\beta\cdot\Pr\left[\Bin(n-1,\rho)<k\right]+(1-\beta)\cdot \frac{\bE[\min\{\Bin(n-1,\rho),k\}]}{k}$.
It suffices to show that \[\min_{\beta\in[0,1]}\sup_{\rho\in(0,1]}\lambda(\beta,\rho)=\sup_{\rho\in(0,1]}\min_{\beta\in[0,1]}\lambda(\beta,\rho).\]
Note that $\lambda(\beta,\rho)$ is linear in $\beta$, from Sion's minimax theorem, it only remains to show that $\lambda(\beta,\rho)$ is quasi-concave over $\rho\in(0,1]$, for each fixed $\beta\in[0,1]$. We now assume $\beta$ is fixed and we show $\lambda(\beta,\rho)$ is a unimodal function over $\rho$, which implies quasi-concavity.
Note that
\begin{eqnarray*}
\lambda(\beta, \rho)&=& \beta \sum_{s=0}^{k-1}   C_{n}^s \rho^s (1-\rho)^{n-s} + (1-\beta)\left\{\frac{1}{k} \sum_{s=1}^{k-1} s C_{n}^s \rho^s (1-\rho)^{n-s}
+\sum_{s=k}^{n} C_{n}^s \rho^s (1-\rho)^{n-s}
\right\} \\
%&=& \beta \sum_{s=0}^{k-1}   C_{n}^s \rho^s (1-\rho)^{n-s} + (1-\beta)\left\{\frac{1}{k} \sum_{s=1}^{k-1} s C_{n}^s \rho^s (1-\rho)^{n-s} + 1- \sum_{s=0}^{k-1} C_{n}^s \rho^s (1-\rho)^{n-s} \right\}  \\
& =& 1-\beta + (2\beta-1)  \sum_{s=0}^{k-1}   C_{n}^s \rho^s (1-\rho)^{n-s}
+ \frac{1-\beta}{k} \sum_{s=1}^{k-1} s C_{n}^s \rho^s (1-\rho)^{n-s}
\end{eqnarray*}
Then,
The derivative of $\lambda (\beta, \rho)$ over $\rho$ is
\begin{eqnarray*}
\frac{\partial}{\partial\rho}\lambda(\beta, \rho) & = & (2 \beta-1) \left\{-n (1-\rho)^{n-1} + \sum_{s=1}^{k-1} [s C_{n}^s \rho^{s-1} (1-\rho)^{n-s} - (n-s) C_{n}^s \rho^s (1-\rho)^{n-s-1}]   \right \} \\
 & & +\frac{1-\beta}{k} \sum_{s=1}^{k-1}\left\{ s^2 C_{n}^s \rho^{s-1} (1-\rho)^{n-s} -
s (n-s) C_{n}^s \rho^s (1-\rho)^{n-s-1}
\right\}
\end{eqnarray*}
Notice that
\begin{eqnarray*}
& & \sum_{s=1}^{k-1} [s C_{n}^s \rho^{s-1} (1-\rho)^{n-s} - (n-s) C_{n}^s \rho^s (1-\rho)^{n-s-1}]   \\
%&= & \sum_{s=0}^{k-2} (s+1) C_{n}^{s+1} \rho^{s} (1-\rho)^{n-s-1} -\sum_{s=1}^{k-1}(n-s) C_{n}^s \rho^s (1-\rho)^{n-s-1}\\
&=& n (1-\rho)^{n-1} - (n-k+1) C_{n}^{k-1} \rho^{k-1} (1-\rho)^{n-k}\\
&&+\sum_{s=1}^{k-2} [(s+1) C_{n}^{s+1} \rho^{s} (1-\rho)^{n-s-1}-(n-s) C_{n}^s \rho^s (1-\rho)^{n-s-1}]\\
&=& n (1-\rho)^{n-1} - (n-k+1) C_{n}^{k-1} \rho^{k-1} (1-\rho)^{n-k}
\end{eqnarray*}
where the last equality holds since
$(s+1) C_{n}^{s+1} =(n-s) C_{n}^s.$
Similarly,
\begin{eqnarray*}
& & \sum_{s=1}^{k-1}\left\{ s^2 C_{n}^s \rho^{s-1} (1-\rho)^{n-s} -
s (n-s) C_{n}^s \rho^s (1-\rho)^{n-s-1} \right\} \\
%&=& \sum_{s=0}^{k-2} (s+1)^2 C_{n}^{s+1} \rho^{s} (1-\rho)^{n-s-1} -\sum_{s=1}^{k-1} s (n-s) C_{n}^s \rho^s (1-\rho)^{n-s-1} \\
%&=& n (1-\rho)^{n-1} - (k-1)(n-k+1) C_{n}^{k-1} \rho^{k-1} (1-\rho)^{n-k} +\sum_{s=1}^{k-2} \left((s+1)^2 C_{n}^{s+1} - s (n-s) C_{n}^s \right ) \rho^s (1-\rho)^{n-s-1}\\
%&=& n (1-\rho)^{n-1} - (k-1)(n-k+1) C_{n}^{k-1} \rho^{k-1} (1-\rho)^{n-k} + \sum_{s=1}^{k-2} (s+1) C_{n}^{s+1} \rho^s (1-\rho)^{n-s-1}\\
&=& \sum_{s=0}^{k-2} (s+1) C_{n}^{s+1} \rho^s (1-\rho)^{n-s-1} -(k-1)(n-k+1) C_{n}^{k-1} \rho^{k-1} (1-\rho)^{n-k}
\end{eqnarray*}
Thus,
\begin{eqnarray*}
\frac{\partial}{\partial\rho}\lambda(\beta, \rho)& =&-(2\beta-1)(n-k+1) C_{n}^{k-1} \rho^{k-1} (1-\rho)^{n-k} \\
& & + \frac{1-\beta}{k} \sum_{s=0}^{k-2} (s+1) C_{n}^{s+1} \rho^s (1-\rho)^{n-s-1}
-\frac{1-\beta}{k}(k-1)(n-k+1) C_{n}^{k-1} \rho^{k-1} (1-\rho)^{n-k} \\
&= & -\left(2\beta-1 + \frac{1-\beta}{k}(k-1) \right)(n-k+1) C_{n}^{k-1} \rho^{k-1} (1-\rho)^{n-k}\\
&& + \frac{1-\beta}{k} \sum_{s=0}^{k-2} (s+1) C_{n}^{s+1} \rho^s (1-\rho)^{n-s-1}
\end{eqnarray*}
By dividing both sides by $\rho^{k-1} (1-\rho)^{n-k}$, we know that
$\frac{\partial}{\partial\rho}\lambda(\rho)=0$ is equivalent to
\begin{equation}\label{eqn:STOST02}
\frac{1-\beta}{k} \sum_{s=0}^{k-2} (s+1) C_{n}^{s+1} (\frac{\rho}{1-\rho})^{s-k+1} =
\left(2\beta-1 + \frac{1-\beta}{k}(k-1) \right)(n-k+1) C_{n}^{k-1}.
\end{equation}
Clearly, when $\beta=1$, \eqref{eqn:STOST02} does not hold, which implies that $\lambda(\beta,\rho)$ is either non-decreasing, or non-increasing, over $\rho$. When $\beta \in [0,1)$,
the function  $\frac{1-\beta}{k} \sum_{s=0}^{k-2} (s+1) C_{n}^{s+1} t^{s-k+1}$
is strictly decreasing in $t \in [0,1]$. Thus, the equation $\frac{\partial}{\partial\rho}\lambda(\rho)=0$ can have at most one solution is $[0,1]$, which implies that $\lambda(\beta,\rho)$ is unimodal in $[0,1]$.
\end{myproof}

\section{Missing Proofs for \Cref{sec:IID}}\label{sec:pfsec4}
\begin{myproof}[Proof of \Cref{thm:iidFramework}]
Denote by $\hat{\vG}$ the distributions such that the type distribution of each agent $i$ is a uniform distribution over $[0,1]$. Then, from definitions, we have that
\[\begin{aligned}
&\IIDLP^{\DP/\Proph}_{k,n}=\InnerLP^{\DP/\Proph}_{k,n}(\hat{\vG})\geq\inf_{\vG:G_{1j}=\cdots=G_{nj}\forall j}\InnerLP^{\DP/\Proph}_{k,n}(\vG)
%&\IIDLP^{\DP/\ExAnte}_{k,n}=\InnerLP^{\DP/\ExAnte}_{k,n}(\hat{\vG})\geq\inf_{\vG:G_{1j}=\cdots=G_{nj}\forall j}\InnerLP^{\DP/\ExAnte}_{k,n}(\vG)\\
\end{aligned}\]
Moreover, denoting by $\{\theta^*, \mathbf{x}^*, \mathbf{y}^*\}$ the optimal solution of $\IIDLP^{\DP/\Proph}_{k,n}$, it is easy to see that $\{\theta^*, \mathbf{x}^*, \mathbf{y}^*\}$ is a feasible solution to $\InnerLP^{\DP/\Proph}_{k,n}(\vG)$ for any $\vG$ satisfying $G_{1j}=\cdots=G_{nj}$ for all $j$. Thus, we know that
\[
\IIDLP^{\DP/\Proph}_{k,n}\leq\inf_{\vG:G_{1j}=\cdots=G_{nj}\forall j}\InnerLP^{\DP/\Proph}_{k,n}(\vG)
\]
which implies that
\[
\IIDLP^{\DP/\Proph}_{k,n}=\inf_{\vG:G_{1j}=\cdots=G_{nj}\forall j}\InnerLP^{\DP/\Proph}_{k,n}(\vG)
\]
Applying the same arguments to $\IIDLP^{\DP/\ExAnte}_{k,n},~\sup_{\tau}\InnerLP^{\OST(\tau)/\Proph}_{k,n}, ~\sup_{\tau}\InnerLP^{\OST(\tau)/\ExAnte}_{k,n}$,\\ $\IIDLP^{\ST/\Proph}_{k,n}$ and $\IIDLP^{\ST/\ExAnte}_{k,n}$, we complete our proof.
\end{myproof}

\begin{myproof}[Proof of \Cref{thm:iidST}]
For any $(\vx,\vy)\in\cP^k_n$, the objective value of $\IIDLP^{\DP/\ExAnte}_{k,n}$ will equal
\begin{align}
\theta
&=\inf_{q\in(0,1]}\frac{\sum_{i=1}^n\sum_{l=1}^k \min\{y^l_i,qx^l_i\}}{\min\{nq,k\}}  \nonumber \\
&=\min\left\{\inf_{q\in(0,k/n]}\frac{\sum_{i=1}^n\sum_{l=1}^k \min\{\frac{y^l_i}{q},x^l_i\}}{n},\inf_{q\in(k/n,1]}\frac{\sum_{i=1}^n\sum_{l=1}^k \min\{y^l_i,qx^l_i\}}{k}\right\} \nonumber
\\ &=\frac{\sum_{i=1}^n\sum_{l=1}^k \min\{y^l_i,\frac{k}{n}x^l_i\}}{k} \label{eqn:2678}
\end{align}
where we note that $k/n<1$ since we are assuming $n>k$.
Thus, the problem of $\IIDLP^{\DP/\ExAnte}_{k,n}$ is equivalent to maximizing expression~\eqref{eqn:2678} subject to $(\vx,\vy)\in\cP^k_n$.
From this it is easy to see that the optimal solution involves setting $y^l_i=\frac{k}{n}x^l_i$ for all $l$ and $i$, equivalent to a static threshold policy with $\tau=k/n$.
Therefore, we have
\begin{align} \label{eqn:binnk}
\IIDLP^{\DP/\ExAnte}_{k,n}=\frac{\frac{k}{n}\sum_i\sum_l x^l_i}{k}=\frac{\bE[\min\{\Bin(n,k/n),k\}]}{k}
\end{align}
where the final equality follows from~\eqref{eqn:lemAnalogues}.

We now show that the same expression results from analyzing $\sup_{\tau}\IIDLP^{\OST(\tau)/\Proph}_{k,n}$ or $\sup_{\tau}\IIDLP^{\OST(\tau)/\ExAnte}_{k,n}$.
For the first one, by~\eqref{eqn:RHSsimplificationOSTIID}, the objective value of $\IIDLP^{\OST(\tau)/\Proph}_{k,n}$ for any OST $\tau$ will equal
\begin{subequations}
\begin{align}
\theta
&=\inf_{q\in(0,1]}\frac{\min\{\tau,q\}\sum_{i=1}^n\sum_{l=1}^k x^l_i}{\bE[\min\{\Bin(n,q),k\}]} \nonumber
=\min\left\{
\inf_{q\in(0,\tau]}\frac{\sum_{i=1}^n\sum_{l=1}^k x^l_i}{\frac{\bE[\min\{\Bin(n,q),k\}]}{q}},
\inf_{q\in(\tau,1]}\frac{\tau\sum_{i=1}^n\sum_{l=1}^k x^l_i}{\bE[\min\{\Bin(n,q),k\}]}
\right\} \nonumber
\\ &=\min\left\{
\frac{\sum_{i=1}^n\sum_{l=1}^k x^l_i}{n},
\frac{\tau\sum_{i=1}^n\sum_{l=1}^k x^l_i}{k}
\right\} \label{eqn:5828}
\\ &=\frac{\bE[\min\{\Bin(n,\tau),k\}]}{\max\{\tau n,k\}} \label{eqn:7529}
\end{align}
\end{subequations}
where the first argument in~\eqref{eqn:5828} results because $\frac{\bE[\min\{\Bin(n,q),k\}]}{q}$ is maximized over $q\in(0,\tau]$ as $q$ approaches 0 from the positive side, and the final equality applies~\eqref{eqn:lemAnalogues} to both arguments after multiplying and dividing the first argument by $\tau$.
From this it is easy to see that the supremum of expression~\eqref{eqn:7529} over $\tau$ is achieved by setting $\tau=k/n$, since ratio $\frac{\bE[\min\{\Bin(n,\tau),k\}]}{\tau}$ is decreasing over $\tau\in[k/n,1]$. To see this, by noting that $\bE[\min\{\Bin(n,0),k\}]=0$, it is enough to show that $\bE[\min\{\Bin(n,\tau),k\}$ is concave over $\tau\in[0,1]$. We define the function $h(x)=\min\{x,k\}$. Then, we have
\[\begin{aligned}
\frac{\partial}{\partial\tau} \bE[\min\{\Bin(n,\tau),k\}%&=\sum_{j=0}^{k-1}(k-j)\cdot C_n^j\cdot(j\tau^{j-1}(1-\tau)^{n-j}-(n-j)\tau^j(1-\tau)^{n-j})\\
%&=n\cdot\sum_{j=0}^{k-2}(k-1-j)\cdot C_{n-1}^j\cdot\tau^j(1-\tau)^{n-1-j}-n\cdot\sum_{j=0}^{k-1}(k-j)\cdot C_{n-1}^j\cdot\tau^j(1-\tau)^{n-1-j}\\
&=n\cdot\bE_{X\sim\Bin(n-1,\tau)}[h(X+1)-h(X)]
\end{aligned}\]
which implies that
\[
\frac{\partial^2}{\partial\tau^2} \bE[\min\{\Bin(n,\tau),k\}=n(n-1)\cdot\bE_{X\sim\Bin(n-2,\tau)}[h(X+2)+h(X)-2h(X+1)]\leq0
\]
by noting that $h(X+2)+h(X)-2h(X+1)\leq0$ for any $X$. Thus, $\bE[\min\{\Bin(n,\tau),k\}$ is concave over $\tau\in[0,1]$ and $\frac{\bE[\min\{\Bin(n,\tau),k\}]}{\tau}$ is decreasing over $\tau\in[k/n,1]$.
Therefore, we have shown that $\sup_{\tau}\IIDLP^{\OST(\tau)/\Proph}_{k,n}$ is identical to~\eqref{eqn:binnk}.

Finally, the same argument can be made to show that $\sup_{\tau}\IIDLP^{\OST(\tau)/\ExAnte}_{k,n}$ is equal to~\eqref{eqn:binnk}, since the same expression~\eqref{eqn:7529} can be derived for $\IIDLP^{\OST(\tau)/\ExAnte}_{k,n}$.  This completes the proof of \Cref{thm:iidST}.
\end{myproof}

\begin{myproof}[Proof of \Cref{thm:iidSTNoBetter}]
Let $\tau$ be the unique value in $(\utau,\otau)$ at which
\begin{align} \label{eqn:iidSTconstruction}
\bE[\min\{\Bin(n,\tau),k\}]=\frac{\bE[\min\{\Bin(n,\utau),k\}]+\bE[\min\{\Bin(n,\otau),k\}]}{2}.
\end{align}
We show that for this value of $\tau$,~\eqref{eqn:iidSTNoBetter} holds for all choices of $q$.

First, if $q>\otau$, then all agents accepted by any static threshold $\tau$, $\utau$, or $\otau$ will have type $q$ or better.  In this case, using the fact that $\min\{y^l_i(\tau),q x^l_i(\tau)\}=y^l_i(\tau)=\tau x^l_i(\tau)$ (due to~\eqref{constr:innerDualSTIIDThreshold}, and an analogous argument can be made for $\utau,\otau$) and applying the second identity in~\eqref{eqn:lemAnalogues} with $i=n$, we see that~\eqref{eqn:iidSTNoBetter} is equivalent to $\bE[\min\{\Bin(n,\tau),k\}]\ge\frac{1}{2}(\bE[\min\{\Bin(n,\utau),k\}]+\bE[\min\{\Bin(n,\otau),k\}])$.
In other words, we need to check that $\tau$ accepts no fewer agents than the average of $\utau$ and $\otau$, which in fact holds as equality due to~\eqref{eqn:iidSTconstruction}.

Next, if $q\in(\tau,\otau]$, then all agents accepted by static thresholds $\tau,\utau$ still have type $q$ or better, while some of the agents accepted by $\otau$ may not.
Since $\tau$ accepts no fewer agents than the average of $\utau$ and $\otau$,~\eqref{eqn:iidSTNoBetter} must still be true.

The third case we consider is $q<\utau$.  Agents of types $q$ or better are accepted by all policies, so the probability of accepting any agent $i$ who has type $q$ or better is simply $q$ times the probability of having a slot available at that time.
Formally, using the fact that $\min\{y^l_i(\tau),q x^l_i(\tau)\}=q x^l_i(\tau)$ (due to~\eqref{constr:innerDualSTIIDThreshold}, and an analogous argument can be made for $\utau,\otau$) and applying the first identity in~\eqref{eqn:lemAnalogues}, we see that~\eqref{eqn:iidSTNoBetter} is equivalent to
\begin{align} \label{eqn:6215}
q\sum_{i=1}^n\Pr[\Bin(i-1,\tau)<k]\ge\frac{q}{2}\left(\sum_{i=1}^n\Pr[\Bin(i-1,\utau)<k]+\sum_{i=1}^n\Pr[\Bin(i-1,\otau)<k]\right).
\end{align}
In words, after canceling out the $q$'s, we need to prove that the total probability of having a slot available is higher for $\tau$, as compared to the average of $\utau$ and $\otau$.

Before proving~\eqref{eqn:6215}, we show that the final case $q\in(\utau,\tau]$ reduces to it.
Indeed, in the final case static threshold $\utau$ may not accept some agents who have type $q$ or better, which only decreases the RHS of~\eqref{eqn:6215}.
Therefore, establishing~\eqref{eqn:6215} would complete the proof.

To establish~\eqref{eqn:6215}, we observe that~\eqref{eqn:iidSTconstruction}, which holds by construction, is equivalent to
\begin{align} \label{eqn:8018}
\tau\sum_{i=1}^n\Pr[\Bin(i-1,\tau)<k]=\frac{1}{2}\left(\utau\sum_{i=1}^n\Pr[\Bin(i-1,\utau)<k]+\otau\sum_{i=1}^n\Pr[\Bin(i-1,\otau)<k]\right).
\end{align}
Recalling that $\utau<\otau$, we now argue two facts.
First, we immediately see that $\Pr[\Bin(i-1,\utau)<k]\ge\Pr[\Bin(i-1,\otau)<k]$ for all $i$.
Second, we argue that $\tau\le(\utau+\otau)/2$.
This is because the function $\bE[\min\{\Bin(n,\tau),k\}]$ is strictly increasing and concave in $\tau$---so if $\tau>(\utau+\otau)/2$, then Jensen's inequality would say
$
\bE[\min\{\Bin(n,\tau),k\}]>\bE[\min\{\Bin(n,\frac{\utau+\otau}{2}),k\}]\ge\frac{1}{2}(\bE[\min\{\Bin(n,\utau),k\}]+\bE[\min\{\Bin(n,\otau),k\}]),
$
which contradicts~\eqref{eqn:iidSTconstruction}.
From these facts we derive
\begin{align*}
\frac{\utau+\otau}{2}\sum_{i=1}^n\Pr[\Bin(i-1,\tau)<k]
&\ge\tau\sum_{i=1}^n\Pr[\Bin(i-1,\tau)<k]
\\ &=\frac{1}{4}\left(2\utau\sum_{i=1}^n\Pr[\Bin(i-1,\utau)<k]+2\otau\sum_{i=1}^n\Pr[\Bin(i-1,\otau)<k]\right)
\\ &\ge\frac{1}{4}(\utau+\otau)\left(\sum_{i=1}^n\Pr[\Bin(i-1,\utau)<k]+\sum_{i=1}^n\Pr[\Bin(i-1,\otau)<k]\right)
\end{align*}
where the equality holds by~\eqref{eqn:8018}, and the inequality holds by applying the rearrangement inequality with $\utau<\otau$, $\sum_{i=1}^n\Pr[\Bin(i-1,\utau)<k]\ge\sum_{i=1}^n\Pr[\Bin(i-1,\otau)<k]$ and then factoring.
This implies~\eqref{eqn:6215} and completes the proof of \Cref{thm:iidSTNoBetter}.
\end{myproof}

\begin{myproof}[Proof of \Cref{prop:DiscreteIID}]
Take an optimal solution $(\vx,\vy)$ to $\IIDLP_n(\cK)$ with objective value $\theta$.  We show that $(\vx,\vy)$ is a feasible solution to $\IIDLP_n((0,1])$ with objective value $(1-\eps)\theta$. It is enough to show that for each $q\in(0,1]$, it holds
$A_n(q,\vx, \vy)\geq(1-\eps)\theta B_n(q)$.
First, if $q\in(q_{j+1},q_j]$ for some $j\in[M-1]$, then we have
\begin{align*}
A_n(q,\vx,\vy)
&\ge A_n(q_{j+1},\vx,\vy) \ge \theta\cdot B_n(q_{j+1}) \ge (1-\eps)\theta B_n(q_j) \ge(1-\eps)\theta B_n(q)
\end{align*}
where the second inequality uses the feasibility of the solution $(\vx,\vy)$ to $\IIDLP_n(\cK)$ with objective value $\theta$, the third inequality uses condition~1 above, and the last inequality uses the monotonicity of $B_n(q)$ in $q$.
Meanwhile, if $q\in(0,q_M]$, then
\begin{align*}
A_n(q,\vx,\vy)
&\ge A_n(q_M,\vx,\vy)\cdot\frac{q}{q_M}\ge \theta B_n(q_M)\cdot\frac{q}{q_M}\ge \theta (1-\eps)n q_M\cdot\frac{q}{q_M}\ge \theta (1-\eps)B_n(q)
\end{align*}
where the first inequality uses the concavity of function $A_n$ in $q$, and the third inequality uses the fact that $B_n(q_M)\ge(1-\eps)n q_M$, which follows from condition 2 in the statement of the lemma.
The latter fact follows from
\begin{align*}
B_n(q_M)
&=n q_M-\bE[\max\{\Bin(n, q_M)-k,0\}]
\ge n q_M-\Pr[\Bin(n,q_M)>k] \\
&\ge n q_M-\Pr[\Pois(n q_M)>k]
\ge n q_M-n q_M\eps.
\end{align*}
Our proof is therefore completed.
\end{myproof}

\begin{myproof}[Proof of \Cref{lem:LPmon}]
We consider further relaxing LP \eqref{lp:monotone}.
It is clear to see that the expression in large parentheses at the LHS of \eqref{eqn:minOverKappaI}, 
\[
(1-(1-\kappa)^n)\theta-\kappa\sum_{i=I}^{n-1}(1-Y_i),
\]
is a concave function over $\kappa$, which implies that the maximum is attained by setting the first derivative to 0.
Letting
\begin{align*}
x_I:=\sum_{i=I}^{n-1}(1-Y_i) \qquad z_I:=\frac{x_I}{n\theta} && \forall I=0,\ldots,n-1
\end{align*}
we have that the maximum for constraint \eqref{eqn:minOverKappaI} for $I=0,\ldots,n-1$ is attained when
\begin{align*}
(1-\kappa)^{n-1}=z_I \qquad \kappa=1-z_I^{1/(n-1)}.
\end{align*}
Note that $Y_I=1+x_{I+1}-x_I=1+n\theta(z_{I+1}-z_I)$ for all $I=0,\ldots,n-1$, with $z_n=0$. Then, constraint \eqref{eqn:contrK} can be transformed into the following constraint over the $z_I$ variables for $I=0,\ldots,n-1$:
\[
z_1-z_0\le-\frac{1}{n\theta}\le z_2-z_1\le\cdots\le z_n-z_{n-1} \le0.
\]
Moreover, the constraints \eqref{eqn:minOverKappaI} are satisfied if for all $I=0,\ldots,n-1$,
\[
 (1-z_I^{n/(n-1)})\theta-\left(1-z_I^{1/(n-1)}\right)x_I \leq 1+x_{I+1}-x_I,
\]
which is equivalent to
\[
    1-\frac{1}{\theta}+(n-1)z_I^{n/(n-1)} \leq n z_{I+1}.
\]
As a result, LP \eqref{lp:monotone} can again be characterized as the following optimization problem over $z$-variables:
\begin{subequations}\label{lp:newmonotone}
\begin{align}
\max\ &\theta& \nonumber
\\ \text{s.t. \ \ }&(n-1)z_I^{n/(n-1)}\le nz_{I+1}+\frac{1}{\theta}-1 &\forall I=0,\ldots,n-1 \label{constraintforz}
\\ &z_n=0 &\nonumber
\\ &z_1-z_0\le-\frac{1}{n\theta}\le z_2-z_1\le\cdots\le z_n-z_{n-1} \le0 & \label{eqn:contoverz}
\end{align}
\end{subequations}
Finally, combining the constraint $z_1-z_0\le-\frac{1}{n\theta}$ from \eqref{eqn:contoverz} and the constraint \eqref{constraintforz} with $I=0$, we get
\[
(n-1)z_0^{n/(n-1)}\leq nz_{1}+\frac{1}{\theta}-1\leq nz_{0}-1\qquad\Longrightarrow\qquad(n-1)z_0^{n/(n-1)}-nz_0\leq -1.
\]
Note that expression $(n-1)z_0^{n/(n-1)}-nz_0$ as a function over $z_0\ge0$ is always at least -1, with equality achieved only when $z_0=1$.  Therefore, we can add the constraint $z_0=1$ to $\LPmon_n$ without changing the objective value. Together with $z_0=1$ and $z_n=0$, the constraint \eqref{eqn:contoverz} can be further relaxed into $z_i\in[0,1]$ for $i=1,\ldots,n-1$, which establishes $\LPre_n$ as a relaxation of LP \eqref{lp:newmonotone}. Thus, our proof is completed.
\end{myproof}

\begin{myproof}[Proof of \Cref{Ostructurelemma}]
Denote by $\{\theta, z_I\}_{I=0}^n$ as an optimal solution of $\LPre_n$ and denote by $I_1$ the smallest index such that
\[
z_{I_1+1}>\max\{\frac{n-1}{n}z_{I_1}^{n/(n-1)}-\frac{1}{n\theta}+\frac{1}{n}, 0\}
\]
Then, we construct another solution $\{\hat{\theta}, \hat{z}_I\}_{I=0}^n$ such that
\[
\hat{\theta}=\theta,~~\hat{z}_{I_1+1}=\max\{\frac{n-1}{n}z_{I_1}^{n/(n-1)}-\frac{1}{n\theta}+\frac{1}{n}, 0\},~~\hat{z}_I=z_I\text{~for~}I=0,\ldots,I_1, I_1+2,\ldots,n
\]
Clearly, since $\hat{z}_{I_1+1}\leq z_{I_1+1}$, we must have $\{\hat{\theta}, \hat{z}_I\}_{I=0}^n$ as an optimal solution to $\LPre_n$. Every time we repeat the above construction procedure, the value of $I_1$ will be increased by at least $1$. Thus, after a finite number of steps, we obtain an optimal solution $\{\hat{\theta}, \hat{z}_I\}_{I=0}^n$ such that
\begin{equation}\label{iteratez}
\hat{z}_{I+1}=\max\{\frac{n-1}{n}\hat{z}_{I}^{n/(n-1)}-\frac{1}{n\hat{\theta}}+\frac{1}{n},0\},~~\forall I=0,\ldots,n-1
\end{equation}
We regard $\hat{z}_I(\theta)$ as a function of $\theta$, for each $I=0,\ldots, n-1$, where $\hat{z}_I(\theta)$ is computed iteratively from \eqref{iteratez}. Note that the RHS of \eqref{iteratez} is non-decreasing over $\theta$. We must have $\hat{z}_I(\theta)$ is a non-decreasing function over $\theta$. If there exists an $I_2\leq n-1$ such that $\frac{n-1}{n}\hat{z}_{I_2}(\theta)^{n/(n-1)}-\frac{1}{n\theta}+\frac{1}{n}<0$, we can always increase the value of $\theta$ by $\eps>0$ such that it still holds $\frac{n-1}{n}\hat{z}_{I_2}(\theta+\eps)^{n/(n-1)}-\frac{1}{n(\theta+\eps)}+\frac{1}{n}<0$, and $\{\theta+\eps, \hat{z}_I(\theta+\eps)\}$ is still feasible to $\LPre_n$. Thus, in order for $\hat{\theta}$ to be optimal, we must have $\hat{z}_{I+1}=\frac{n-1}{n}\hat{z}_{I}^{n/(n-1)}-\frac{1}{n\hat{\theta}}+\frac{1}{n}$ for all $I=0,\ldots,n-1$, which completes our proof. 
\end{myproof}
\begin{myproof}[Proof of \Cref{lem:LPequivalent}]
Denote $\{\theta, z_I\}_{I=0}^n$ as the solution constructed in \Cref{Ostructurelemma}. Then, we denote
\[
Y_I=1+n\theta(z_{I+1}-z_I)\text{~for~}I=0,1,\ldots,n-1.
\]
Clearly, it holds that
\[
\max_{\kappa\in(0,1]}\left((1-(1-\kappa)^n)\theta-\kappa\sum_{i=I}^{n-1}(1-Y_i)-Y_I\right)=0\text{~for~}I=0,1,\ldots,n-1.
\]
We also denote $y_i=Y_i-Y_{i-1}$ for $i=1,\ldots, n-1$. Note that $\sum_{i=1}^{n-1}y_i=Y_{n-1}\leq1$, we also denote $y_n=1-\sum_{i=1}^{n-1}y_i$. Then, we have
\begin{equation}\label{eqn:maximumequal0}
\max_{\kappa\in(0,1]}\left((1-(1-\kappa)^n)\theta-\kappa\sum_{i=I+1}^{n}(1-\sum_{i'=1}^{i-1}y_{i'})-\sum_{i=1}^I y_i\right)=0\text{~for~}I=0,1,\ldots,n-1.
\end{equation}
In order to show that $\{\theta, y_i\}_{i=1}^n$ is a feasible solution to $\IIDLP^{\DP/\Proph}_{1,n}$, it suffices to show that 
\begin{equation}\label{eqn:supermodular}
(1-(1-\kappa)^n)\theta\leq \sum_{i\in S}y_i+\kappa\cdot\sum_{i\in[n]\setminus S}(1-\sum_{i'=1}^{i-1}y_{i'}),~\text{for~any~}\kappa\in[0,1]\text{~and~}S\subset [n]
\end{equation}
where the constraint $y_i\geq0$ also follows from \eqref{eqn:supermodular} by setting $\mathcal{S}=\{j\}$ and $\kappa=0$.

We now proceed to prove \eqref{eqn:supermodular}. 
For any fixed $S$, clearly, the left hand side of constraint \eqref{eqn:supermodular} is a concave function over $\kappa$. Thus, after we maximize over $\kappa$ in \eqref{eqn:supermodular}, we get that $\{\theta, y_i\}_{i=1}^n$ is a feasible solution to $\IIDLP^{\DP/\Proph}_{1,n}$ if 
\begin{equation}\label{eqn:supermodularfinal}
f(S)\leq0,~~\forall S\subset[n]
\end{equation}
where $f(S)$ is a set function defined for any subset $S\subset[n]$ as follows
\begin{equation}\label{eqn:functionsupermodular}
    f(S):=1+(n-1)\cdot\left(\frac{\sum_{i\in[n]\setminus S}\alpha_i}{n\theta}\right)^{\frac{n}{n-1}}-\frac{\sum_{i\in[n]\setminus S}\alpha_i}{\theta}-\frac{\sum_{i\in S}y_i}{\theta},~\text{for~any~}S\subset[n]
\end{equation}
and we denote $1-\sum_{i'=1}^{i-1}y_{i'}$ by $\alpha_i$ for notation brevity. Note that from \eqref{eqn:maximumequal0} by setting $\kappa=0$, we have $\sum_{i=1}^Iy_i\geq0$ for each $I=0,1,\ldots,n-1$. Also, we have \[
\sum_{i=1}^Iy_i=Y_I-Y_0=1+n\theta(z_{I+1}-z_I)\leq1,~\forall I=0,1,\ldots,n-1
\]
by noting that $z_I$ is a decreasing sequence in $I$. Thus, we claim that $\alpha_i\in[0,1]$ for each $i\in[n]$.

The condition \eqref{eqn:maximumequal0} can also be expressed via the function $f(\cdot)$. We denote by $E_i=\{1,2,\ldots,i\}$ for each $i\in[n]$. Note that the left hand side of \eqref{eqn:maximumequal0} is a concave function over $\kappa$. Then after maximizing over $\kappa$ in \eqref{eqn:maximumequal0}, we have that
\begin{equation}\label{eqn:equal0overE}
f(E_i)=0\text{~for~}i=1,\ldots,n-1\text{~and~}f(E_n)\leq0
\end{equation}
We now use the condition \eqref{eqn:equal0overE} to prove \eqref{eqn:supermodularfinal}. A key step is to show the set function $f(\cdot)$ to be supermodular.
This allows us to ultimately show that it is maximized when $S$ takes the form of an interval $\{1,\ldots,I\}$, for which we already knew by the construction in \eqref{eqn:equal0overE}. 

Note that we have 
\[
f(S)=g(\sum_{i\in S}\alpha_i)-\frac{\sum_{i\in[n]\setminus S}\alpha_i}{\theta}-\frac{\sum_{i\in S}y_i}{\theta}\] 
where
\[
g(x)=1+(n-1)\cdot\left(\frac{\sum_{i\in[n]}\alpha_i-x}{n\theta}\right)^{\frac{n}{n-1}}
\]
It is clear to see that $g(x)$ is a convex function over $x$. Then, it is well-known (e.g. Lemma 2.6.2 in \cite{topkis2011supermodularity}) that $f(S)$ is a supermodular function.

%The supermodularity is formally proved in the following claim, where the proof can be found at the end of the proof of \Cref{lem:LPequivalent}. 
%\begin{claim}\label{claim:supermodular}
%For fixed $\{\theta, y_i, \alpha_i\}$, the function $f(S)$ is a supermodular function of $S$, i.e., 
%\[
%f(S_1\cup S_2)+f(S_1\cap S_2)\geq f(S_1)+f(S_2)
%\]
%for any $S_1, S_2\subset\mathcal{N}$.
%\end{claim}
For any set $S\subset\mathcal{N}$, we assume without loss of generality that the elements in $S$ are sorted in an increasing order. We denote $S(i)$ as the $i$-th element of $S$ and denote by $\sigma(S)$ the number of $i$ such that $S(i+1)>S(i)+1$. Then, for any $k$, we denote
\[
T_k=\{S: \sigma(S)\leq k \}
\]
Clearly, we have $T_0=\{E_1,\ldots,E_n\}$, which implies that $\max_{S\in T_0}f(S)\leq0$. Now we will prove \eqref{eqn:supermodularfinal} by induction. Suppose that there exists an integer $l$ such that 
\[
\max_{S\in T_l}f(S)\leq0
\]
For any set $\hat{S}\subset[n]$ such that $\sigma(\hat{S})=l+1$, we denote $\hat{i}$ as the largest index such that $\hat{S}(\hat{i}+1)>\hat{S}(\hat{i})+1$, and we denote $\hat{S}(\hat{j})$ as the last element of the set $\hat{S}$. We denote $\hat{T}=E_{\hat{S}(i)}$. Clearly, it holds that 
\[
\hat{T}\cup\hat{S}=E_{\hat{S}(\hat{j})}\text{~and~}\sigma(\hat{T}\cap\hat{S})=l
\]
From the supermodularity of $f(S)$, we have
\[
f(\hat{T})+f(\hat{S})\leq f(\hat{T}\cup\hat{S})+f(\hat{T}\cap\hat{S})
\]
From the induction hypothesis, we know $f(\hat{T}\cap\hat{S})\leq 0$. Also, from \eqref{eqn:equal0overE}, we know $f(\hat{T}\cup\hat{S})=f(E_{\hat{S}(\hat{j})})\leq0$ and $f(\hat{T})=f(E_{\hat{S}(\hat{i})})=0$ since $\hat{S}(\hat{i})\leq n-1$. Thus, we concludes that $f(\hat{S})=0$, which implies that
\[
\max_{S\subset T_{l+1}}f(S)\leq0
\]
From the induction, we know that 
\[
\max_{S\subset T_n}f(S)=\max_{S\subset[n]}f(S)\leq0
\]
which completes our proof.
\end{myproof}

%\begin{myproof}[Proof of \Cref{claim:supermodular}]
%From \citet{oxley2006matroid}, an equivalent condition for $f(\cdot)$ to be supermodular is that for any $S_1\subset S_2\subset\mathcal{N}$ and any $j\in\mathcal{N}\backslash S_2$, it holds that
%\begin{equation}\label{eqn:supermodularproof1}
%f(S_2\cup\{j\})-f(S_2)\geq f(S_1\cup\{j\})-f(S_1)
%\end{equation}
%Note that \eqref{eqn:supermodularproof1} is equivalent to
%\[
%(\beta_2-\alpha_j)^{\frac{n}{n-1}}-\beta_2^{\frac{n}{n-1}}\geq (\beta_1-\alpha_j)^{\frac{n}{n-1}}-\beta_1^{\frac{n}{n-1}}
%\]
%where $\beta_1=\sum_{i\in\mathcal{N}\backslash S_1}\alpha_i$ and $\beta_2=\sum_{i\in\mathcal{N}\backslash S_2}\alpha_i$. Denote by $g(x)$ the function
%\[
%g(x):=(x-\alpha_j)^{\frac{n}{n-1}}-x^{\frac{n}{n-1}}.
%\]
%We have
%\[
%g'(x)=\frac{n}{n-1}\cdot((x-\alpha_j)^{\frac{1}{n-1}}-x^{\frac{1}{n-1}})\leq0,~~\text{when~}x\geq\alpha_j
%\]
%Note that $S_1\subset S_2$ implies that $\beta_1\geq\beta_2$, we must have $g(\beta_2)\geq g(\beta_1)$, which completes our proof.
%\end{myproof}

\begin{myproof}[Proof of \Cref{Incnlemma}]
We denote $\{\theta, z_I\}_{I=0}^{2n}$ as an optimal solution to $\LPre_{2n}$. We construct a feasible solution to $\LPre_n$, denoted by $\{\hat{\theta}, \hat{z}_I\}_{I=0}^n$, To be specific, we set
\[
\hat{\theta}=\theta, ~~\hat{z}_I=z_{2I}\text{~for~}I=0,1,\dots,n
\]
It only remains to show that
\[
\hat{z}_{I+1}=z_{2I+2}\geq \frac{n-1}{n}z_{2I}^{n/(n-1)}-\frac{1}{n\theta}+\frac{1}{n},~~\forall I=0,\ldots, n-1
\]
Note that we have
\[
z_{2I+2}\geq \frac{(2n-1)\left[\frac{(2n-1)z_{2I}^{\frac{2n}{2n-1}}-\frac{1}{\theta}+1}{2n} \right]^{\frac{2n}{2n-1}}-\frac{1}{\theta}+1}{2n}
\]
Denote by $\alpha=\frac{1}{\theta}-1$. It only remains to show that
\[
f(\alpha)=(2n-1)\left[\frac{(2n-1)z_{2I}^{\frac{2n}{2n-1}}-\alpha}{2n}\right]^{\frac{2n}{2n-1}}-2(n-1)z_{2I}^{\frac{n}{n-1}}+\alpha\geq0
\]
Note that
\[
f'(\alpha)=1-\left[ \frac{(2n-1)z_{2I}^{\frac{2n}{2n-1}}-\alpha}{2n} \right]^{\frac{1}{2n-1}}
\]
Further note that
\[
1\geq z_{2I+1}\geq \frac{(2n-1)z_{2I}^{\frac{2n}{2n-1}}-\alpha}{2n}
\]
It holds that $f'(\alpha)\geq0$. Thus, in order to show that $f(\alpha)\geq0$, it suffices to show that $f(0)\geq0$, i.e.,
\[
\left[\frac{(2n-1)z_{2I}^{\frac{2n}{2n-1}}}{2n} \right]^{\frac{2n}{2n-1}}\geq\frac{2n-2}{2n-1}\cdot z_{2I}^{\frac{n}{n-1}}
\]
which is equivalent to showing
\[
(\frac{2n-1}{2n})^{\frac{2n}{2n-1}}\geq\frac{2n-2}{2n-1}\cdot z_{2I}^{\frac{n}{(n-1)(2n-1)^2}}
\]
Thus, it only remains to show that
\[
(1-\frac{1}{2n})^{2n}\geq(1-\frac{1}{2n-1})^{2n-1}\cdot z_{2I}^{\frac{n}{(n-1)(2n-1)}}
\]
The above inequality holds by noting that $z_{2I}\in[0,1]$ and $(1-\frac{1}{2n})^{2n}\geq(1-\frac{1}{2n-1})^{2n-1}$, which completes our proof.
\end{myproof}
\begin{myproof}[Proof of \Cref{theorem:limit}]
We denote function
\[
f(x)=x(\ln x-1)\text{~~and~~}f_n(x)=(n-1)\cdot x^{n/(n-1)}-nx\text{~for~each~}n
\]
Then, for each $n$, we define a sequence of values $\{H_{n,i}\}_{i=0}^n$ such that
\[
H_{n,0}=1,~ H_{n,i}=H_{n,i-1}+\frac{1}{n}\cdot (f_n(H_{n,i-1})-\frac{1}{\theta_n}+1)\text{~for~}i=1,\ldots,n
\]
where $\theta_n$ is selected such that $H_{n,n}=0$. Clearly, from \Cref{Ostructurelemma}, we know that $\{\theta_n, H_{n,i}\}_{i=0}^n$ is an optimal solution to $\LPre_n$. Then, from \Cref{lem:LPequivalent}, we know that $\theta_n=\IIDLP^{\DP/\Proph}_{1,n}$. Moreover, \Cref{Incnlemma} implies that $\inf_{n}\theta_n=\liminf_{n\rightarrow\infty}\theta_n$. Thus, it only remains to show that 
\[
\lim_{n\rightarrow\infty}\theta_n=\theta^*
\]
The remaining part mainly follows the proof of Lemma 6.2 in \citet{kertz1986stop}. Here, we include the whole proof for completeness.
For any $i=1,\ldots,n$, it holds that
\[
H_{n,i}-H_{n,i-1}=\frac{1}{n}\cdot(f(H_{n,i-1})-\frac{1}{\theta^*}+1)+\frac{1}{n}\cdot(f_n(H_{n,i-1})-f(H_{n,i-1}))+\frac{1}{n}\cdot(\frac{1}{\theta^*}-\frac{1}{\theta_n})
\]
which implies that
\[
\frac{1}{n}\cdot\frac{ \frac{1}{\theta^*}-\frac{1}{\theta_n} }{ f(H_{n,i-1})-\frac{1}{\theta^*}+1 }=\frac{H_{n,i-1}-H_{n,i}}{\frac{1}{\theta^*}-1-f(H_{n,i-1})}-\frac{1}{n}-\frac{1}{n}\cdot\frac{f_n(H_{n,i-1})-f(H_{n,i-1})}{f(H_{n,i-1})-\frac{1}{\theta^*}+1}
\]
Note that $f(H_{n,i-1})-\frac{1}{\theta^*}+1\in[-\frac{1}{\theta^*}, 1-\frac{1}{\theta^*}]$.\\
Case I. If $\theta^*\leq\theta_n$, we have
\begin{equation}\label{eqn:summH1}
\begin{aligned}
\frac{\theta^*}{n}\cdot |\frac{1}{\theta^*}-\frac{1}{\theta_n}|&\leq -\frac{1}{n}\cdot\frac{ \frac{1}{\theta^*}-\frac{1}{\theta_n} }{ f(H_{n,i-1})-\frac{1}{\theta^*}+1 } =-\frac{H_{n,i-1}-H_{n,i}}{\frac{1}{\theta^*}-1-f(H_{n,i-1})}+\frac{1}{n}+\frac{1}{n}\cdot\frac{f_n(H_{n,i-1})-f(H_{n,i-1})}{f(H_{n,i-1})-\frac{1}{\theta^*}+1} \\
&\leq -\frac{H_{n,i-1}-H_{n,i}}{\frac{1}{\theta^*}-1-f(H_{n,i-1})}+\frac{1}{n}+\left|\frac{1}{n}\cdot\frac{f_n(H_{n,i-1})-f(H_{n,i-1})}{f(H_{n,i-1})-\frac{1}{\theta^*}+1}  \right|\\
&\leq -\frac{H_{n,i-1}-H_{n,i}}{\frac{1}{\theta^*}-1-f(H_{n,i-1})}+\frac{1}{n}+\frac{\frac{1}{\theta^*}-1}{n}\cdot\|f_n-f\|_{\infty} 
\end{aligned}
\end{equation}
Sum over both sides of \eqref{eqn:summH1} for $i=1,\ldots,n$, we have
\[
\theta^*\cdot|\frac{1}{\theta^*}-\frac{1}{\theta_n}|\leq-\sum_{i=1}^n\frac{H_{n,i-1}-H_{n,i}}{\frac{1}{\theta^*}-1-f(H_{n,i-1})}+1+(\frac{1}{\theta^*}-1)\cdot\|f_n-f\|_{\infty}
\]
Case II. If $\theta^*>\theta_n$, we have
\[\begin{aligned}
\frac{\theta^*}{n}\cdot |\frac{1}{\theta^*}-\frac{1}{\theta_n}|&\leq \frac{1}{n}\cdot\frac{ \frac{1}{\theta^*}-\frac{1}{\theta_n} }{ f(H_{n,i-1})-\frac{1}{\theta^*}+1 }=\frac{H_{n,i-1}-H_{n,i}}{\frac{1}{\theta^*}-1-f(H_{n,i-1})}-\frac{1}{n}-\frac{1}{n}\cdot\frac{f_n(H_{n,i-1})-f(H_{n,i-1})}{f(H_{n,i-1})-\frac{1}{\theta^*}+1} \\
&\leq \frac{H_{n,i-1}-H_{n,i}}{\frac{1}{\theta^*}-1-f(H_{n,i-1})}-\frac{1}{n}+\frac{1}{n\theta^*}\cdot\|f_n-f\|_{\infty} 
\end{aligned}\]
Sum over both sides for $i=1,\ldots,n$, we have
\[
\theta^*\cdot|\frac{1}{\theta^*}-\frac{1}{\theta_n}|\leq\sum_{i=1}^n\frac{H_{n,i-1}-H_{n,i}}{\frac{1}{\theta^*}-1-f(H_{n,i-1})}-1+\frac{1}{\theta^*}\cdot\|f_n-f\|_{\infty}
\]
Thus, on both cases, we have that
\[
\theta^*\cdot|\frac{1}{\theta^*}-\frac{1}{\theta_n}|\leq \left| 1-\sum_{i=1}^n\frac{H_{n,i-1}-H_{n,i}}{\frac{1}{\theta^*}-1-f(H_{n,i-1})} \right| +\frac{1}{\theta^*}\cdot\|f_n-f\|_{\infty} 
\]
Further note that we have $|H_{n,i-1}-H_{n,i}|\leq \frac{1}{n\theta_n}$, which implies
\[
\lim_{n\rightarrow\infty}\sum_{i=1}^n\frac{H_{n,i-1}-H_{n,i}}{\frac{1}{\theta^*}-1-f(H_{n,i-1})}=\int_{h=0}^1\frac{dh}{\frac{1}{\theta^*}-1-f(h)}=1.
\]
Also, note that $\|f_n-f\|_{\infty}\leq\frac{1}{en}$, then we must have 
\[
\lim_{n\rightarrow\infty}\theta_n=\theta^*
\]
which completes our proof.
\end{myproof}

\end{APPENDICES}

\end{document}